\documentclass[reprint, amsmath,amssymb,aps,prx,nofootinbib]{revtex4-2}

\usepackage{bm}
\usepackage{color}
\usepackage{graphicx}
\usepackage{hyperref}
\usepackage{braket}
\usepackage{bbm}
\usepackage{tikz}
\usepackage{nicefrac}
\usepackage{comment}
\usepackage{enumerate}
\usepackage{empheq}

\usepackage{stmaryrd}

\newcommand{\be}{\begin{equation}}
\newcommand{\ee}{\end{equation}}
\newcommand{\ba}{\begin{aligned}}
\newcommand{\ea}{\end{aligned}}

\newcommand{\tr}{\mathrm{tr}}

\DeclareMathOperator{\sgn}{sgn}

\usepackage[normalem]{ulem}

\begin{document}
\title{
Kicking Quantum Fisher Information out of Equilibrium 
}
\author{Florent Ferro}
\affiliation{%
 Universit\'e Paris-Saclay, CNRS, LPTMS, 91405, Orsay, France%
}
\author{Maurizio Fagotti}
\affiliation{%
 Universit\'e Paris-Saclay, CNRS, LPTMS, 91405, Orsay, France%
}
\begin{abstract}
Quantum Fisher Information (QFI) is a ubiquitous quantity with applications ranging from quantum metrology and resource theories to condensed matter physics. In equilibrium local quantum many-body systems, the QFI of a subsystem with respect to an extensive observable is typically proportional to the subsystem's volume. 
Specifically, in large subsystems at equilibrium, the QFI per unit volume squared becomes negligible.
We reveal a natural mechanism that amplifies the QFI in a quantum spin chain with a zero-temperature ordered phase. At zero or sufficiently low temperatures, a transient localized 
perturbation enhances the QFI, causing it to scale quadratically with the subsystem’s length. Furthermore, this enhancement can be controlled through more general localized kicking protocols.  
We also revisit the behavior of the quantum Fisher information after a global quench in the thermodynamic limit, focusing on the generation of localized---confined within compact subsystems---multipartite entanglement. We show that the density of localized multipartite entanglement approaches zero at late times, but there is an optimal time frame proportional to the  length in which subsystems fall into macroscopic quantum states.
We test our predictions against numerical data obtained using a novel technique, based on a remarkable identity between quantum Fisher information and Wigner-Yanase-Dyson skew information, that allows one to compute the quantum Fisher information with respect to the order parameter in noninteracting spin chains. 

\end{abstract}
\maketitle

\tableofcontents

\section{Introduction}
Quantum information science continues to provide innovative tools for exploring the properties of quantum many-body systems. Among these, entanglement measures such as the von Neumann entropy~\cite{Bennett1996Concentrating} and negativity~\cite{Eisert1999A} have become central to the study of extended quantum systems. Originating within the framework of quantum resource theory---specifically, the theory of quantum entanglement~\cite{Plenio2014An}---these measures hold significance that extends far beyond their initial role in quantifying entanglement. Their utility now spans a wide range of applications, offering deep insights into topological properties and critical phenomena in many-body physics both in and out of equilibrium---see, e.g., Refs~\cite{Latorre2004Ground,Kitaev2006Topological,Hastings2007An,Amico2008Entanglement,Caraglio2008Entanglement,Cardy2008Form,Calabrese2009Entanglement,Casini2009Entanglement, Ercolessi2011Essential, Calabrese2012Entanglement,Eisler2015On,Alba2017Entanglement,Alba2019Entanglement,Bertini2022Growth,Dalmonte2022Entanglement,Maric2024Entanglement}.

Theoretical condensed matter physics has also seen growing interest in quantum resources beyond entanglement~\cite{Chitambar2019Quantum}, such as magic, asymmetry, and quantum coherence. From such a general perspective, it stands out the strong message of Ref.~\cite{Tan2021Fisher}: Fisher information and its quantum counterpart identify resourceful quantum states across a wide range of quantum resource theories. 
Quantum Fisher information (QFI) plays a fundamental role in quantum metrology, where it defines a Riemannian metric that quantifies the distinguishability of quantum states~\cite{PhysRevLett.72.3439}.  It sets the  precision limit in parameter estimation via the Cram\'er-Rao bound \cite{cramer_rao}. 
Beyond metrology, QFI per unit volume, when computed with respect to an extensive operator, serves as a lower bound on the entanglement depth, 
making it a witness of multipartite entanglement~\cite{Pezze2009Entanglement,Hyllus2012Fisher,Toth2012Multipartite,Froewis2012Measures}. Furthermore, QFI is closely related to other information measures, such as the Wigner-Yanase skew information (WYI)~\cite{Wigner1963INFORMATION}, with which it is bounded above and below by a constant factor. 
Notably, Ref.~\cite{PhysRevLett.127.260501} introduced a sequence of lower bounds that converge to the QFI and make it experimentally accessible.
We also remark that the QFI was  identified (up to a constant) with the convex roof of the variance~\cite{Toth2013Extremal,yu2013quantum}, offering an intuitive explanation of why it is equivalent to a  quantity dubbed ``quantum variance''~\cite{Frerot2016Quantum}. 

The thermodynamic limit is not typically considered a fundamental aspect of quantum information theory, but it holds particular importance in the study of many-body systems, in which, strictly speaking, only subsystems are accessible. 
While new phenomena, such as spontaneous symmetry breaking, emerge in the thermodynamic limit, other features may become trivial. For instance, multipartite entangled states like those proposed by Greenberger, Horne, and Zeilinger cannot be distinguished from separable states when examining only subsystems. 
In this respect, we emphasize that 
Ref.~\cite{Frerot2016Quantum} provided evidence that the quantum variance of an extensive observable 
remains proportional to the subsystem's volume even at thermal phase transitions. This behavior contrasts with the anomalous scaling of classical fluctuations, which are characterized by nontrivial critical exponents.   As we will confirm in the following, a similar result emerges in quantum spin chains at low temperatures in the thermodynamic limit, particularly in systems where a symmetry is spontaneously broken at zero temperature. While the order-parameter correlation length grows indefinitely as the temperature decreases, the subsystem QFI per unit volume remains bounded.
To summarize, the quantum properties of subsystems in equilibrium many-body systems with local interactions are essentially determined by a collection of underlying microscopic quantum effects. 
This raises an intriguing question: How far from equilibrium must a system be driven to manifest more interesting quantum features at the level of subsystems, such as the quantum macroscopicity firstly envisioned by Leggett~\cite{Leggett1980Macroscopic}?

We are going to show that the answer is \emph{just a bit}. 
We consider the effect of a localised perturbation that can be represented by the action of a unitary operator with finite support on the equilibrium state. After the perturbation the state is let to evolve with the same Hamiltonian with respect to which it was originally at equilibrium. Since the quantum Fisher information is bounded from above by four times the variance, a large QFI can be found only in states with a large correlation length. The natural settings for a QFI enhancement are therefore  phase transitions, in which even at equilibrium the system is prone to develop large fluctuations. We focus on quantum spin chains, in which discrete symmetries can be broken at zero temperature. Refs~\cite{Zauner2015Time,Eisler2020Front,Delfino2022Space} showed that localised perturbations can have strong effects on the expectation values of local observables in the presence of spontaneous symmetry breaking. Specifically, order parameters remain affected by the perturbation at the Euler scale in which the distance from the perturbation scales with time. Ref.~\cite{Eisler2020Front} investigated also the behavior of two-point functions, from which one can readily infer that the order-parameter connected correlations remain $\sim 1$ at distances proportional to the time. The question has remained open as to whether or not such breakdown of clustering properties at Euler scales is a purely classical phenomenon. And, if not, how large is the quantum contribution. We fill this gap and show that the subsystem quantum Fisher information is sensitive to generic unitary transformations with finite support. Especially in the field of chaos, transient perturbations on top of a more regular unitary time evolution are known as kicking~\cite{Haake2018Quantum,Prosen1998Time,Prosen2002General}. We thus propose localized kicking in systems sufficiently close to ordered phases as a mechanism to enhance the quantum Fisher information of subsystems. In addition, we will show that even sparsely distributed localized kicking boosts the QFI, which provides a compelling reason to revisit the nonequilibrium time evolution after a quantum quench with translational symmetry. We will provide evidence that even after a global quench there is a time interval proportional to the  length in which the subsystem QFI becomes proportional to the volume squared, provided that the length does not exceed a threshold inversely proportional to the density of excitations. 

Our investigation combines both analytical and numerical approaches. The analytical predictions are mainly derived using a semiclassical theory of excitations in ferromagnetic phases. 
On the numerical side, we take advantage of the Gaussian structure of the states under investigation to efficiently compute the Wigner-Yanase-Dyson skew information. We then estimate the quantum Fisher information using a novel identity that we establish, which, to the best of our knowledge, has remained unnoticed until now. This identity expresses the QFI as an average of the Wigner-Yanase-Dyson skew information over a complex parameter.

\subsection{Organization of the paper}
The remainder of the paper is structured into six largely self-contained sections, followed by the conclusions. Readers interested in a specific topic are encouraged to navigate directly to the relevant section.

\noindent The first two sections are mainly introductory, but there are insights that could interest also the expert reader:

\begin{description}
\item[Section~\ref{s:QFI}] This is a gentle introduction to the quantum Fisher information and the Wigner-Yanase-Dyson skew information from the perspective of multipartite entanglement, intended for physicists working on quantum many-body physics. The main results are the bounds in eq.~\eqref{eq:QFIbounds}, the identity~\eqref{eq:QFIidentity}, and the numerical procedure outlined in subsection~\ref{ss:numerical}. 
\item[Section~\ref{s:equilibrium}] This is an overview of the behavior of the quantum Fisher information in quantum spin chains in thermal equilibrium both at finite and at zero temperature. We provide a critical reinterpretation of recent findings in symmetry protected topological phases~\cite{Pezze2017Multipartite,DellAnna2023Quantum} and in ordered phases with spontaneous symmetry breaking~\cite{Qu2025Quantum}. We draw the conclusion that macroscopic quantum states do no emerge in equilibrium quantum spin chains with local interactions. 
\end{description}
The next two sections overview our findings and the methods used:
\begin{description}
\item[Section~\ref{s:offequilibrium}] This presents the nonequilibrium dynamics considered, i.e., the kicking protocol and the global quench, and summarizes the main results. We exhibit two of our predictions: eq.~\eqref{eq:chibasic} is a semiclassical formula for the QFI after a local perturbation; eq.~\eqref{eq:predictionquench} is a remarkable  relation between QFI and order parameter correlations after global quenches with a quasiparticle pair structure.
\item[Section~\ref{s:model}] This brefly reviews the quantum XY model, in which we carry out numerical checks, as well as the free fermion techniques---based on Wick's theorem---that allow us to express the WYDI in terms of the two-point fermionic correlation functions---eq.~\eqref{eq:Ialphafree}.
\end{description}
The  core of the paper, focused on nonequilibrium systems, is split in two sections:
\begin{description}
\item[Section~\ref{s:kicking}] This details the investigation into kicking protocols. In particular, we explain how to generalize eq.~\eqref{eq:chibasic} to more complex protocols: simultaneous localized perturbations at different positions, eq.~\eqref{eq:qfi_pred_DW}, and localized perturbations at periodic times, eq.~\eqref{eq:qfi_pred_difftimes}.
\item[Section~\ref{s:global}] This details the investigation into global quenches and derive eq.~\eqref{eq:predictionquench} within a semiclassical theory.
\end{description}
Conclusive remarks are collected in Section~\ref{s:conclusion}, where we also show the stability of our findings under a small increase of temperature.

\section{Quantum Fisher information and Wigner-Yanase-Dyson skew information}\label{s:QFI}%

\subsection{Overview}
We indicate the variance of $ O$ in the state with density matrix $\rho$ by
\begin{equation}\label{eq:skewinfo}
\Delta_\rho  O^2=\mathrm{tr}(\rho { O}^2)-\mathrm{tr}(\rho  O)^2
\end{equation} 
In the following  $ O=\sum_j O_j$ will be an extensive operator with a local density $ O_j$.
In a weakly correlated system, $\Delta_\rho  O^2$ scales with the volume. This happens, in particular, when correlations decay exponentially. The variance $\Delta_\rho  O^2$ per unit volume is then proportional to the typical volume of the region correlated with a given point. In a pure state correlations are purely quantum, so the variance per unit volume allows one to estimate the size of the quantum correlated clusters. 
If this typical size is comparable to the entire system's volume, quantum effects extend to macroscopic scales (assuming the system itself is macroscopically large). In such cases, we say the system is in a \emph{macroscopic quantum state}. A well-known example in lattice spin models is the generalized GHZ (Greenberger, Horne, and Zeilinger) state, which represents a superposition of two ferromagnetic states with spins aligned in opposite directions. A more conceptual example comes from the famous Schr\"odinger's cat Gedankenexperiment, which results in the superposition of two macroscopically distinct states that, according to classical intuition, should be mutually exclusive.

However, we point out that, as the system size increases, describing it through a pure state becomes more and more artificial. This is analogous to attempting to describe a classical many-body system without statistical physics, assuming that all particle positions and velocities are precisely known. In many-body systems, it is more natural to shift focus from the entire system to its subsystems. This shift can be seen as an implicit change in perspective when taking the thermodynamic limit. In that process, information about the state is inevitably lost, and the accessible portion of the state becomes mixed.
As a result, quantum correlations involving inaccessible parts of the system are turned into classical ones, and the variance $\Delta_\rho  O^2$ per unit volume has both quantum and classical contributions. 
In the attempt to extract  quantum  correlations, one could envisage expressing the mixed state as an incoherent superposition of pure states
\begin{equation}
\rho=\sum_i p_i \ket{\Psi_i}\bra{\Psi_i}\, .
\end{equation}
This can be physically interpreted as the state being prepared in the pure state $\ket{\Psi_i}$ with classical probability $p_i$. Since at each preparation the state would be pure, the average fraction $\chi$ of the system that is quantum correlated with respect to $O$ would be given by
\begin{equation}
\chi(\{p,\Psi\};O)= \frac{1}{\parallel O\parallel^2}\sum_i p_i \Delta_{\ket{\Psi_i}\bra{\Psi_i}}O^2
\end{equation}
The decomposition in pure states, however, is not unique---the states do not need to be orthogonal (and, in the presence of degeneracies, one could also  make an arbitrary change of basis in the degenerate eigenspaces)---indeed a mixed state can be prepared in, generically, infinitely many ways. Since the information about the preparation of the state is arguably classical, assuming one decomposition causes  $\chi(\{p,\Psi\};O)$ to overestimate quantum correlations. 
From this perspective, it is reasonable to minimize $\chi(\{p,\Psi\};O)$ over all possible decompositions of $\rho$, resulting in the definition
\begin{empheq}[box=\fbox]{equation}\label{eq:chi0}
\chi(\rho;O)= \inf_{\{p,\Psi\}}\frac{1}{\parallel O\parallel^2}\sum_i p_i \Delta_{\ket{\Psi_i}\bra{\Psi_i}}O^2
\end{empheq}
Typically, $\chi(\rho;O)$ approaches zero as  the subsystem size increases. Indeed, the average variance is bounded from above by the variance~\cite{Toth2013Extremal}, and the latter, as mentioned earlier, is  proportional to $\parallel O\parallel$  in the case of exponentially decaying correlations; in fact, even with algebraically decaying correlations, it can scale at most as some power of  $\parallel O\parallel$ strictly smaller than $2$.
A necessary condition for $\chi(\rho;O)$ to remain finite is indeed the breakdown of cluster decomposition properties, which is already quite atypical in a quantum many-body system. But that is not even sufficient, as in many cases clustering properties are broken at the classical level. For instance, every subsystem of a generalized GHZ state experiences the breakdown of cluster decomposition, but this is simply due to the fact that the state is in an incoherent superposition of two macroscopically different (separable) states, which is fully separable. 
In a quantum many-body system, a nonzero value of $\chi(\rho;O)$ in a subsystem so large to be considered macroscopic is highly unusual; it can occur only if the subsystem exhibits full-range correlations that cannot be attributed to a mere lack of information. When that happens, we say  that the subsystem is in a macroscopic quantum state. 

We mention that some authors prefer the more conservative term ``(extensive) multipartite entangled'' even when referring to a macroscopic quantum state, while others use ``macroscopically entangled''. In this work, we use these terms interchangeably, always implying the existence of an observable $O$ with local density such that $\chi(\rho;O)$ remains finite in a large subsystem described by~$\rho$.

\subsection{Inequalities}
Besides its conceptual significance, \eqref{eq:chi0} presents a significant technical challenge: how can that minimization be efficiently performed---either numerically or analytically---in a large subsystem of a quantum many-body system?
A partial answer to this question comes from Refs~\cite{Toth2013Extremal,yu2013quantum}, who identified the minimal averaged variance with a quarter of the quantum Fisher information $F(\rho,O)$, that is to say,
\begin{multline}\label{eq:chi}
\chi(\rho;O)=\frac{F(\rho,O)}{4\parallel O\parallel^2}=\\
\sum_{i,j}\tfrac{e^{-2\epsilon_i}+e^{-2\epsilon_j}}{2}\tanh^2(\epsilon_i-\epsilon_j)\tfrac{|\braket{\phi_i|O|\phi_j}|^2}{\parallel O\parallel^2}
\end{multline}
\begin{equation}\label{eq:chialt}
=\sum_i e^{-2\epsilon_i}\tfrac{\Delta_{\ket{\phi_i}\bra{\phi_i}}O^2}{\parallel O\parallel^2}-\sum_{i\neq j}\tfrac{2}{e^{2\epsilon_i}+e^{2\epsilon_j}}\tfrac{|\braket{\phi_i|O|\phi_j}|^2}{\parallel O\parallel^2}\, ,
\end{equation}
where $e^{-2\epsilon_i}$ are the eigenvalues of $\rho$ with eigenvectors $\ket{\phi_i}$. From now on we will refer to $\chi(\rho;O)$ as the \emph{normalized quantum Fisher information} of $\rho$ with respect to $O$. Unfortunately, despite being explicit, this representation involves two sums over the eigenstates, which are very hard if not impossible to compute in an extended system. The good news is that there are equivalent quantities that can be computed more easily. To that aim, we introduce the Wigner-Yanase-Dyson (WYDI) skew information~\cite{Wigner1963INFORMATION}
\begin{multline}\label{eq:WYDI}
I_{\alpha}(\rho, O)=\mathrm{tr}[\rho O^2]-\mathrm{tr}[\rho^{\alpha
}O\rho^{1-\alpha}O]=\\
\sum_{i,j}\tfrac{e^{-2\epsilon_i}+e^{-2\epsilon_j}}{2}[1-\tfrac{\cosh((1-2\alpha)(\epsilon_i-\epsilon_j))}{\cosh(\epsilon_i-\epsilon_j)}]|\braket{\phi_i|O|\phi_j}|^2\, .
\end{multline}
From the explicit representation in the eigenbasis, it is evident that $I_{\alpha}(\rho,O)$ is a concave function of $\alpha$ with maximum at $\alpha=\frac{1}{2}$, as originally proven by Lieb in Ref.~\cite{Lieb1973Convex}. This special value corresponds to the Wigner-Yanase skew information (WYI), for which we will also use the shorthand  $I(\rho,O)\equiv I_{\frac{1}{2}}(\rho,O)$. The WYDI shares many properties with the QFI; in particular, as shown in Ref.~\cite{luoshulongwigvsqfi}, they all arise as natural quantizations of the classical Fisher Information \cite{Fisher_1925}. In that regard, we note that, since $1-\frac{1}{\cosh x}\leq \tanh^2 x\leq 2-\frac{2}{\cosh x}$, the WYI is equivalent to the QFI 
\begin{equation}\label{eq:WYSbounds}
I(\rho, O) \leq \tfrac{1}{4}F(\rho, O)\leq 2 I(\rho, O)\, ,
\end{equation}
and similar bounds are obtained for the Wigner-Yanase-Dyson skew information 
\begin{equation}\label{eq:WYDbounds}
I_\alpha(\rho, O) \leq \tfrac{1}{4}F(\rho, O)\leq \frac{I_\alpha(\rho, O)}{2\alpha(1-\alpha)} \, ,
\end{equation}
manifesting the equivalence of all the quantities introduced. Specifically, if QFI witnesses macroscopic entanglement, then also the Wigner-Yanase-Dyson skew information does for any $0<\alpha\leq \frac{1}{2}$. 

In order to check analytic predictions, however, even factors are important, so the confidence interval provided by \eqref{eq:WYDbounds} is unsatisfactory. 
In that respect, we note that the relative errors associated with approximating $\tfrac{1}{4}F(\rho, O)$ with the lower bounds $I_\alpha(\rho, O)$ are generally smaller than those associated with the upper bounds\footnote{For example, for $\alpha=\frac{1}{2}$ the relative error of the lower bound becomes (much) smaller than the other as long as $|\epsilon_i-\epsilon_j|\gtrsim 1.32$. In addition,  the relative error of the lower bounds approach zero as $|\epsilon_i-\epsilon_j|\rightarrow\infty$, whereas the others approaches $\frac{1}{2\alpha(1-\alpha)}-1$.}. Alternative upper bounds can be constructed considering linear combination of WYDI with different values of $\alpha$; a simple one is given by
\begin{empheq}[box=\fbox]{equation}
\label{eq:QFIbounds}
I(\rho, O) \leq \tfrac{1}{4}F(\rho, O)\leq 10 I(\rho, O)-9 I_{\frac{1}{3}}(\rho, O)\, ,
\end{empheq}
which can be proven again using the explicit representation in the eigenbasis. We will see that, in the systems that we consider, this upper bound is tighter than $2I(\rho, O)$ in \eqref{eq:WYSbounds}. We stress again that both lower and upper bounds correspond to different quantizations  of the classical Fisher information.

Even if it is not generally saturated, another lower bound has recently attracted attention~\cite{Girolami_2014}
\begin{equation}\label{eq:Irho2}
    \tfrac{1}{4} I(\rho^2, O)<  I(\rho, O) \qquad\Bigl(\leq \tfrac{1}{4}F(\rho, O) \Bigr)\, .
\end{equation}
More generally, Ref.~\cite{PhysRevLett.127.260501} showed that this is the first term of a sequence of lower bounds that rapidly converge to $F(\rho, O)$ and that are potentially measurable in experiments as well as in numerical simulations.
Incidentally, we note that the bound~\eqref{eq:Irho2} can be made tighter
\begin{equation}
\tfrac{1}{4}e^{S_\infty[\rho]} I(\rho^2, O)\leq I(\rho, O)
\end{equation}
where $S_\infty$ denotes the R\'enyi-$\infty$ entropy, i.e., $e^{-S_\infty[\rho]}$ is the maximal eigenvalue of $\rho$.

We conclude this section by highlighting a key property: QFI is  \emph{convex}
\begin{equation}\label{eq:QFIconvex}
F(p\rho_1+(1-p)\rho_2,O)\leq p F(\rho_1,O)+(1-p)F(\rho_2,O)\, .
\end{equation}
Consequently, the QFI of an incoherent superposition of states does not exceed the maximum QFI of the individual states. This property  also holds for the WYDI for $0< \alpha <1$, as proven by Lieb in Ref.~\cite{LIEB1973267}.

\subsection{Identities}
There are situations in which the WYDI are identical and equal to a quarter of the QFI.
\begin{itemize}
\item The state is pure 
\begin{equation}
\Rightarrow \frac{1}{4}F(\rho,O)=I_\alpha(\rho,O)=\Delta_\rho O^2\, .
\end{equation} 
\item The operator $O$ links an eigenstate only with degenerate eigenstates or with an eigenstate of $\rho$ with zero eigenvalue
\begin{equation}\label{eq:FeqI}
\Rightarrow \frac{1}{4}F(\rho,O)=I_\alpha(\rho,O)=\sum_{i}e^{-2\epsilon_i}\Delta_{\ket{\phi_i}\bra{\phi_i}}O^2\, ,
\end{equation}
where $\ket{\phi_i}$ is the eigenbasis of $\rho$ that diagonalizes the restriction of $O$ within each degenerate eigenspace of $\rho$.  
\end{itemize}
This second case is particularly interesting as in some settings that we study the underlying assumption is satisfied, either asymptotically or approximately.

We are not aware of scientific works pointing out identities between QFI and WYDI that hold in general. We have however discovered a simple relation that links the QFI with the 
WYDI for $\alpha\in \mathbb C$:
\begin{empheq}[box=\fbox]{equation}\label{eq:QFIidentity}
\frac{1}{4}F(\rho,O)=\int_{-\infty} ^\infty \frac{d \beta}{\cosh(\pi \beta)}I_{\frac{1}{2}+i \beta}(\rho, O)\, .
\end{empheq}
This can be readily proven using the explicit representation provided in \eqref{eq:WYDI}. Indeed, for $\alpha=\frac{1}{2}+i \beta$ the hyperbolic cosine is turned into a cosine---$I_{\frac{1}{2}+i \beta}(\rho, O)$ is real for every $\beta\in \mathbb R$---and the integral in \eqref{eq:QFIidentity} essentially becomes the Fourier transform of the hyperbolic secant.
From now on we will use the notation
\begin{equation}
J_{\beta}(\rho, O)\equiv I_{\frac{1}{2}+i \beta}(\rho, O)
\end{equation}
and call $J_{\beta}(\rho, O)$ the rotated Wigner-Yanase-Dyson skew information (rWYDI). 

\subsection{On the numerical evaluation}\label{ss:numerical}
Beyond its intrinsic interest, \eqref{eq:QFIidentity} provides a means to overcome the technical challenge of computing the QFI in a large subsystem.
Specifically, since $|\cos(k x)|\leq \cosh(x)$ for every $k,x\in \mathbb R$, 
$J_{\beta}(\rho, O) $ is bounded from above by the variance, and hence the integrand in \eqref{eq:QFIidentity} approaches zero exponentially in $\beta$. The integral can then be evaluated with high accuracy using the following procedure:
\begin{enumerate}
\item $J_{\beta}(\rho, O)$ is computed at enough values of $\beta$ that allow for a good interpolation of $J_{\beta}(\rho, O)$ in a sufficiently large domain $\beta\in(-B, B)$ with $B>0$ ($B=2$ could be enough).
\item The integral of the interpolation is evaluated in $N$ domains $(-B_j,B_j)$ with $j=1,\dots,N$, $B_j<B_{j+1}$ and $B_N=B$. 
\item The $N$ data points are fitted to a curve describing an exponentially fast saturation to a constant.
\item The constant provides an estimation of $\frac{1}{4}F(\rho,O)$.
\end{enumerate}
The reader might have noticed that we have assumed that computing WYDI or rWYDI numerically is feasible. This is not necessarily true in a generic quantum spin chain, but the structure of the (r)WYDI resonates with the structure of the tensor networks that are routinely employed to simulate such systems both in and out of equilibrium. This is evident if one considers the full density matrix at finite temperature, as the evaluation of the (r)WYDI would simply require to compute a finite number of contractions between thermal states (possibly with complex temperatures) and local operators. The situation becomes even better when considering noninteracting spin chains in which the reduced density matrices are Gaussian.  A power of a Gaussian is indeed a Gaussian, hence the (r)WYDI can be computed efficiently using free-fermion techniques, which will be discussed in Section~\ref{ss:free_fermion}.

\section{Critical overview of QFI at equilibrium in quantum spin chains}\label{s:equilibrium}

This section is a brief overview of the behavior of the quantum Fisher information at thermal equilibrium with respect to extensive observables. In particular, we will comment on some recent studies to clarify the point of view that we will follow in the rest of the paper.

\paragraph{Thermal equilibrium.}
\begin{figure}[t]
\includegraphics[width=0.45\textwidth]{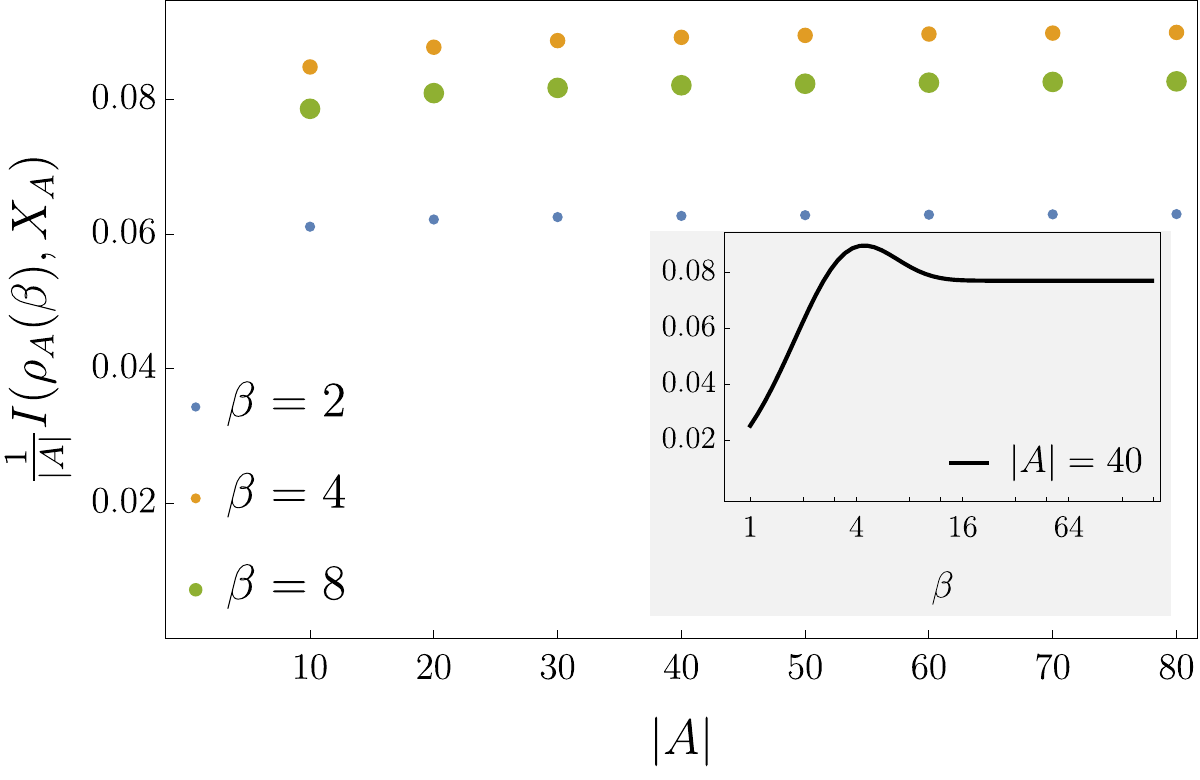}
\caption{The equilibrium Wigner-Yanase skew information per unit length of a spin block  $A$ with respect to $X_A=\sum_{\ell\in A} \sigma_\ell^x$ as a function of the block's length at different values of the inverse temperature $\beta$. The model is the transverse-field Ising chain with $h=1/2$ (ferromagnetic at zero temperature). The inset shows the temperature dependence at fixed length.}\label{f:finiteT}
\end{figure}
We start with considering systems with local interactions in equilibrium at nonzero temperature.
Local observables satisfy cluster decomposition properties
\begin{equation}
\braket{O(x)O(y)}-\braket{O(x)}\braket{O(y)}\xrightarrow{|x-y|\rightarrow\infty}0\, .
\end{equation}
This implies that the variance of an extensive operator with a local density cannot scale as the square of the volume. Since variance serves as an upper bound for the QFI, clustering properties hinder multipartite entanglement. However, this upper bound weakens as the system approaches a thermal phase transition where a symmetry is spontaneously broken. 
Analyzing, in particular, thermal states of hardcore bosons on a square lattice, Ref.~\cite{Frerot2016Quantum} pointed out that the breakdown of clustering properties as the critical temperature is approached from a disordered phase is not accompanied by an analogous increase in the subsystem quantum Fisher information (in fact, Ref.~\cite{Frerot2016Quantum} studied the quantum variance, but the latter is equivalent to QFI). 

Since in one dimensional systems  phase transitions occur only at zero temperature, we validate those findings by investigating the low-temperature behavior of the Wigner-Yanase skew information of a subsystem with respect to an extensive order parameter, which, by definition, experiences the breakdown (in the zero temperature limit) and restoration (at exactly zero temperature) of cluster decomposition properties. 
The data reported in Fig.~\ref{f:finiteT} supplement the findings of Ref.~\cite{Frerot2016Quantum} that the divergence of the order parameter correlation length in the zero temperature limit is not associated with  subsystems falling in macroscopic quantum states.

\paragraph{Critical system.}
\begin{figure}[t]
\includegraphics[width=0.45\textwidth]{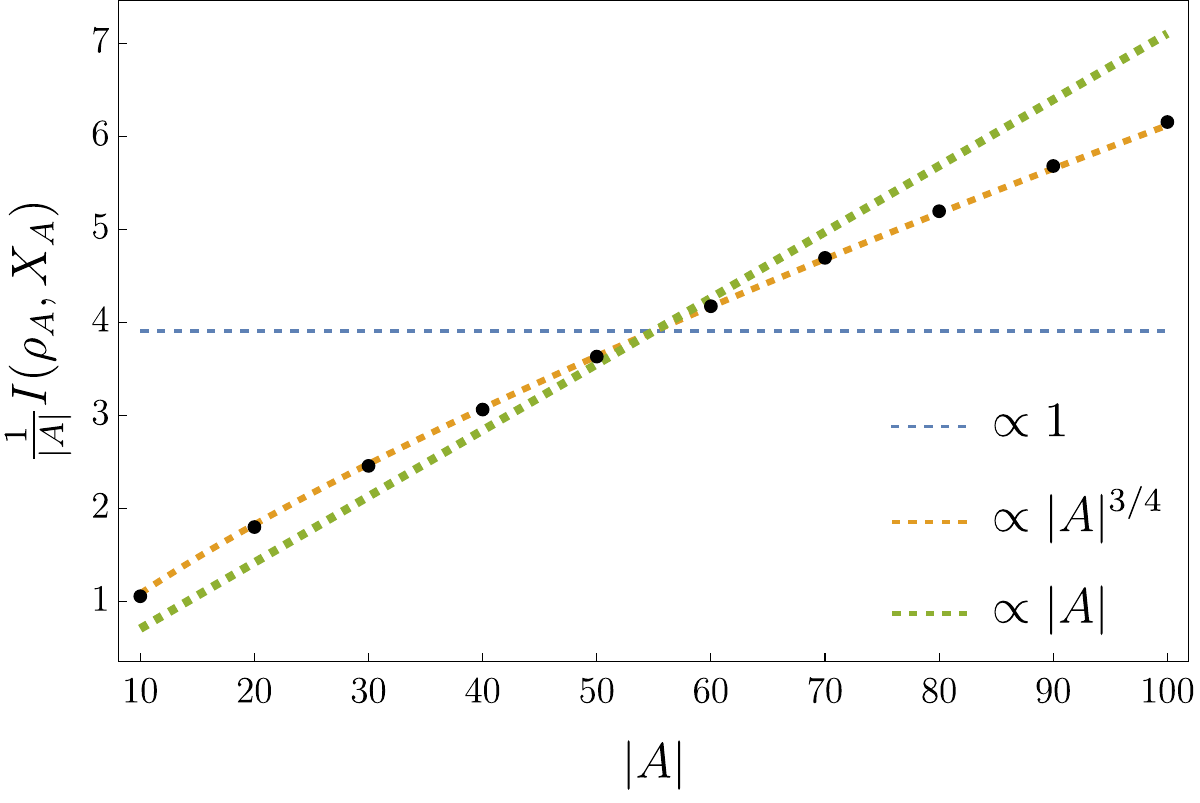}
\caption{The  Wigner-Yanase skew information per unit length of a spin block  $A$ with respect to $X_A=\sum_{\ell\in A} \sigma_\ell^x$ as a function of the block's length. The model is the critical Ising chain ($h=1$) in the thermodynamic limit. The dashed curves are power laws aimed at guiding the eye.}\label{f:QFIcritical}
\end{figure}

At strictly zero temperature, a paradigmatic example of a system with enhanced multipartite entanglement is a gapless system, where correlations decay algebraically. As a result, the QFI with respect to operators with slowly decaying correlations can scale with a power of the system volume greater than one~\cite{Lambert2019Estimates, Liu2013Quantum}, which has also been indirectly observed in experiments~\cite{Hauke2016Measuring}. 
For example, in the critical Ising chain separating a ferromagnetic phase from a paramagnetic one, the order parameter correlation decays to $0$ as $|x-y|^{-\frac{1}{4}}$, hence the corresponding variance behaves as the volume to the power of $7/4$.  Fig.~\ref{f:QFIcritical} shows that also the subsystem QFI exhibits the same scaling behavior.
Since also gapless systems obey weak cluster decomposition properties, however, the QFI per unit volume squared---measured with respect to an extensive operator---will always vanish in the limit of large system size. Thus, the QFI does not witness macroscopic quantum states at criticality. 

\paragraph{Symmetry-protected topological order.}
Equilibrium and clustering properties are deeply connected, making it seemingly impossible for the normalized QFI---\eqref{eq:chi}--- to remain nonzero in large subsystems under equilibrium conditions. Nevertheless, symmetry protected topological order in 1D gapped phases has been identified as a driver of multipartite entanglement~\cite{Pezze2017Multipartite,DellAnna2023Quantum}. 
To the best of our understanding, this paradox arises because the definition of `part of the system' is left implied, and, as a result, risks being misunderstood. 
When QFI is used to witness multipartite entanglement, the choice of the operator with respect to which the QFI is computed is not arbitrary but should reflect the structure of the Hilbert space. 
We focus exclusively on spin chains, where the composite Hilbert space has a tensor product structure, $\mathcal{H} = \bigotimes_n \mathcal{H}_n$. It is worth noting, in passing, that this structure is absent in fermionic systems~\cite{Vidal2021Quantum}, which complicates the relationship between  QFI and multipartite entanglement. Unfortunately, Ref.~\cite{Pezze2017Multipartite} seems to have overlooked this issue, leading to questionable conclusions and potentially misleading strategies (those strategies, indeed, were subsequently applied to spin chains~\cite{DellAnna2023Quantum}, resulting in similarly problematic conclusions---see below).

A state is fully separable if no entanglement exists between the subsystems associated with $\mathcal H_n$. However, the entanglement properties of a state depend on how the Hilbert space is partitioned. A state that is separable under one partitioning can become maximally entangled under another, and vice versa.
In a finite chain, indeed, pure states can always be transformed into one another via unitary operations. Consequently, it is always possible to construct nonlocal extensive operators that exhibit maximal variance. This is true, in particular, for product states---the paradigm of separable states---which can be mapped, e.g., into the (generalized) GHZ state---the paradigm of multipartite entangled state. The operator that is transformed into the extensive operator with local density that maximizes the variance in the GHZ state will, in turn, maximize its variance in the product state.   
This fundamental observation suggests that the QFI  with respect to a generic extensive nonlocal operator cannot serve as a reliable witness for multipartite entanglement. Therefore, it is essential to examine the properties of the operators proposed for detecting extensive multipartite entanglement in SPT phases. As we clarify below, the operators studied in Ref.~\cite{DellAnna2023Quantum} are, in fact, as generic as those obtained through a transformation that maps a GHZ state into a product state.
Specifically, Ref.~\cite{DellAnna2023Quantum} computed the QFI with respect to operators of the form $\tilde O=\sum_n \tilde O_n$, with semilocal density $\tilde O_n$. This means that $\tilde O_n$ commutes with every operator that acts as the identity on   $\mathcal H_n$ and that has the symmetries underlying SPT order (e.g., it commutes with the Hamiltonian density far enough from the position associated with $\mathcal H_n$).
It does not commute, however, with non-symmetric local operators, irrespectively of their distance. Thus, operators such as $\tilde O_n$ represent local observables provided that non-symmetric local operators are somehow excluded from the theory 
(see also Ref.~\cite{Fagotti2024Nonequilibrium}). 
On the other hand, it is well established that the ground state of a system in an SPT phase can be mapped onto the ground state of a system in a trivial phase via a local unitary transformation~\cite{Zeng2019book}. While a precise definition of extensive multipartite entanglement (also referred to as macroscopic entanglement or macroscopic quantumness) remains elusive~\cite{Frowis2018Macroscopic}, a fundamental requirement should be that a macroscopic quantum state cannot be transformed into a conventional quantum state through local unitary operations\footnote{Here we define a local unitary transformation as one that is generated by an extensive operator with a local or quasilocal density.}---transformations that are also used to define phases of matter. Consequently, the multipartite entanglement properties of SPT ground states should not differ significantly from those of ground states in trivial (e.g., paramagnetic) phases. 

For clarity, we consider an example: the transverse-field cluster model in zero field, described by the Hamiltonian
\begin{equation}
H=-\sum_{\ell=1}^{L-2} \sigma_{\ell}^x\sigma_{\ell+1
}^z\sigma_{\ell+2}^x\, .
\end{equation}
This model has a $Z_2\times Z_2$ symmetry generated by $\prod_{j=1}^{L/2}\sigma_{2j}^z$ and $\prod_{j=1}^{L/2}\sigma_{2j-1}^z$, where we assumed that $L$ is even. The ground state is four-fold degenerate and is in a  topologically ordered phase protected by the  $Z_2\times Z_2$ symmetry~\cite{Zeng2019book}.
It is pedagogical to simultaneously consider an alternative representation that uses spin variables $\tau$ that are dual to the $\sigma$ through the following Kramers-Wannier transformation
\begin{equation}
\begin{aligned}
\tau_j^x=&\begin{cases}
\sigma_j^x \sigma_{j+1}^x &j<L\\
-\sigma_1^y\prod_{n=2}^{L-1}\sigma_n^z\ \sigma_L^y&j=L
\end{cases}\\
\tau_j^y=&\begin{cases}
\sigma^z_1 \sigma^x_{2}&j=1\\
-\sigma_1^y\prod_{n=2}^{j-1}\sigma_n^z\ \sigma_j^y\sigma_{j+1}^x &1<j<L\\
-\sigma_L^x&j=L
\end{cases}\\
\tau_j^z=&
-\sigma_1^y\prod_{n=2}^{j}\sigma_n^z \, .
\end{aligned}
\end{equation}
In the $\tau$ representation the Hamiltonian reads
\begin{equation}
H=\sum_{\ell=1}^{L-2} \tau_\ell^y\tau_{\ell+1}^y\, .
\end{equation}
Note that $H$ acts as the identity on the last site of the $\tau$ chain. The generators of the  $Z_2\times Z_2$ symmetry are now represented by $\prod_{j=1}^L\tau_j^z$ and $(-1)^{L/2}\prod_{j=1}^L \tau_j^y$. 
In this representation it is clear that the operators $\tau_\ell^y$ do not cluster without breaking the symmetry, and 
the operator $\tilde O=\sum_\ell(-1)^\ell\tau_\ell^y$ is the best candidate to witness multipartite entanglement through the quantum Fisher information. 
Let us then pick a ground state that respects the symmetry, for example,
\begin{equation}
\ket{\textsc{gs}_1}=\frac{1}{\sqrt{2}}\Bigl(\ket{\rightarrow_y\leftarrow_y}^{\otimes \frac{L}{2}}+\ket{\leftarrow_y\rightarrow_y}^{\otimes \frac{L}{2}}\Bigr)\, ,
\end{equation}
where $\rightarrow_y$ (or $\leftarrow_y$) indicates the eigenstate of $\tau^y$ with eigenvalue $1$ (or $-1$).
Returning to the original representation, we can readily conclude that the quantum Fisher information of this symmetric ground state with respect to the operator $\tilde O=\sum_{j=1}^L \tilde O_j$, with
\begin{equation}\label{eq:semilocal}
\tilde O_j=\begin{cases}
-\sigma_1^z\sigma_2^x&j=1\\
-(-1)^j\sigma_1^y\prod_{n=2}^{j-1}\sigma_n^z\ \sigma_j^y\sigma_{j+1}^x&1<j<L\\
-\sigma_L^x&j=L\, ,
\end{cases}
\end{equation}
equals $4L^2$, which is the maximal value achievable.  Operator $\tilde O$ has a semilocal density $\tilde O_j$ (with respect to the symmetry underlying SPT order) and is an example of the operators that have been proposed to probe multipartite entanglement in symmetry-protected topological phases. It is easy to see that every ground state of $H$ is linked to $\ket{\textsc{gs}_1}$ by a unitary transformation acting only on the spins $\{1,L-1,L\}$. In particular we have
\begin{equation}\label{eq:statesSPT}
\begin{aligned}
e^{-i\frac{\pi}{4}\sigma_1^x\sigma_{L-1}^x \sigma_{L}^y}\ket{\textsc{gs}_1}=&\ket{\rightarrow_y\leftarrow_y}^{\otimes\frac{L}{2}}\\
e^{i\frac{\pi}{4}\sigma_1^x\sigma_{L-1}^x \sigma_{L}^y}\ket{\textsc{gs}_1}=&\ket{\leftarrow_y\rightarrow_y}^{\otimes\frac{L}{2}}\\
e^{-i\frac{\pi}{4}\sigma_1^x\sigma_{L-1}^x \sigma_{L}^y}\sigma_{L-1}^x \sigma_{L}^y\ket{\textsc{gs}_1}=&\ket{\rightarrow_y\leftarrow_y}^{\otimes (\frac{L}{2}-1)}\otimes\ket{ \rightarrow_y\rightarrow_y}\\
-e^{i\frac{\pi}{4}\sigma_1^x\sigma_{L-1}^x \sigma_{L}^y}\sigma_{L-1}^x \sigma_{L}^y\ket{\textsc{gs}_1}=&\ket{\leftarrow_y\rightarrow_y}^{\otimes(\frac{L}{2}-1)}\otimes\ket{ \leftarrow_y\leftarrow_y}
\end{aligned}
\end{equation}
These transformations  (which are local if we imagine to fold the chain and identify site $L+1$ with site $1$) should not convert a macroscopic quantum state into a conventional one, or vice versa. Thus, all ground states of $H$ exhibit essentially the same multipartite entanglement properties (in the $\sigma$ representation). 
We now provide evidence that they should not be regarded as macroscopically entangled. 
There is indeed a local unitary transformation bringing $\ket{\textsc{gs}_1}$ into a $\mathbb Z_2\times \mathbb Z_2$ symmetric product state (in the $\sigma$ representation)---the state $\ket{\Uparrow}$ with all spins aligned in the $z$ direction:
\begin{equation}\label{eq:trivialization}
\ket{\Uparrow}=e^{i\frac{\pi}{4}(\sigma_1^x+\sigma_L^y)}e^{i\frac{\pi}{4}\sum_{\ell=1}^{L-1}(-1)^\ell \sigma_\ell^x\sigma_{\ell+1}^x}e^{-i\frac{\pi}{4}\sigma_1^x\sigma_{L-1}^x \sigma_{L}^y}\ket{\textsc{gs}_1}\, .
\end{equation}
This means that, for any ordered pair of sites $u,v\in \llbracket 1,L\rrbracket$, the density matrix $\ket{\textsc{gs}_1}\bra{\textsc{gs}_1}$ can be decomposed as follows
\begin{equation}
\ket{\textsc{gs}_1}\bra{\textsc{gs}_1}=(U_u U_v)(\bar \rho_{ u}\otimes\bar \rho_{\llbracket u+1,v-1\rrbracket}\otimes \bar \rho_{v}\otimes\bar \rho_{\overline{\llbracket u,v\rrbracket}}) (U_uU_v)^\dag
\end{equation}
where $U_u$ and $U_v$ are unitary acting like the identity outside $\llbracket u-1,u+1\rrbracket$ and $\llbracket v-1,v+1\rrbracket$ (with $L+1\equiv 1$ and $0\equiv L$), respectively, and $\bar \rho_{A}$ is defined in subsystem $A$. 
In addition, an analogous decomposition holds for the density matrices $\bar \rho_{\llbracket u+1,v-1\rrbracket}$ and $\bar \rho_{\overline{\llbracket u,v\rrbracket}}$.
Since any subsystem can be disentangled from the rest with a unitary transformation localized at its edges, we do not interpret the maximal variance of the semilocal operator $\tilde O$ as evidence of multipartite entanglement. Instead, the state is more appropriately characterized as approximately separable.

\paragraph{Spontaneous symmetry breaking.}

\begin{figure}[t]
\includegraphics[width=0.45\textwidth]{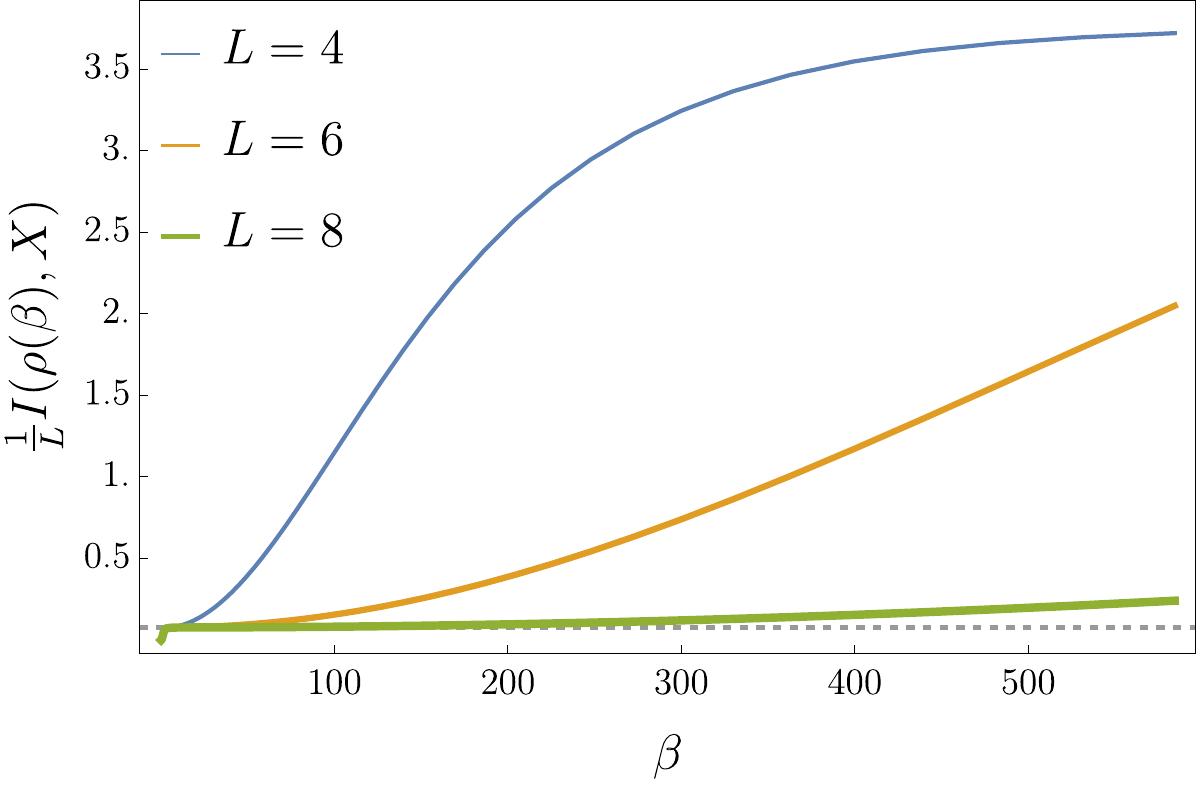}
\caption{The finite-temperature Wigner-Yanase skew information per unit length with respect to $X=\sum_{\ell=1}^L \sigma_\ell^x$ as a function of the inverse temperature $\beta$ in small chains of up to $L=8$ spins. The model is the transverse-field Ising chain with $h=1/2$ (ferromagnetic at zero temperture). The (underlying) dashed horizontal line is the value in a symmetry-breaking ground state as $L\rightarrow\infty$.}\label{f:fi niteTsmallL}
\end{figure}

Different types of subtleties appear when finite systems are investigated in phases with spontaneous symmetry breaking. The choice becomes between artificially breaking a symmetry and considering a state that is not even stationary, or overlooking spontaneous symmetry breaking and considering the actual ground state of the finite system. Choosing the second option results in spurious parabolic large-size behaviors of the QFI~\cite{Liu2013Quantum, Qu2025Quantum}. However, the physical significance of the finite-size ground state is questionable. For example, given that there is no spontaneous symmetry breaking in finite chains, one could imagine to prepare the ground state by lowering the temperature. Very quickly the variance of an order parameter would exhibit a quadratic scaling with the volume. Much different, however, would be the behavior of the quantum Fisher information. Indeed, the gaps between the energy eigenstates that become ground states in the thermodynamic limit 
close very quickly with the chain's length $L$. In the transverse-field Ising chain, for example, they are exponentially small in $L$ so only at an exponentially small temperature the state will become close to the finite-chain ground state. At less extreme low temperatures, instead, the state would approach the incoherent superposition of those eigenstates, which can be reinterpreted as the incoherent superposition of the physical symmetry-breaking ground states. Consequently, the QFI of the superposition is bounded from above by the average of the QFI of the symmetry-breaking ground states (by virtue of \eqref{eq:QFIconvex}), which scales linearly with the volume. Fig.~\ref{f:fi niteTsmallL}  shows this phenomenon very clearly: even in ridiculously small chains of $4$ or $6$ spins the normalized QFI with respect to an extensive order parameter with one-site density remains smaller than its norm for impressively low temperatures.

The physical conclusion of this brief overview of quantum spin chains at equilibrium is that  quantum macroscopic states 
should not be expected at equilibrium.  
We have emphasized some controversial results not only for clarifying our perspective, but also
to make evident some  benefits of focusing on subsystems in the bulk rather than on the entire system. First, the semilocal operators that had been proposed to probe multipartite entanglement in topologically ordered phases would not be represented in the subsystem. Second, in a conventional ordered phase, the scaling behavior of the QFI would remain linear in the volume independently of whether a symmetry is spontaneously broken or artificially preserved. 

\section{QFI of subsystems out of  equilibrium: overview and results}\label{s:offequilibrium}
Preserving locality of interactions but bringing the system out of equilibrium can be effective in producing macroscopic quantum states in time scales that, as shown in Ref.~\cite{Chu2023Strong}, are at least proportional to the (sub)system's length. 
Our starting point is again to identify 
scenarios in which correlations do not decay over arbitrarily large scales.
For example, at sufficiently low temperatures, it has been observed that a glimmer of long-range order survives a global quench in the form of prethermalization, manifesting as a state with an effectively infinite correlation length~\cite{Alba2017Prethermalization}. This phenomenon becomes more pronounced in the zero-temperature limit, where the initial state evolves from an incoherent superposition of symmetry-breaking ground states. Under time evolution with a symmetry-breaking Hamiltonian, the system locally relaxes into a collection of distinct generalized Gibbs ensembles, each corresponding to a macroscopically different state.
However, by convexity of the QFI~\eqref{eq:QFIconvex}, we can readily infer that such a form of prethermalization does not enhance the QFI with respect to the more standard quench protocol in which the system is prepared in a ground state breaking the symmetry---see also Section~\ref{s:conclusion}. 
The latter global quench was studied in Ref.~\cite{Pappalardi2017Multipartite}. Remarkably, Ref.~\cite{Pappalardi2017Multipartite} found that the full chain variance per unit length of the order parameter can become arbitrarily large after a quench in the ferromagnetic phase of the Ising model in the limit in which the post-quench Hamiltonian approaches the pre-quench one. 
Similar observations were also reported in Ref.~\cite{Collura2019Relaxation}.
The question of whether multipartite entanglement could survive the thermodynamic limit (in which the equivalence between QFI and variance breaks down as information about the system becomes incomplete) was however left open. Sections~\ref{ss:global} and \ref{s:global} provide an answer to this question.

Relaxing translational invariance, several scenarios have been recently identified where a perturbation localized deep in the bulk leads to the growth of QFI up to macroscopic scales. This phenomenon occurs, for instance: when the system is prepared in a quantum jammed state of a kinetically constrained model \cite{Fagotti2024Quantum},
in low-entangled many-body quantum scars~\cite{Bocini2023Growing},
in systems with semilocal conservation laws~\cite{Fagotti2022Global}.
Multipartite entanglement has been observed in those settings when time evolution remains confined within a lightcone, considering subsystems  surrounding the lightcone. Since those subsystems remain in nearly pure states, the QFI can be effectively approximated by the variance, making its macroscopic scaling evident. However, it remained unclear whether the phenomenon required the subsystem to be pure.
We have recently investigated  another setting where the subsystems are instead in mixed states, yet the QFI  scales as the square of the subsystem length. Specifically, we considered a semi-infinite chain at finite temperature brought into point contact with a semi-infinite chain in a symmetry-breaking ground state. This setup leads to the propagation of a disorder-order interface at maximal velocity, and we found that large subsystems within the interfacial region fall into macroscopic quantum states~\cite{Maric2024Macroscopic, Maric2024Disorder}.
However, while conceptually intriguing, the quantitative impact of this result is somewhat underwhelming. The examples  analyzed in Ref.~\cite{ Maric2024Disorder} confirm that $\chi(\rho,O)$ remains nonzero for large subsystems, but its magnitude is relatively small and the typical length of the multipartite entangled subsystems grows only as $t^{1/3}$, limiting the practical implications.

We propose in the following more effective protocols.

\subsection{Kicking protocol}

\begin{figure}[t]
\includegraphics[width=0.48\textwidth]{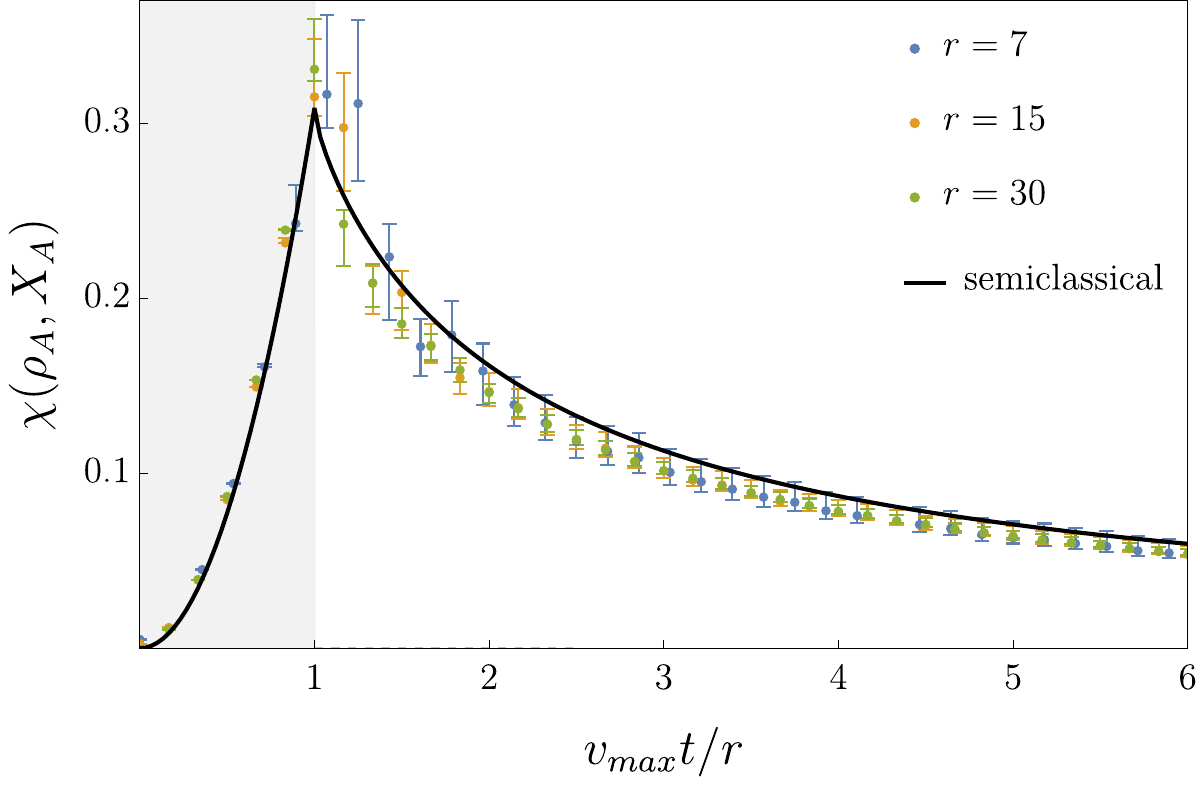}
\caption{The normalized QFI  $\chi(\rho_A,X)$~\eqref{eq:chi} of a spin block $A = \llbracket -r, r\rrbracket$ after a spin flip at site $0$ evolving under the Ising hamiltonian with $h=0.5$. The error bars correspond to the bounds \eqref{eq:QFIbounds}, and the solid line is the semiclassical approximation~\eqref{eq:chibasic}, which does not take into account the interference between nearby excitations. The shaded region corresponds to times for which the subsystem is pure in the semiclassical theory.}
\label{f:QFI_singlespinflip}
\end{figure}

We imagine a quantum spin chain that is sufficiently isolated from the environment and is described by a Hamiltonian with a symmetry that is spontaneously broken at zero temperature. We consider the following protocol:
\begin{enumerate}
\item The system is prepared in equilibrium at low temperature.
\item \label{s:perturbation}The state is locally perturbed (by a unitary transformation with finite support) for a finite time, which we assume to be instantaneous.  
\item The system undergoes unitary time evolution with the equilibrium Hamiltonian.
\item Possible repetition from step~\ref{s:perturbation}. 
\end{enumerate}

Studying this protocol  in a quantum spin chain that can be mapped to free fermions, such as the transverse-field Ising chain, offers significant practical advantages.  The Gaussian nature of the state enables highly efficient numerical simulations and allows for the exact computation of the Wigner-Yanase(-Dyson) skew information, as firstly shown in Refs~\cite{Maric2024Macroscopic,Maric2024Disorder}.

However, focusing on a specific model carries the risk of encountering exceptional behaviors unique to the system. To ease this issue, we carry out  an asymptotic  expansion for the QFI, which is valid also in the presence of interactions, and reinterpret the exact results derived from explicit calculations and numerical computations within a broader semiclassical framework, ensuring the insights remain applicable beyond the peculiarities of the model. The asymptotic expansion is outlined in Section~\ref{ss:beyond}. 
Here we anticipate some results of the semiclassical approximation. Even if we do not expect qualitative changes in generic models---the lower the number of excitations and the stronger the effect---we assume the system to be integrable, which means there are stable quasiparticle excitations.

First of all, we point out 
that the dynamics strongly depend on the localized perturbations applied to a symmetry-breaking ground state. Such feature was already evident from the previous studies on the behavior of correlation functions carried out by Eisler and Maislinger~\cite{Eisler2020Front}: the details of the perturbation do not become irrelevant as the time approaches infinity. 
Simplifications occur when the perturbation generates only single-particle excitations of the same species, allowing the system to be treated as noninteracting. This kind of perturbations in ferromagnetic phases are not strictly local, as they create domain walls (see also Ref.~\cite{capizzi2024entanglementcontentkinkexcitations}). A truly local perturbation must involve at least two elementary excitations. These excitations either manifest themselves as magnons, characterized by model-dependent scattering phases, or may form bound states. Generically, the probability of forming bound states decreases as the initial excitations become more spatially separated. Additionally, greater separation reduces interference as well as the effects of phase shifts acquired during scattering.   

We focus on systems with a spontaneously broken $\mathbb Z_2$ symmetry and consider first the case of a single localized perturbation that satisfies the conditions above. Our semiclassical analysis is consistent with the previous investigations and reveals that the order-parameter correlation functions can be expressed in terms of a monotonous continuous scaling function $\mathcal M$ of the rescaled positions that depends only on the dispersion relation of the single-particle excitations. Specifically, we find
\begin{equation}\label{eq:12points}
\begin{aligned}
\braket{O_\ell }\sim&
\braket{\textsc{gs}|O_\ell |\textsc{gs}}[\mathcal M(\tfrac{\ell-j^0}{\bar v_{M} t})]^2\\
&+O(\tfrac{R}{t})+O(\text{interference})+O(\text{other species})\\
\braket{O_\ell O_n}\sim&
\braket{\textsc{gs}|O_\ell O_n|\textsc{gs}}\left[1-\Bigl|\mathcal M(\tfrac{n-j^0}{\bar v_{M} t})-\mathcal M(\tfrac{\ell-j^0}{\bar v_{M} t})\Bigr|\right]^2\\
&+O(\tfrac{R}{t})+O(\text{interference})+O(\text{other species})
\end{aligned}
\end{equation}
where $j^0$ is the position of the perturbation, $R$  its  range and $\bar v_M$ is the maximal group velocity of the single-particle dispersion relation. We have been explicit about the presence of a correction $O(\tfrac{R}{t})$ because it represents a contribution that can be  easily understood within the semiclassical theory.  The source of error that we labeled as ``interference'' is instead  beyond the  theory, as it  depends on additional system details that are expected to become less and less important as the distance between the original positions of the particles is increased (the error is expected to decay algebraically with $r$---see Section~\ref{ss:beyond}). 
The error labeled as ``other species'' includes the possible formation of bound states mentioned earlier. 
Function $\mathcal M(x)$ describes the magnetization after a domain-wall perturbation generating only single-particle excitations, hence---by the Lieb-Robinson bounds~\cite{Lieb1972The}---satisfies $\mathcal M(x)=\mathrm{sgn}(x)$ for $|x|>1$.  From \eqref{eq:12points}, one can then readily conclude that subsystems of extent $\sim t $ do not possess cluster decomposition properties. This simple observation, which could also be inferred from the previous studies, opens the door to the possibility of subsystems falling in macroscopic quantum states. We have tested it within the semiclassical theory, which suggests \eqref{eq:FeqI} to hold  asymptotically. We find
\begin{multline}\label{eq:chibasic}
    \chi(\rho_A, O_A) =\frac{1}{\parallel O_A\parallel ^2}\tr[\rho_AO_A^2] \\
    - \braket{\textsc{gs}|O_\ell |\textsc{gs}}^2 \iint_{\mathcal A_t}\frac{d^2 x}{|\mathcal A_t|^2} \left(p_{\mathcal A_t} \sigma_{\mathcal A_t}(x_1)\sigma_{\mathcal A_t}(x_2)+p_{\overline{\mathcal A_t}}\right)^2\, ,
\end{multline}
where $\mathcal A_t$ is the rescaled region associated with $A$ ($j\in A\Leftrightarrow \frac{j-j^0}{\bar v_{M}t}\in \mathcal A_t$), and we defined
\begin{equation}
\begin{aligned}
\sigma_{B}(x)=&\frac{\int_{B}d y\, \mathrm{sgn}(x-y)\mathcal M'(y)}{\int_{B} d y\mathcal M'(y)}\\
p_B=&\frac{1}{2}\int_{B} d y\mathcal M'(y)\, .
\end{aligned}
\end{equation}

Fig.~\ref{f:QFI_singlespinflip} shows a comparison between numerical data and semiclassical prediction in the transverse-field Ising model in which the ground state is perturbed, at time $t=0$, with a spin flip at site $0$ represented by the unitary $\sigma^z_0$. In the TFIM there is a single species of excitations and this perturbation produces a superposition of excited states with two excitations and the ground state. 
The small but visible disagreement between data and the semiclassical approximation is due to the residual overlap with the ground state and to 
the aforementioned interference terms---see Section~\ref{s:scQFI}.

Regardless of the subsystem's length, there exists a time window during which the subsystems remain macroscopically entangled. However, as the time per unit length grows significantly larger than the inverse of the maximal velocity, the normalized QFI decays as $1/t$. This decay occurs because, at large times, only the slowest particles remain within the subsystem, leading to the eventual suppression of macroscopic entanglement. Consequently, the duration of the macroscopic entanglement regime is proportional to the subsystem length.

\begin{figure}[t]
\includegraphics[width=0.48\textwidth]{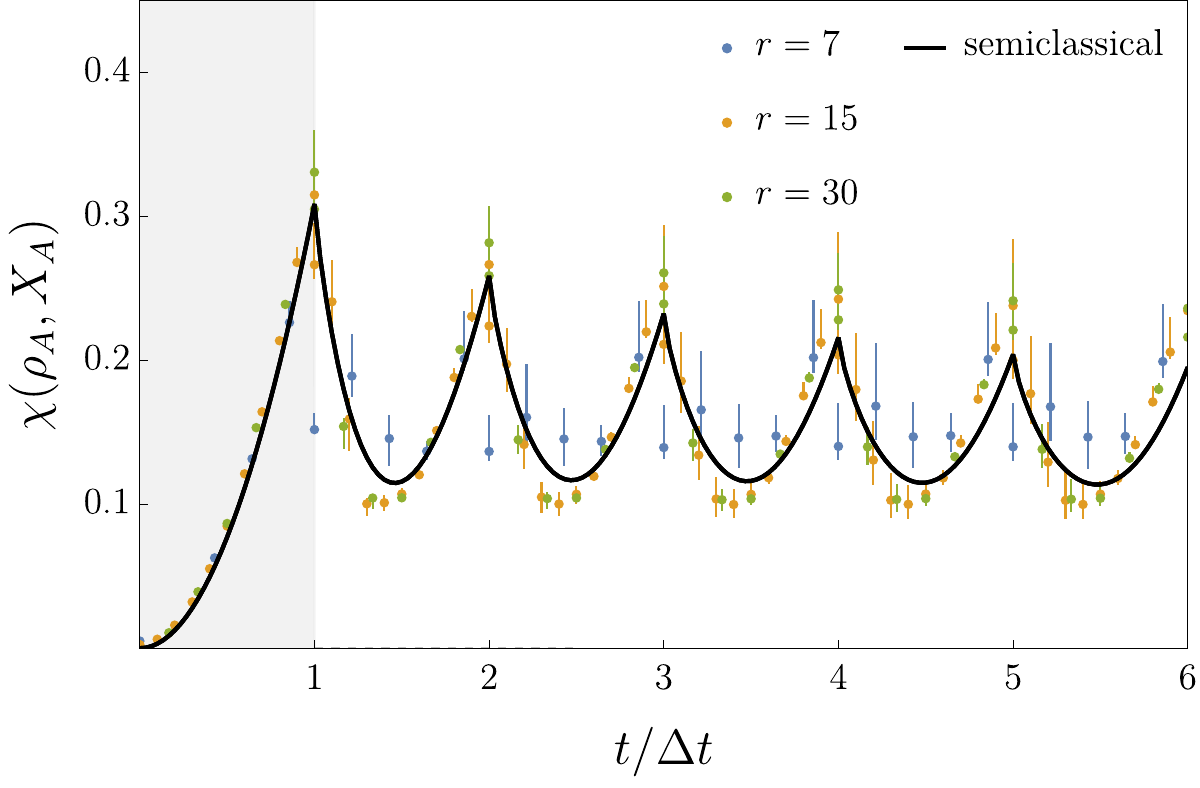}
\caption{The normalized QFI  $\chi(\rho_A,X)$~\eqref{eq:chi} of a spin block $A = \llbracket -r, r\rrbracket$ under a kicking protocol: every $\Delta t=r/v_{max}$, the spin at site $0$ is flipped, while evolving at all times under the Ising hamiltonian with $h=0.5$. The error bars correspond to the bounds \eqref{eq:QFIbounds}, and the solid line is the semiclassical approximation~\eqref{eq:qfi_pred_DW} with~\eqref{eq:qfi_pred_difftimes}, which does not take into account the interference between nearby excitations. The shaded region corresponds to times for which the subsystem is semiclassically pure.}
\label{f:QFI_kick_spinflip}
\end{figure}

In fact, the protocol can be easily enhanced. If the influence of the slowest particles could be disregarded, then resetting the protocol---i.e., returning to Step~\ref{s:perturbation}---would continuously regenerate macroscopic entanglement, leading to a sustained enhancement of QFI. While this assumption is admittedly strong, Fig.~\ref{f:QFI_kick_spinflip} illustrates that such a local periodic driving scheme can be highly effective in maintaining a nonzero $\chi$ indefinitely.
Moreover, by tuning the period of the local kicking, one can effectively control the characteristic length of subsystems that remain in macroscopic quantum states, as this length is directly proportional to the driving period.

\begin{figure}[t]
\includegraphics[width=0.48\textwidth]{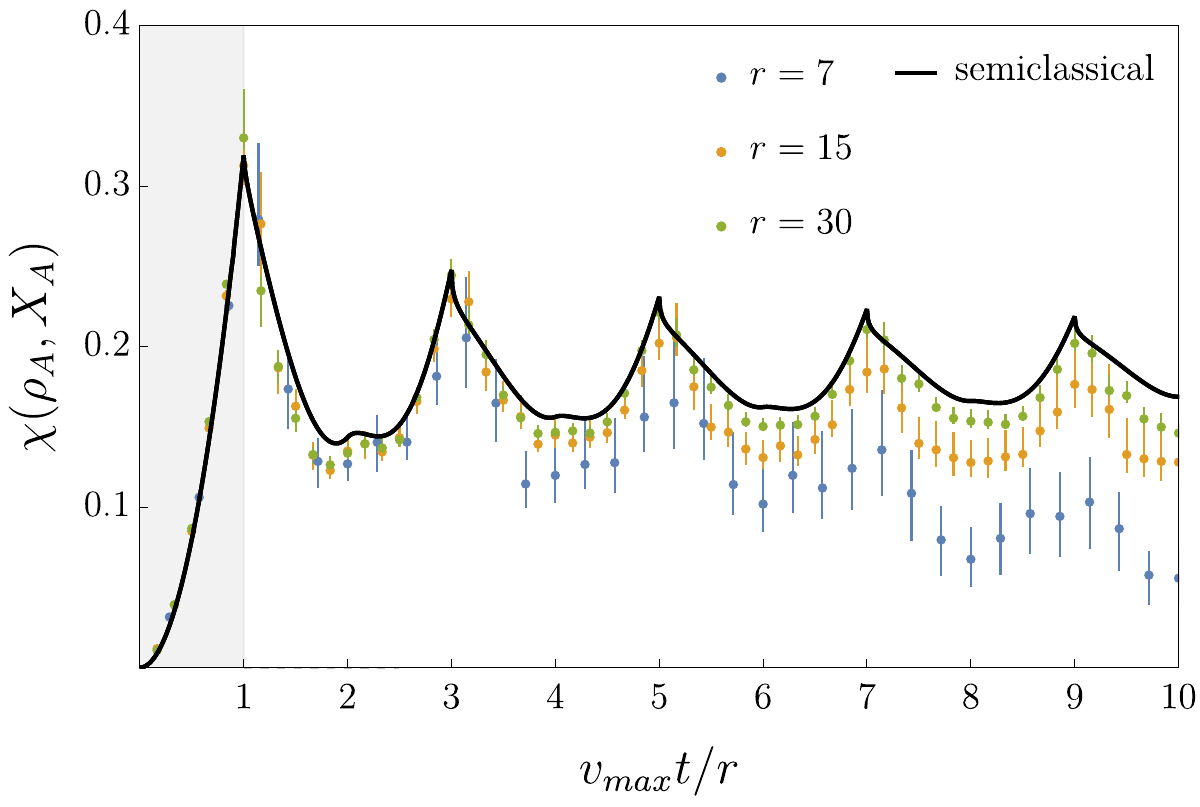}
\caption{The normalized QFI  $\chi(\rho_A,X)$~\eqref{eq:chi} of a spin block $A = \llbracket -r, r\rrbracket$ for an initial state consisting of the ground state with spin flips at positions $2rz$ with $z\in \mathbb{Z}$  with respect to the order parameter $X_A=\sum_{\ell\in A}\sigma_\ell^x$ as a function of $\bar v_{M}t/r$. The error bars correspond to the bounds \eqref{eq:QFIbounds}, and the solid line is the semiclassical approximation~\eqref{eq:qfi_pred_DW}, which does not take into account the interference between nearby excitations. The shaded region corresponds to times for which the subsystem is semiclassically pure.}
\label{f:QFI_spinflip}
\end{figure}

The effectiveness of this protocol makes one wonder if exchanging time for space in the localized kicking could yield a similar enhancement. In other words, what would happen if the initial state were perturbed with a grid of localized perturbations applied simultaneously at $t=0$?
If the distance between perturbations is kept fixed, this setup can be interpreted as a global quench with translational invariance---specifically, invariance under shifts by the number of sites needed to align one perturbation with the next. However, the dynamical properties at times and length scales comparable to the perturbation spacing may exhibit unconventional behavior, as conventional studies typically focus on systems that are invariant under few site shifts.
Fig.~\ref{f:QFI_spinflip} shows that the QFI exhibits an enhancement similar to the other case, but the effect is temporary, indeed, at late times, the normalized QFI seems to approach zero. Indeed, it is reasonable to expect that at late times the distance between the perturbations becomes an irrelevant length and the behavior corresponds to that of more standard global quenches, discussed in the following section.

\subsection{Global quench}\label{ss:global}

\begin{figure*}[t]
\includegraphics[width=0.5\textwidth]{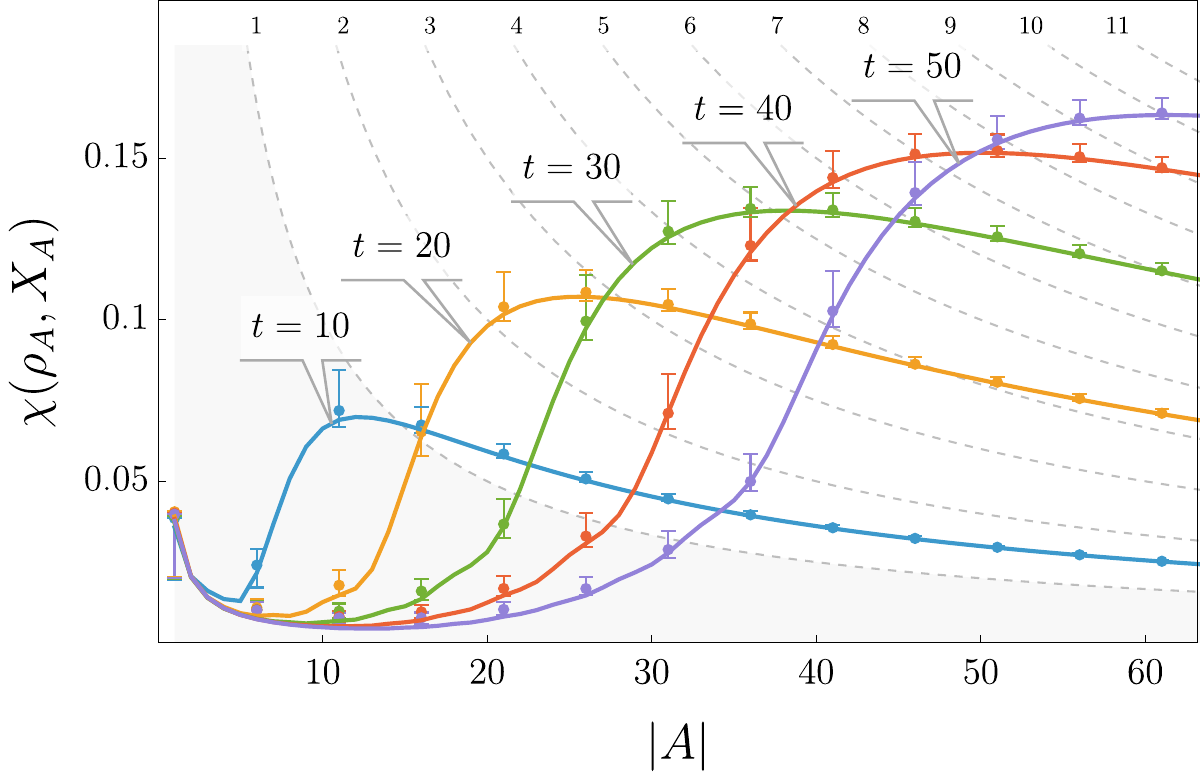}\includegraphics[width=0.47\textwidth]{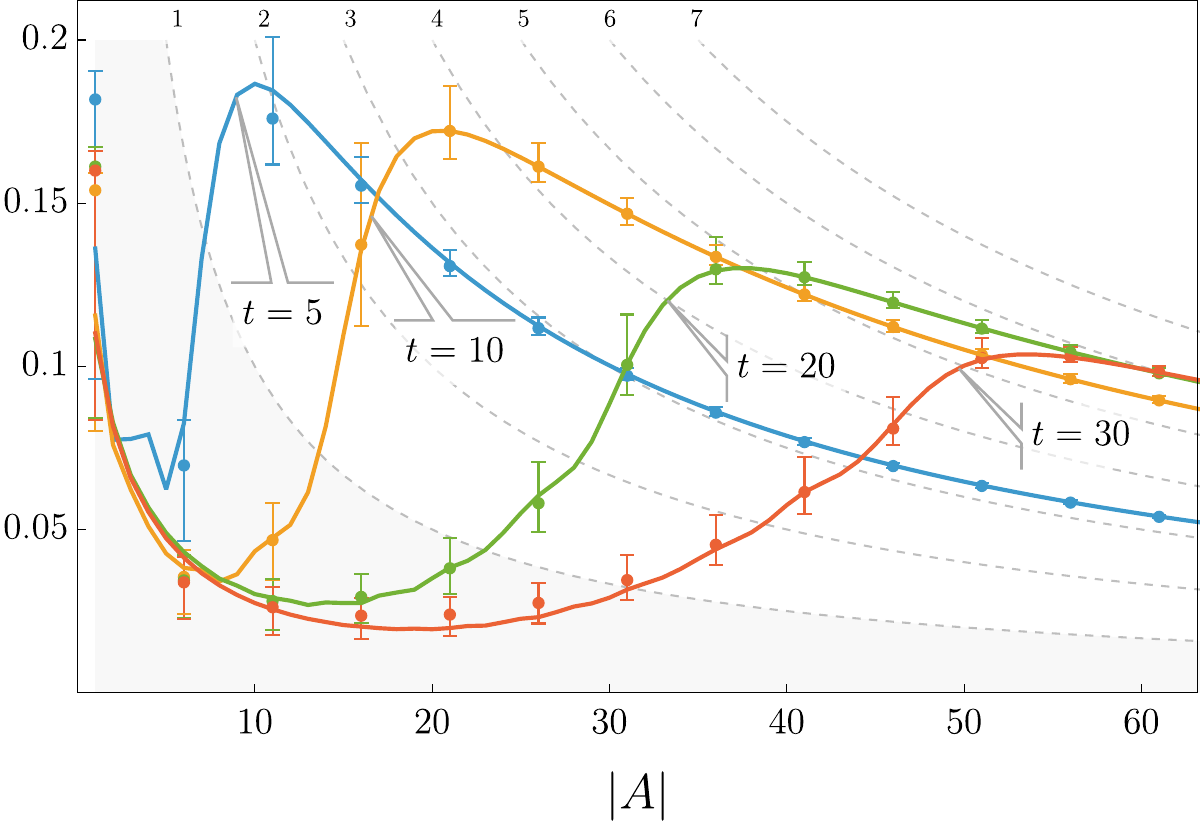}
\caption{The normalized QFI~\eqref{eq:chi}---a quarter of the quantum Fisher information per unit norm squared---of a spin block $A$ after a global quench in the transverse-field Ising model with $h=0\rightarrow 0.4$ (left) and $h=0\rightarrow 0.8$ (right)  with respect to the order parameter $X_A=\sum_{\ell\in A}\sigma_\ell^x$ at several times. The error bars correspond to the bounds \eqref{eq:QFIbounds}. The solid curves are the predictions \eqref{eq:predictionquench}. The grid of dashed curves shows integer multiples of $|A|^{-1}$. Left: The quench is rather small and the QFI shows the signature of multipartite entanglement. Right: Multipartite entanglement is extinguished very quickly.}
\label{f:quench0to0dot4or8}
\end{figure*}

As anticipated earlier, 
Ref.~\cite{Pappalardi2017Multipartite} found indications that the variance of the order parameter in a finite transverse-field Ising chain after a quench within the ferromagnetic phase can become arbitrarily large in the limit of small quench. 
In the thermodynamic limit the equivalence between QFI and variance breaks down as information about the system becomes incomplete. Consequently, calculating the QFI becomes significantly more challenging. 
As detailed in Section~\ref{s:global}, however, in noninteracting spin chains such as the transverse-field Ising model, the Wigner-Yanase-Dyson skew information can be computed numerically also in the thermodynamic limit with a complexity that scales polynomially with the subsystem's length.  This allows us to bound the QFI using, e.g., the inequalities in \eqref{eq:QFIbounds}, or even to estimate it, using the remarkable identity~\eqref{eq:QFIidentity} and the procedure described in Section~\ref{ss:numerical}.
  
A priori, global quenches in translationally invariant integrable systems are not promising settings where to find significant multipartite entanglement in the thermodyamic limit. 
Indeed, quasiparticle pictures have proved very effective, and they are generally based on the interpretation of the state as a collection of particles that are correlated only within small groups (often, just pairs). One might  then be led to think that large subsystems will never be in macroscopic quantum states.  
However, this intuition overlooks that elementary excitations could have nontrivial topological properties.  In this respect, Ref.~\cite{Maric2024Macroscopic} showed  that domain-wall excitations in a ferromagnetic phase can trigger multipartite entanglement.  The caveat is that this  happens when the density of particles approaches zero. 
After a global quench, such a density remains nonzero, thereby undermining multipartite entanglement in arbitrarily large subsystems. However, this does not preclude the possibility that subsystems significantly smaller than the correlation length may, at some point, fall into macroscopic quantum states.

Our numerical analysis shows that after quenches within a ferromagnetic phase, remarkably large subsystems can become multipartite entangled at intermediate times---see Fig.~\ref{f:quench0to0dot4or8}.
We understand this behavior in the framework of a semiclassical theory of domain-wall excitations assuming that the system is prepared in a squeezed state consisting of pairs of stable quasiparticles with opposite momenta. The calculations are done assuming that the both the subsystem's length and the time are large. Our explicit example is the transverse-field Ising model, but the structure of the state is common to other models, even in the presence of interactions~\cite{Alba_2017, Piroli_2019}. Thus, we present the result in a form that could be tested in a generic quench within a ferromagnetic phase. 
If $O_\ell$ is a local order parameter, we argue that, at given time $t$, there is an interval of lengths $|A|=r-l+1$ in which the QFI of  subsystem $A=\llbracket l,r \rrbracket$ with respect to $O_A=\sum_{\ell\in A} O_\ell$ is captured by the following expression
\begin{multline}\label{eq:predictionquench}
\tfrac{1}{4}F_Q(\rho_{A},O_A)\sim\\
\sum_{\ell,n\in A}\braket{O_\ell O_n}-\tfrac{\braket{O_{x_l-1}O_{\max(\ell,n)}}\braket{O_{\min(\ell,n)} O_{x_r+1}}}{\braket{O_{x_l-1}O_{x_r+1}}}
\end{multline}
Since the semiclassical calculation assumes large subsystems and large times, one could expect this formula to  hold asymptotically in the scaling limit $|A|,t\rightarrow\infty$ at fixed ratio $|A|/t$. However, as mentioned above, similar arguments are expected to break down when $|A|$ approaches the correlation length, which remains finite at any time. While the nature and magnitude of corrections to this approximation remain theoretically unclear, Fig.~\ref{f:quench0to0dot4or8} demonstrates a striking agreement between numerical results and predictions---exceeding even the most optimistic expectations. In Section~\ref{s:global} we exhibit also an almost exact integral formula  for the asymptotic behavior of \eqref{eq:predictionquench} in the transverse-field Ising model, which is based on the results of Ref.~\cite{Calabrese2012Quantum1} and is not exact only because of a subtle issue pointed out in Ref.~\cite{Granet2020Finite}.

\begin{figure}[t]
\includegraphics[width=0.48\textwidth]{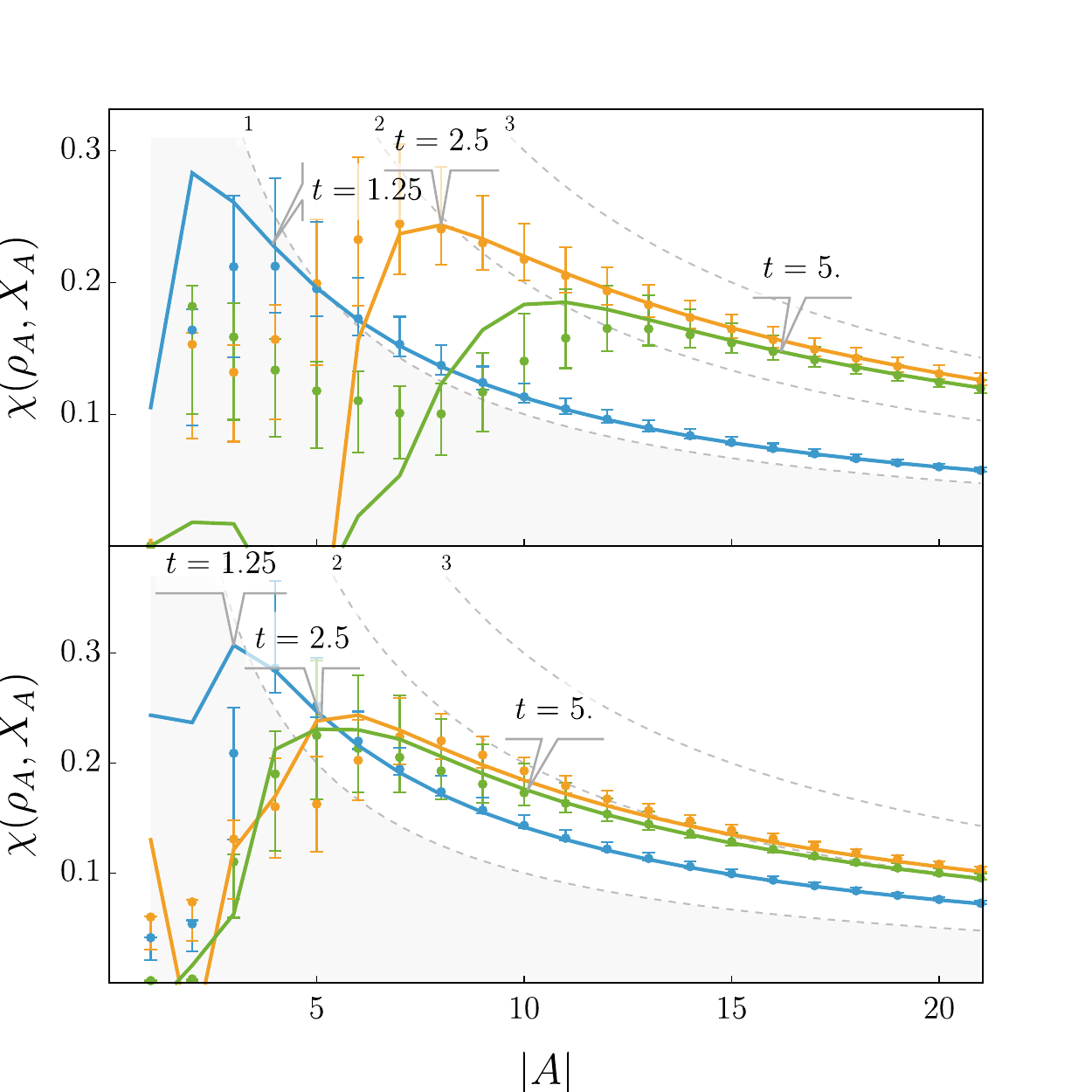}
\caption{The normalized QFI  $\chi(\rho_A,X)$~\eqref{eq:chi} of a spin block $A$  with respect to the order parameter $X_A=\sum_{\ell\in A}\sigma_\ell^x$ as a function of $|A|$ after a ferro-to-para quench $(h,\gamma)=(0,1)\rightarrow (2,0)$  (top) and after a ferro$(x)$-to-ferro$(y)$ quench $(h,\gamma)=(0,1)\rightarrow (0.7,-1)$ (bottom) in the quantum XY model. 
The points, the error bars and the lines have the same meaning as in Fig.~\ref{f:quench0to0dot4or8}}
\label{f:limitations}
\end{figure}

We remark that \eqref{eq:predictionquench} is meaningful only if the two-point function of the order parameter is nonzero, which already highlights certain limitations in its applicability. For example, we  see issues when considering quantum quenches in which the  post-quench Hamiltonian describes a different phase of matter, such as after quenches from ferromagnetic to paramagnatic phases or  between two different ferromagentic phases---Fig.~\ref{f:limitations}. 
We do not have a sensible explanation for why also in those cases the formula seems to fairly capture the behavior for sufficiently large $|A|$; we only note that in the limit of large $|A|$ both the QFI and the right hand side of \eqref{eq:predictionquench} approach the variance.

\section{The model}\label{s:model}
As the paradigmatic spin-chain model exhibiting a zero-temperature ferromagnetic phase, the transverse-field Ising model (TFIM) serves as our reference system. In one dimension, it holds a special status for several reasons, including:
\begin{enumerate}
\item It is integrable;
\item It has a trivial scattering phase ($-1$);
\item It is reflection symmetric, and the quasiparticle velocity has a single local maximum.
\end{enumerate}
The first two properties arise because its Hamiltonian can be mapped onto free fermions via a Jordan-Wigner transformation.
 The TFIM is often introduced as the special case $\gamma=1$ of the more general quantum XY model, which is an exactly solvable noninteracting spin-chain system with Hamiltonian
\begin{equation}\label{eq:HXY}
H(h,\gamma)=-\sum_{\ell\in \mathbb Z}\tfrac{1+\gamma}{2}\sigma_\ell^x\sigma_{\ell+1}^x+\tfrac{1-\gamma}{2}\sigma_\ell^y\sigma_{\ell+1}^y+h\sigma_\ell^z
\end{equation}
The XY system exhibits a spin-flip symmetry, given by $\mathcal P( \sigma_\ell^{x,y})=- \sigma_\ell^{x,y}$, $\mathcal P( \sigma_\ell^z)= \sigma_\ell^z$. At zero temperature, this symmetry is spontaneously broken in the ferromagnetic phase ($|h|<1$ and $\gamma\neq 0$). In contrast, for $|h|>1$, the ground state of $H(h,\gamma) $ corresponds to a paramagnetic phase. 
The XY system undergoes two distinct quantum phase transitions:
\begin{itemize}
\item At $|h|=1$, marking the boundary between the ferromagnetic and paramagnetic phases, it is described by the  CFT of the critical Ising model.
\item At $\gamma=0$ with $|h|<1$, which separates two symmetry-broken phases, it is described by the CFT of a  free boson compactified on a circle.
\end{itemize}
For $\gamma>0$ ($\gamma<0$),  the conventional order parameter is the longitudinal magnetization, given by the expectation value $\braket{ \sigma^x}$ (or $\braket{ \sigma^y}$, respectively). 

The quantum XY model shares the noninteracting structure with  the TFIM. While the explicit examples that we present are specific to the TFIM, all the exact formulas that we exhibit are written in a way that can be applied also in more general noninteracting spin chains, such as the XY model. 

Finally, an alternative,  simplified model that allows for exact calculations and can be easily generalized to interacting spin chains is presented in Section~\ref{ss:beyond}.
\subsection{Fermionic representation}\label{ss:free_fermion}
The Jordan-Wigner transformation is a duality mapping from spins $\frac{1}{2}$ onto spinless fermions. We express it in terms of Majorana fermions $a_j=a_j^\dag$ satisfying $a_j a_n=2\delta_{jn}-a_n a_j$:
\begin{equation}
a_{2l-1}=\prod_{j<l}\sigma_j^z\, \sigma_l^x\qquad a_{2l}=\prod_{j<l}\sigma_j^z\, \sigma_l^y\, .
\end{equation}
We call a state Gaussian if the Wick's theorem holds with respect to the Jordan-Wigner fermions, hence all expectation values can be written in terms of the fermionic 2-point functions $\Gamma=\mathrm I-\braket{\vec a\otimes \vec a}$. Such states are described by density matrices of the form $\frac{1}{Z_W} \exp(\frac{1}{4}\vec a\cdot  W\vec a)$, where $W$ is skew-symmetric. 

A product of Pauli matrices is a product of Majorana fermions, hence Wick's theorem provides a Pfaffian representation for its expectation value. Specifically, the 2-point function of $\sigma^x$ reads
\begin{equation}
\braket{\sigma_\ell^x \sigma_n^x}=(-i)^{n-\ell} \braket{a_{2\ell}\cdots a_{2n-1}}=i^{n-\ell}\mathrm{pf}(\Gamma_{\llbracket 2\ell,2n-1\rrbracket})
\end{equation}
where $\Gamma_{\llbracket 2\ell,2n-1\rrbracket}$ is the finite section of $\Gamma$ in which the indices are restricted to the set $\llbracket 2\ell,2n-1\rrbracket$.

The first technical insight that allows us to compute the quantum Fisher information is that the product of Gaussian density matrices is proportional to a Gaussian density matrix. Specifically, denoting by $\rho[\Gamma]$ the Gaussian density matrix with correlation matrix $\Gamma$, we have~\cite{Fagotti2010disjoint}
\begin{equation}\label{eq:Gaussprod}
\begin{aligned}
\rho[\Gamma_1]\rho[\Gamma_2]=&\sqrt{\det\tfrac{\mathrm I+\Gamma_1\Gamma_2}{2}}\rho[\Gamma_1\times \Gamma_2]\\
\Gamma_1\times\Gamma_2=&1-(1-\Gamma_2)\frac{\mathrm I}{\mathrm I+\Gamma_1\Gamma_2}(1-\Gamma_1)\, .
\end{aligned}
\end{equation}
We also have
\begin{equation}
\begin{aligned}
(\rho[\Gamma])^\alpha =& \kappa_\alpha(\Gamma) \rho[\Gamma^{(\alpha)}]\\
\Gamma^{(\alpha)}=&\tanh(\alpha\, \mathrm{artanh}(\Gamma))\, ,
\end{aligned}
\end{equation}
and from \eqref{eq:Gaussprod} it follows
\begin{equation}
\kappa_\alpha(\Gamma)\kappa_{1-\alpha}(\Gamma)=
1/\sqrt{\det\tfrac{\mathrm I+\Gamma^{(\alpha)}\Gamma^{(1-\alpha)}}{2}}\, .
\end{equation}

The second technical insight is that a unitary transformation represented by a string of Pauli matrices preserves Gaussianity. In particular we have
\begin{equation}
\sigma^x_\ell \rho[\Gamma]\sigma^x_\ell=\rho[D^x_{\ell}\Gamma D^x_{\ell}]
\end{equation}
where $D^{x}_\ell$ denotes the diagonal matrix with elements 
\begin{equation}
[D^x_{\ell}]_{j j}=\begin{cases}
-1&j\leq 2\ell-1\\
1&\text{otherwise.}
\end{cases}
\end{equation}
Thus, we can readily express the Wigner-Yanase-Dyson skew information of $\rho_A$---describing subsystem $A$---with respect to $X_A=\sum_{\ell\in A} \sigma_\ell^x$ as a double sum  
\begin{multline}\label{eq:Ialphafree}
I_\alpha(\rho_A,X_A)=2\sum_{\ell\leq n\in A}i^{n-\ell}\mathrm{pf}(\Gamma_{\llbracket 2\ell,2n-1\rrbracket})\\
-2\sum_{\ell\leq n\in A}\sqrt{\tfrac{\det|\mathrm I+\Gamma_A^{(\alpha)}D^x_{\ell}\Gamma_A^{(1-\alpha)}D^x_{\ell}|}{\det|\mathrm I+\Gamma_A^{(\alpha)}\Gamma_A^{(1-\alpha)}|}}\\
\times \mathrm{Re} \Bigl[i^{n-\ell}
\mathrm{pf}([\Gamma_A^{(\alpha)}\times(D^x_{\ell} \Gamma_A^{(1-\alpha)}D^x_{\ell})]_{\llbracket 2\ell,2n-1\rrbracket})\Bigr]\, ,
\end{multline}
where, by abuse of notation, we set $\mathrm{pf}(M_{\llbracket 2\ell,2\ell-1\rrbracket})=\frac{1}{2}$ whatever $M$ is. Formula~\eqref{eq:Ialphafree} is used in our numerical investigations and serves as the key component for computing the QFI via identity~\eqref{eq:QFIidentity}.  
\section{Kicking protocol} \label{s:kicking}%
In this section, we present the standard semiclassical theory~\cite{PhysRevLett.78.2220} that qualitatively describes the behavior of correlation functions following local perturbations to a ferromagnetic ground state. We will show that, in certain limits, the semiclassical approximation provides not only qualitative but also quantitatively accurate predictions. Building on this framework, we will extend the approach to describe the behavior of the subsystem quantum Fisher information with respect to an extensive order parameter. 

\subsection{Semiclassical picture}
We focus on spin-$\frac{1}{2}$ chains in a ferromagnetic phase in which a spin-flip symmetry is spontaneously broken. Without loss of generality, we can assume that the symmetry is generated by the spin-flip operator $\Pi^z=\prod_\ell\sigma_\ell^z$. 
Let us assume that the symmetry breaking ground states $\ket{\pm}$ are locally equivalent to the product states where all spins are aligned along $x$ or in the opposite direction
\begin{equation}
\ket{+}=U\ket{\Rightarrow_x}\qquad \ket{-}=U\ket{\Leftarrow_x}
\end{equation}
with $[U,\Pi^z]=0$. In the trasverse-field Ising chain, for example, $U$ is the Bogoliubov transformation, which indeed is quasilocal within the ferromagnetic phase~\cite{Bocini2024No}.  
Arguably, the best quasilocal order parameter is $M_\ell=U\sigma_\ell^x U^\dag$, in that its expectation value in the symmetry-breaking ground state is maximal. 

Dynamical properties at low energy are characterized by quasiparticles that can be visualized as domain walls interpolating the two symmetry-breaking ground states. 
We refer the reader to Ref.~\cite{PhysRevLett.78.2220} and Ref.~\cite{Rieger_2011} for the use of
this semiclassical picture in thermal equilibrium and in out-of-equilibrium dynamics, respectively.  The (semiclassical) quasiparticles move along straight lines with a constant velocity depending on the low-energy dispersion relation. We indicate their group velocity by $v_k$. In the quantum XY model~\eqref{eq:HXY}, we have $v_k=\frac{2h\sin k+2(\gamma^2-1)\sin k\cos k}{\sqrt{(h-\cos k)^2+\gamma^2\sin^2 k}}$. When a quasiparticle crosses a position $\ell$, the expectation values of order parameters such as $M_\ell$ reverse their sign. This observation is already sufficient to estimate the order-parameter correlations functions after localized perturbations. 

Let us apply the semiclassical picture to our specific protocol. We consider an initial state with 
$n$ domain walls localized at positions
$j_1^0, \cdots, j_n^0$ on top of the symmetry-broken ground state $\ket{+}$. For the Ising model, this setup corresponds to the protocol studied in Ref.~\cite{Eisler2020Front}.  A local order parameter $O_\ell$ satisfies the same conditions as the ideal order parameter $M_\ell$ but does not attain a maximal expectation value in the ground state. In the semiclassical picture, the expectation value of $O_\ell$ with a single domain wall at $j_1^0$ is given by---see Fig.~\ref{f:domain_wall}
\begin{equation}\label{eq:1partmag}
    \braket{O_\ell} \sim  \braket{+|O_\bullet|+} \int_{-\pi}^\pi 
    d k \sigma(k)\sgn(\ell-j_1^0-v_k t)\, ,
\end{equation}
where $O_\bullet$ emphasizes that the expectation value is translationally invariant and $\sigma(k)$ is the cross section of production of the quasiparticle with momentum $k$. This result follows from counting the quasiparticles on either side of the operator's position. 
Similarly, the two-point function can be estimated. For distances $|n-\ell|$ much larger than the ground-state correlation length, we obtain
\begin{equation}
   \braket{O_\ell O_n} \sim \braket{+|O_\bullet|+}^2 \!\!\!\int_{-\pi}^\pi 
   dk \sigma(k)\sgn(\ell-v_k t)\sgn(n-v_k t)\, .
\end{equation}
This expression accounts for the fact that the two-point function changes sign only when a domain wall lies between the two operators.

For multiple well-separated domain walls, interference effects are negligible, and their contributions factorize as
\begin{equation}\label{eq:correlations}
\begin{aligned}
    \braket{O_\ell} \sim & \braket{+|O_\bullet|+} \prod_{i=1}^n \mathcal M(\tfrac{\ell-j_i^0}{\bar v_{M} t}),\\
    \braket{O_\ell O_m} \sim & \braket{+|O_\bullet|+} ^2 \prod_{i=1}^n \bigl(1-|\mathcal M(\tfrac{m-j_i^0}{\bar v_{M}t})-\mathcal M(\tfrac{\ell-j_i^0}{\bar v_{M}t})|\bigr)\, ,
\end{aligned}
\end{equation}
where
\begin{equation}\label{eq:Mcal}
\mathcal M(\zeta) = \int_{-\pi}^\pi 
dk \sigma(k)\sgn(\zeta-\tfrac{v_k}{\bar v_{M}})\, .
\end{equation}
In the specific case of the Ising model perturbed by a  Majorana fermion $a_\ell$, we have $\mathcal M(\zeta) = \frac 2 \pi \arcsin \zeta$ for $\zeta\in(-1, 1)$, and $\mathcal M(\zeta)= \sgn \zeta$ otherwise (see also Ref~\cite{Eisler2016Universal}). 

Fig.~\ref{f:local_perturbation} depicts the processes associated with a local perturbation that excites only two quasiparticles. Interference comes from the lack of orthogonality between the states in which the quasiparticles are interchanged, and it is therefore also sensitive to phases that go beyond the semiclassical theory---see Section~\ref{ss:beyond}.  

\subsection{Semiclassical QFI}\label{s:scQFI}

\begin{figure}[t]
\includegraphics[width=0.48\textwidth]{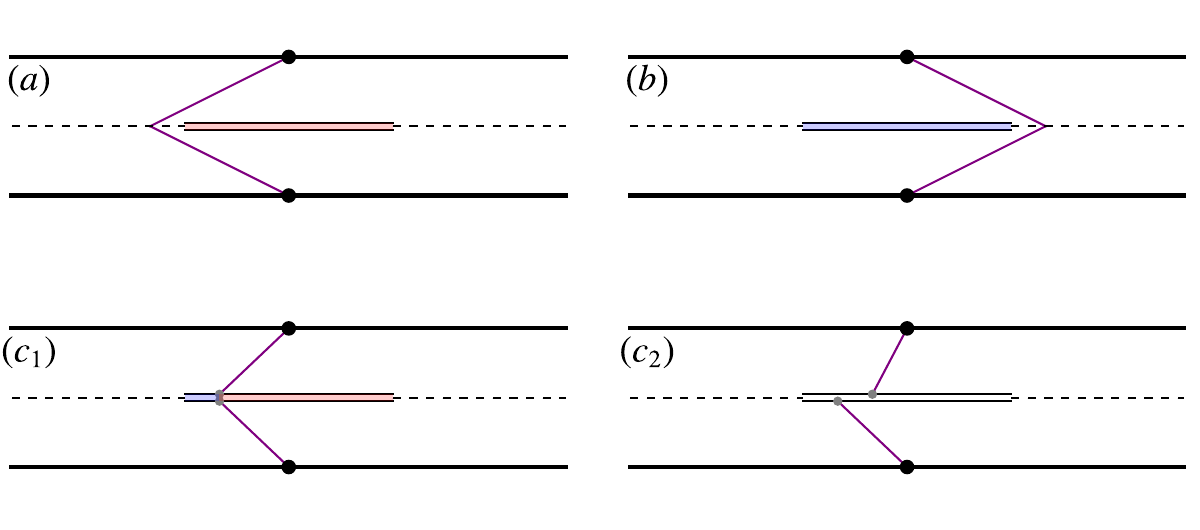}
\caption{List of the processes triggered by a domain-wall excitation. The solid parallel lines represent the ket  and the bra associated with the state. The black circle is the semilocal perturbation. The purple straight lines represent the propagating particles. The subsystem at time $t$ is depicted as the nearby parallel segments. The dashed lines represent instead the partial trace over the rest of the system. Processes $(a)$-$(b)$-$(c_1)$ describe diagonal element of the RDM, whereas $(c_2)$ describes  off-diagonal terms. The sum of the contributions  associated with $(c_1)$ and $(c_2)$ is coherent. The result of it sums incoherently with the ground states contributions associated with $(a)$ and $(b)$.
Being diagonal in this basis, we also exhibited the sign of  a local order parameter associated with each process by coloring the area in between the segments in light red, standing for $+$, and light blue, standing for $-$.
}
\label{f:domain_wall}
\end{figure}

\begin{figure}[t]
\includegraphics[width=0.48\textwidth]{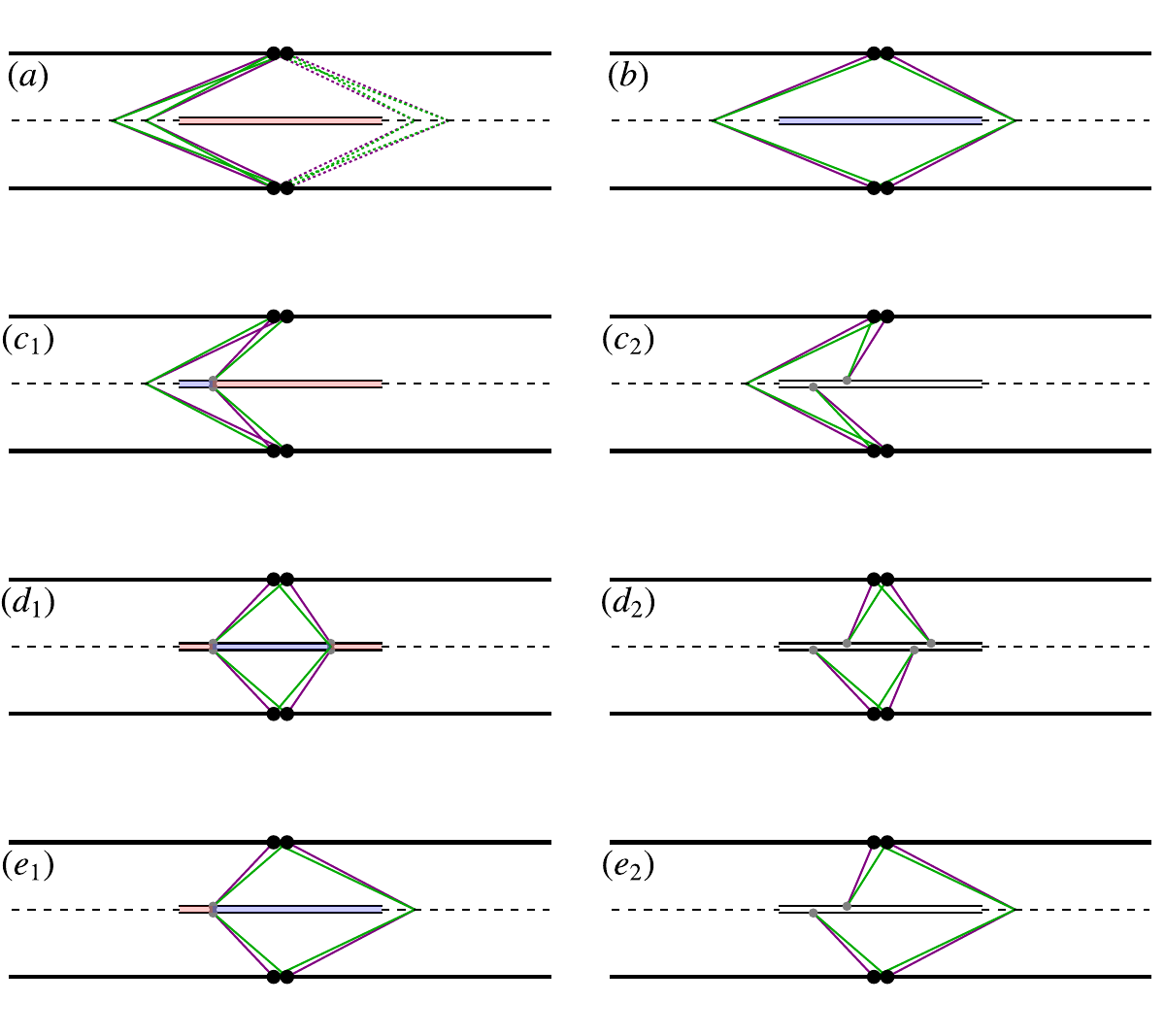}
\caption{The same as in Fig.~\ref{f:domain_wall} for a local perturbation consisting of two nearby domain-wall excitations. Each process shows two events corresponding to interchanging the particles. This causes  interference and is not captured by the semiclassical theory. Processes with the same letter---e.g., $(c_1)$ and $(c_2)$---are summed coherently, whereas processes with different letters are incoherent.}
\label{f:local_perturbation}
\end{figure}

Let us consider first a domain-wall perturbation exciting a single quasiparticle. Within the semiclassical approximation, the reduced density matrix of a subsystem can be decomposed in the contributions depicted in Fig.~\ref{f:domain_wall}.
If, at time $t$, the quasiparticle ends up to the left of the subsystem, then the subsystem remains in the ground state.  If, instead, it ends up to the right of it, the subsystem falls in the other ground state. The remaining contribution corresponds to processes in which the quasiparticle is found in the subsystem. As a whole, the latter  is coherent, i.e., it can be interpreted as a rank-$1$ term. In conclusion, within the semiclassical approximation, the RDM has rank $3$, and can be suggestively written as
\begin{multline} \label{eq:RDM_3}
\rho_A\sim p^{(0)_+}_{A|(j_1^0)}\ket{+}\bra{+}+p^{(0)_-}_{A|(j_1^0)}\ket{-}\bra{-}\\
+p_{A|(j_1^0)}^{(1)_+}\ket{(1)_+,(j_1^0)}\bra{(1)_+,(j_1^0)}\, ,
\end{multline}
where $p_{A|(j_1^0)}^{(0)_+}$ ($p_{A|(j_1^0)}^{(0)_-}$) denotes the probability that the quasiparticle originated at $j_1^0$ is on the left (right) of the subsystem at time $t$, and $p_{A|(j_1^0)}^{(1)_+}=1-p_{A|(j_1^0)}^{(0)_+}-p_{A|(j_1^0)}^{(0)-}$ is the probability that it is found within the subsystem. Note that, for writing \eqref{eq:RDM_3} we are confusing the reduced density matrix of a large subsystem in the ground state with the ground state itself, which is allowed because we are only interested in macroscopic entanglement, which is supposed to be unaffected by boundary corrections (the ground state has exponentially decaying connected correlations and satisfies the area law). 
The order parameter does not have nonzero matrix elements between the three eigenstates, hence \eqref{eq:FeqI} holds and we have
\begin{equation}
\chi(\rho,O)\sim p_{A|(j_1^0)}^{(1)_+} \frac{\Delta_{\ket{(1)_+,(j_1^0)}\bra{(1)_+,(j_1^0)}} O^2}{\parallel O\parallel^2}
\end{equation}
where we used that the ground state variance scales as the volume, and, hence, it gives a  subleading contribution.
To compute $p_{A|(j_1^0)}^{(1)_+}$ let us first define $\mathbb{I}_{(\zeta, \zeta')}(k) = \theta(\zeta'-\frac{v_k}{\bar v_{M}})\theta(\frac{v_k}{\bar v_{M}}-\zeta)$, where $\theta$ is the Heaviside function: $\theta(x)=1$ if $x\geq 0$ and $\theta(x)=0$ otherwise. The probability that the domain wall originated at $j$ be in $A=\llbracket l,r\rrbracket$ is given by
\begin{multline}\label{eq:p_A}
    p_{A|(j_1^0)}^{(1)_+} = \int_{-\pi}^\pi 
    d k\sigma(k)\mathbb{I}_{(
    \frac{l-j_1^0}{\bar v_{M}t}, \frac{r-j_1^0}{\bar v_{M}t})}(k) =
    \\
    \tfrac{1}{2}[\mathcal M(\tfrac{r-j_1^0}{\bar v_{M}t})-\mathcal M(\tfrac{l-j_1^0}{\bar v_{M}t})]\, .
\end{multline}
The variance in $\ket{(1)_+,(j_1^0)}$ is readily obtained by redefining the one-particle scaling function \eqref{eq:Mcal} with the additional condition that the excitation is in the subsystem
\begin{multline}\label{eq:McalA}
    \mathcal M(\zeta|\zeta_l,\zeta_r) = \tfrac{2\int_{-\pi}^\pi dk\, \sigma(k)\sgn(\zeta-\frac{v_k}{\bar v_{M}})\mathbb{I}_{(\zeta_l, \zeta_r)}(k) }{\mathcal M(\zeta_r)-\mathcal M(\zeta_l)}=\\
    \tfrac{2 \mathcal M(\zeta)-\mathcal M(\zeta_l)-\mathcal M(\zeta_r)}{\mathcal M(\zeta_r)-\mathcal M(\zeta_l)}
\end{multline}
where we used \eqref{eq:p_A} for the probability to be in the subsystem. The variance in $A=\llbracket l,r\rrbracket$ is therefore given by
\begin{multline}
\tfrac{\Delta_{\ket{(1)_+,(j_1^0)}\bra{(1)_+,(j_1^0)}}O^2}{\braket{+|O_\bullet|+}^2}=
\\
\sum_{m,n=l'}^{r'} \Bigl[1-
|\mathcal M(\tfrac{n}{ \bar v_{M}t}|\tfrac{l'}{ \bar v_{M}t},\tfrac{r'}{ \bar v_{M}t})-\mathcal M(\tfrac{m}{ \bar v_{M}t}|\tfrac{l'}{ \bar v_{M}t},\tfrac{r'}{ \bar v_{M}t})|\\
-\mathcal M(\tfrac{m}{ \bar v_{M}t}|\tfrac{l'}{ \bar v_{M}t},\tfrac{r'}{ \bar v_{M}t})\mathcal M(\tfrac{n}{ \bar v_{M}t}|\tfrac{l'}{ \bar v_{M}t},\tfrac{r'}{ \bar v_{M}t})\Bigr]
\end{multline}
where $l'=l-j_1^0$ and $r'=r-j_1^0$. This can be simplified using the explicit expression shown in \eqref{eq:McalA} and the Euler-Maclaurin formula to turn the sums into integrals over rescaled positions. Finally, we find 
\begin{multline}\label{eq:qfi_pred_singleDW}
\chi(\rho_A,O)\sim\tfrac{\mathrm{tr}[\rho_A O^2]}{\parallel O\parallel^2}-\kappa^2(1-\tfrac{\mathcal M(\delta \zeta_r)-\mathcal M(\delta \zeta_l)}{2})\\
-\tfrac{2\kappa^2}{\mathcal M(\delta \zeta_r)-\mathcal M(\delta \zeta_l)}\bigl(\tfrac{\mathcal M(\delta \zeta_l)+\mathcal M(\delta\zeta_r)}{2} -\tfrac{\bar v_{M} t}{|A|}\smallint_{\delta \zeta_l}^{\delta \zeta_r} d\zeta \mathcal M(\zeta)\bigr)^2
\end{multline}
where $\kappa=\frac{|A|\braket{+|O_\bullet|+}}{\parallel O\parallel}$,  $\delta \zeta_{l}=\frac{l-j_1^0}{\bar v_M t}$, and $\delta \zeta_{r}=\frac{r-j_1^0}{\bar v_M t}$. One of the most interesting properties of this formula is that it depends only on the scaling function~\eqref{eq:Mcal} characterizing the correlation functions. 

\begin{figure}[t]
\includegraphics[width=0.48\textwidth]{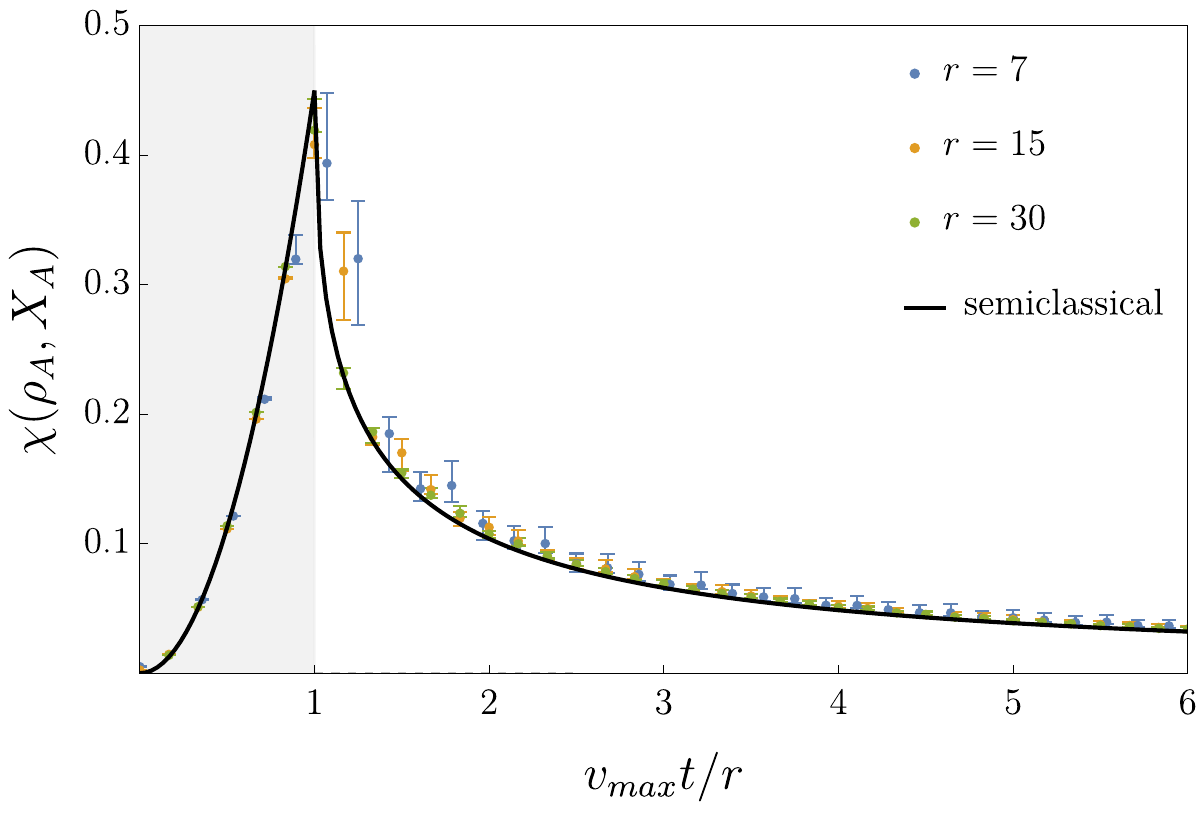}
\caption{The normalized QFI  $\chi(\rho_A,X)$~\eqref{eq:chi} of a spin block $A = \llbracket -r, r\rrbracket$ for an initial state consisting of a domain wall at site $0$ evolving under the Ising hamiltonian with $h=0.5$. The error bars correspond to the bounds \eqref{eq:QFIbounds}, and the solid line is the semiclassical approximation~\eqref{eq:qfi_pred_singleDW}, which becomes exact for large times and lengths in the absence of interference between the excitations. The shaded region corresponds to times for which the subsystem is pure in the semiclassical theory.}
\label{f:QFI_domainwall}
\end{figure}
The generalization to more excitations---which is necessary as we aim at studying local perturbations---is straightforward, provided that the interference between different excitations can be ignored. First of all, as depicted in Fig.~\ref{f:local_perturbation} for the specific case of two excitations,  \eqref{eq:RDM_3} is generalized by the following incoherent superposition
\begin{equation}
\rho_A\sim \sum_{b=\pm 1}\sum_{s_1,\dots, s_n=0}^1 p_{A|\underline{j^0}}^{\underline s_b}\ket{\underline s_b,\underline{j^0}}\bra{\underline s_b,\underline{j^0}}\, .
\end{equation}
Here $\underline s = (s_1, \cdots, s_n)$, $\underline{j^0} = (j^0_1, \cdots, j^0_n)$, $p_{A|\underline{j^0}}^{\underline s_b}$ is the probability that the particles originated at $j_i^0$ are in the subsystem if and only if $s_i = 1$, $b$ represents the parity of the number of particles on the right hand side of the subsystem, and $\ket{\underline s_b,\underline{j^0}}$ is the pure state in $A$ associated with that configuration of particles. Interference comes from the fact that the states $\ket{\underline s_b,\underline{j^0}}$ are generally not orthogonal, as one can immediately infer by permuting the original positions of the excitations $j_1^0,\dots, j_n^0$ (cf. Fig~\ref{f:local_perturbation}). Within the semiclassical approximation that we propose, we overlook this issue and treat the states as if they were orthogonal. This allows us to use \eqref{eq:FeqI}
again, which gives
\begin{multline}\label{eq:chigen}
\chi(\rho,O)\sim\\
\frac{1}{\parallel O\parallel^2}\sum_{b=\pm 1}\sum_{s_1,\dots, s_n=0}^1 p_{A|\underline{j^0}}^{\underline s_b}\Delta_{\ket{\underline s_b,\underline{j^0}}\bra{\underline s_b,\underline{j^0}}}O^2=\\
\frac{1}{\parallel O\parallel^2}\sum_{s_1,\dots, s_n=0}^1 p_{A|\underline{j^0}}^{\underline s}\Delta_{\ket{\underline s_+,\underline{j^0}}\bra{\underline s_+,\underline{j^0}}}O^2\, ,
\end{multline}
where $p_{A|\underline{j^0}}^{\underline s}=p_{A|\underline{j^0}}^{\underline s_+}+p_{A|\underline{j^0}}^{\underline s_-}$ and we used that $\ket{\underline s_\pm,\underline{j^0}}$ are mapped to one another by the spin flip transformation, under which the variance in invariant.  
Since the particle trajectories are independent, we have 
\begin{equation}
p_{A|\underline{j^0}}^{\underline s}=\prod_{\jmath=1}^n (p_{A|(j^0_\jmath)}^{(1)_+})^{s_\jmath}(1-p_{A|(j^0_\jmath)}^{(1)_+})^{1-s_\jmath}\, .
\end{equation}
In addition, the correlation functions in the states $\ket{\underline s_b,\underline{j^0}}$ factorize as in \eqref{eq:correlations}, and we find 
\begin{multline}
\tfrac{\Delta_{\ket{\underline s_+,\underline{j^0}}\bra{\underline s_+,\underline{j^0}}}O^2}{\braket{+|O_\bullet|+}^2}=\sum_{\ell,n=l}^{r}\Bigl[
\\ \prod_{\jmath=1}^n\left(1-
|\mathcal M(\tfrac{n-j_\jmath^0}{ \bar v_{M}t}|\tfrac{l-j_\jmath^0}{ \bar v_{M}t},\tfrac{r-j_\jmath^0}{ \bar v_{M}t})-\mathcal M(\tfrac{\ell-j_\jmath^0}{ \bar v_{M}t}|\tfrac{l-j_\jmath^0}{ \bar v_{M}t},\tfrac{r-j_\jmath^0}{ \bar v_{M}t})|\right)^{s_{\jmath}}\\
-\prod_{\jmath=1}^n\mathcal M^{s_\jmath}(\tfrac{\ell-j_\jmath^0}{ \bar v_{M}t}|\tfrac{l-j_\jmath^0}{ \bar v_{M}t},\tfrac{r-j_\jmath^0}{ \bar v_{M}t})\mathcal M^{s_\jmath}(\tfrac{n-j_\jmath^0}{ \bar v_{M}t}|\tfrac{l-j_\jmath^0}{ \bar v_{M}t},\tfrac{r-j_\jmath^0}{ \bar v_{M}t})\Bigr]\, .
\end{multline}
The simple structure that the probabilities and the variances exhibit allow us to carry out the sum in \eqref{eq:chigen} explicitly. We finally find 
\begin{multline}\label{eq:qfi_pred_DW}
\chi(\rho_A, O)\sim\tfrac{\mathrm{tr}[\rho_A O^2]}{\parallel O\parallel^2}-\kappa^2(\tfrac{\bar v_{M}t}{|A|})^2 \iint\limits_{\mathcal A_t} d \zeta d\zeta'\prod_{\jmath=1}^n\Bigl[\\
p_{A|(j^0_\jmath)}^{(1)_+}\mathcal M(\delta_\jmath\zeta|\delta_\jmath \zeta_l,\delta_\jmath\zeta_r)\mathcal M(\delta_\jmath\zeta'|\delta_\jmath \zeta_l,\delta_\jmath\zeta_r)+1-p_{A|(j^0_\jmath)}^{(1)_+}\Bigr],
\end{multline}
where $\delta_{\jmath}\zeta=\zeta-\frac{j^0_\jmath}{\bar v_M t}$, $\delta_{\jmath}\zeta_l=\frac{l-j^0_\jmath}{\bar v_M t}$, $\delta_{\jmath}\zeta_r=\frac{r-j^0_\jmath}{\bar v_M t}$, and $\mathcal A_t=(\frac{l}{\bar v_M t},\frac{r}{\bar v_M t})$. If $|A|$ and $t$ are large enough compared to the distances between the excitations at time zero, this can be further simplified and we get
\begin{multline}\label{eq:qfi_pred_DW2}
\chi(\rho_A,O)\sim\tfrac{\mathrm{tr}[\rho_A O^2]}{\parallel O\parallel^2}-\kappa^2(\tfrac{\bar v_{M}t}{|A|})^2 \iint\limits_{\mathcal A_t} d \zeta d\zeta'\Bigl[\\
\tfrac{4\mathcal M(\zeta)\mathcal M(\zeta')-2(\mathcal M(\zeta)+\mathcal M(\zeta'))(\mathcal M(\zeta_l)+\mathcal M(\zeta_r))}{2[\mathcal M(\zeta_r)-\mathcal M(\zeta_l)]}\\
+\tfrac{4 \mathcal M(\zeta_l)\mathcal M(\zeta_r)+2(\mathcal M(\zeta_r)-\mathcal M(\zeta_l))}{2[\mathcal M(\zeta_r)-\mathcal M(\zeta_l)]}\Bigr]^n\, .
\end{multline}
For $n=2$ this gives \eqref{eq:chibasic}.
As in the case of a single excitation, also 
in this general case the semiclassical formula for the QFI depends only on the scaling function~\eqref{eq:Mcal}. 
We observe that, at large times for a given $A$ following a perturbation that excites a finite number of quasiparticles, $\zeta$ asymptotically approaches zero across the entire integration domain. Consequently, the expansion $\mathcal M(\zeta)= \mathcal M(0)+\zeta \mathcal M'(0)+O(\zeta^2)$ holds, and  $\chi$ decays as $1/t$. This implies that the macroscopic quantum effect of a local perturbation is transient within any finite subsystem. However, an arbitrarily large subsystem will eventually evolve into a macroscopic quantum state within some time frame.

\begin{figure}[t]
\includegraphics[width=0.48\textwidth]{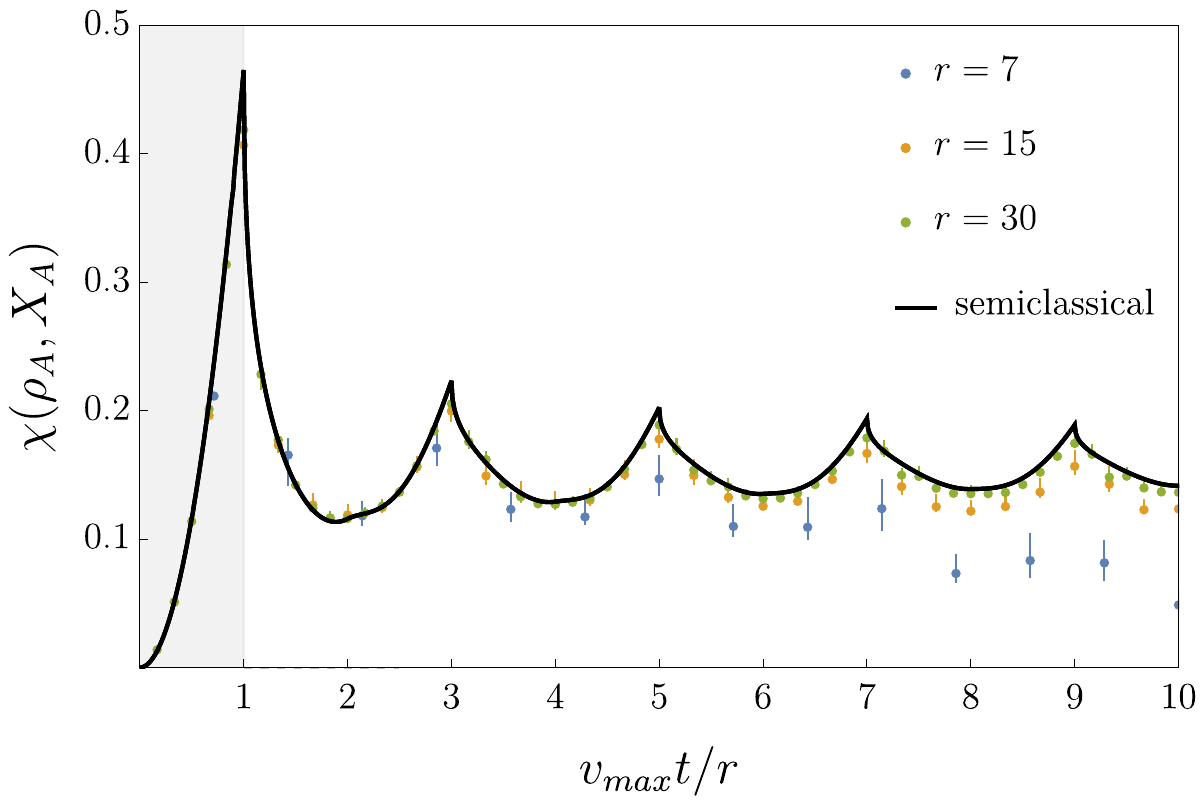}
\caption{The normalized QFI  $\chi(\rho_A,X)$~\eqref{eq:chi} of a spin block $A = \llbracket -r, r\rrbracket$ for an initial state consisting of domain walls at positions $2rz$ with $z\in \mathbb{Z}$  with respect to the order parameter $X_A=\sum_{\ell\in A}\sigma_\ell^x$ as a function of $\bar v_{M}t/r$. The error bars correspond to the bounds \eqref{eq:QFIbounds}, and the solid line corresponds to equation \eqref{eq:qfi_pred_DW}. The shaded region corresponds to times for which the subsystem is semiclassically pure. }
\label{f:QFI_domain_walls}
\end{figure}

Figure~\ref{f:QFI_domain_walls} compares the semiclassical prediction with numerical data obtained by perturbing the ground state of the transverse field Ising model using a uniform grid of domain walls, spaced proportionally to the subsystem length. This spacing ensures that, (i), the predicted value of the normalized QFI remains independent of subsystem size, and, (ii), interference contributions—neglected in our analysis—vanish in the infinite-length limit. The plot clearly shows that the numerical data converge to the semiclassical prediction at fixed rescaled time $t/|A|$ as $|A|\rightarrow\infty$. However, for interference effects to remain subleading, $t/|A|$ should not be comparable to $|A|$. At large times, interference is expected to dominate, ultimately driving $\chi$ to zero. This trend is already visible in Figure~\ref{f:QFI_domain_walls} for the smallest lengths. 
Section~\ref{s:global}, which examines standard quench protocols, will confirm that such eventual decay is generally expected after global quenches in shift-invariant systems.

\begin{figure}[t]
\includegraphics[width=0.48\textwidth]{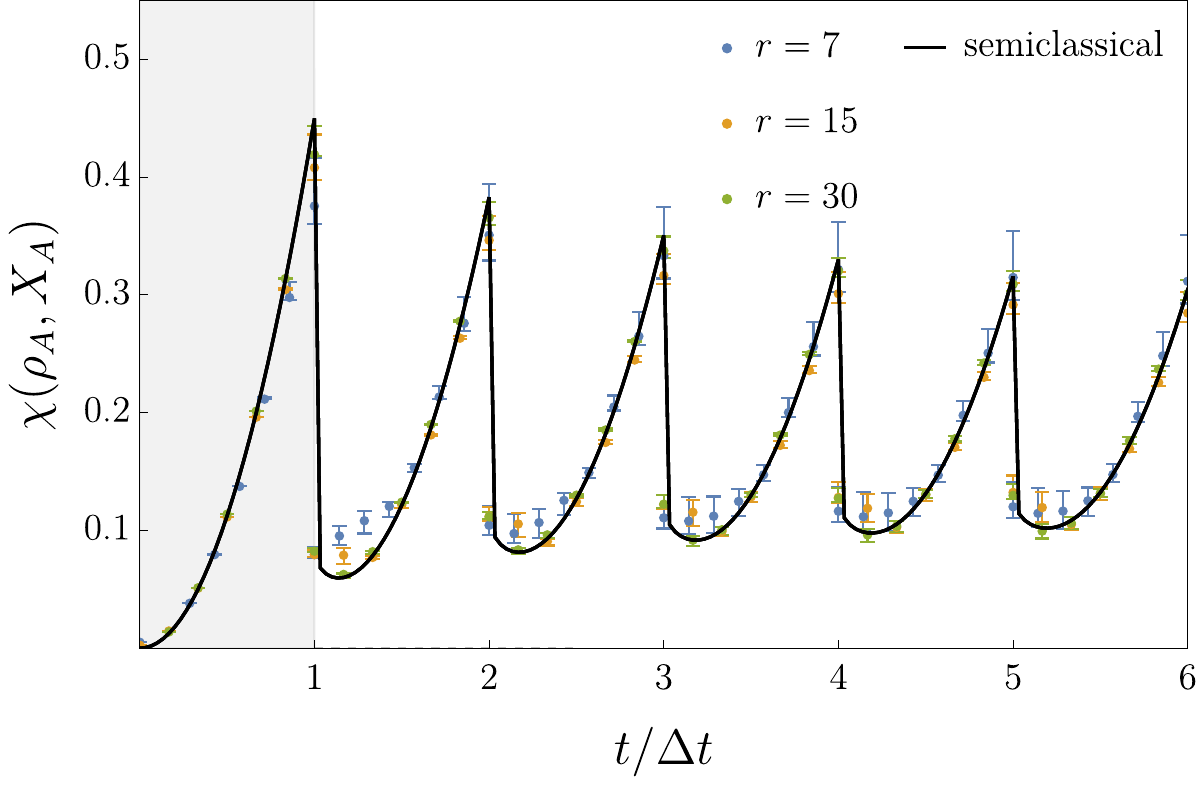}
\caption{The normalized QFI  $\chi(\rho_A,X)$~\eqref{eq:chi} of a spin block $A = \llbracket -r, r\rrbracket$ under a kicking protocol: every $\Delta t$, the Majorana fermion $a_1$ is applied (corresponding to putting a domain wall at site $0$), while evolving at all times under the Ising hamiltonian with $h=0.5$. The error bars correspond to the bounds \eqref{eq:QFIbounds}, and the solid line is the semiclassical approximation~\eqref{eq:qfi_pred_DW} with~\eqref{eq:qfi_pred_difftimes}. The shaded region corresponds to times for which the subsystem is semiclassically pure.}
\label{f:QFI_kick_domain_walls}
\end{figure}
In all the cases considered so far, the normalized QFI  approaches zero in a finite subsystem in the limit of infinite time. Remarkably, this can be prevented still acting just locally on the state. Indeed, we find that interference is not as destructive as before if the uniform grid of perturbations extends over time rather than over space. That is to say, a protocol in which the system is  kicked by a local or semilocal unitary perturbation turns out to be fairly captured by the semiclassical theory even at large times. The semiclassical prediction is the same as for simultaneous perturbations, provided to redefine the rescaled variables as 
$
\frac{\ell-j_\jmath^0}{\bar v_M t}\rightarrow \frac{\ell-j_\jmath^{t_{\jmath}}}{\bar v_M (t-t_{\jmath})}
$, 
where $j_\jmath^{t_{\jmath}}$ is the position of the perturbation made at time $t_{\jmath}< t$. Specifically, we order the perturbations so that $t_{\jmath}\leq t_{\jmath+1}$ and define $n_t$ as the number of excitations created up to time $t$.  The semiclassical prediction is then given by equation~\eqref{eq:qfi_pred_DW} with the following redefinitions
\begin{equation}\label{eq:qfi_pred_difftimes}
\begin{aligned}
n\rightarrow & n_t\\
\delta_{\jmath}\zeta\rightarrow&\tfrac{t}{t-t_{\jmath}}\zeta-\tfrac{j_\jmath^{t_{\jmath}}}{\bar v_M (t-t_{\jmath})}\\
\delta_{\jmath}\zeta_l\rightarrow&\tfrac{l-j_\jmath^{t_{\jmath}}}{\bar v_M (t-t_{\jmath})}\\
\delta_{\jmath}\zeta_r\rightarrow&\tfrac{r-j_\jmath^{t_{\jmath}}}{\bar v_M (t-t_{\jmath})}\\
p_{A|(j^0_\jmath)}^{(1)_+}\rightarrow&\tfrac{1}{2}[\mathcal M(\tfrac{r-j_\jmath^{t_{\jmath}}}{\bar v_{M}(t-t_\jmath)})-\mathcal M(\tfrac{l-j_\jmath^{t_{\jmath}}}{\bar v_{M}(t-t_\jmath)})]
\end{aligned}
\end{equation}
Figures~\ref{f:QFI_kick_domain_walls} and \ref{f:QFI_kick_spinflip} show a comparison between numerical data and semiclassical approximations. Even if the semiclassical approximation is not an exact prediction---especially in the spin-flip case reported in Fig.~\ref{f:QFI_kick_spinflip}---it captures the qualitative behavior even at late times, including the fact that the normalized QFI does not approach zero.

\subsection{Beyond the semiclassical framework}\label{ss:beyond}

To grasp the physics beyond the semiclassical approximation, we consider a simplified model where the calculation can be carried out exactly. Let the Hamiltonian be ferromagnetic with ground states  $\ket{\Uparrow}$ and $\ket{\Downarrow}$; we also assume that the number of domain walls, $\sum_\ell \sigma_\ell^z\sigma_{\ell+1}^z$, is conserved. A spin flip is a local perturbation that brings the ground state into the subspace with two domain walls. Such a subspace 
 is exactly solvable via the Bethe Ansatz~\cite{Bethe1931,Essler2005book}, quite independently of the Hamiltonian's details. 
Specifically, imposing periodic boundary conditions, the eigenstates of $H$ in the relevant sector can be parametrized as
\begin{equation}\label{eq:Bethe}
\ket{k,p;\eta}=\sum_{\ell=-\frac{L}{2}+1}^{\frac{L}{2}-1}\sum_{n=\ell+1}^{\frac{L}{2}}c_{\ell,n}(k,p)[\ket{(\ell,n)^+}+\eta \ket{(\ell,n)^-}]
\, ,
\end{equation}
where $\eta\in\{-1,1\}$ is the eigenvalue of the spin-flip operator and 
we used the notations
\begin{equation}
\begin{aligned}
\ket{(\ell,n)^+}=&\ket{\cdots\uparrow\uparrow \downarrow_{\ell+1}\cdots \downarrow_{n}\uparrow\uparrow \cdots}\\
\ket{(\ell,n)^-}=&\ket{\cdots\downarrow\downarrow \uparrow_{\ell+1}\cdots \uparrow_{n}\downarrow\downarrow \cdots}\, .
\end{aligned}
\end{equation}
If interactions are effectively nearest neighbor, $c_{\ell,n}(k,p)$ are captured by the Ansatz\footnote{Otherwise, it might be necessary to enlarge the local Hilbert space by grouping sites into macrosites.}
\begin{equation}\label{eq:celln}
c_{\ell,n}(k,p)=Z(k,p)[e^{i\ell k+i n p}+S(k,p)e^{i\ell p+i n k}]\, .
\end{equation}
Here $S(k,p)$ is the scattering phase and $Z(k,p)\equiv -S(p,k)Z(p,k)$ ensures the normalization.
Imposing periodic boundary conditions, the momenta satisfy the Bethe equations
$c_{n-L,\ell}(k,p)=\eta c_{\ell,n}(k,p)$, i.e. $
e^{i Lp}=\eta S(k,p)
$ and $
e^{i L(k+p)}=1
$ (note that $S(k,p)S(p,k)=1$).

Since, in more general ferromagnetic models, we can not expect the local perturbation to involve a fixed number of excitations, we generalize the perturbation and replace the spin flip by  $U_0=e^{i\alpha\sigma_0^x}$, which couples the ground state with the two-particle sector.
The overlaps are immediately obtained 
\begin{equation}
\begin{aligned}\label{eq:locpert}
\braket{\Uparrow|U_0|\Uparrow}= &\cos\alpha\\
\braket{k,p;\eta|U_0|\Uparrow}= &i\sin\alpha c^\ast _{-1,0}(k,p)
\end{aligned}
\end{equation} 
The state at time $t$ is therefore given by
\begin{multline}
e^{i E_{GS} t}\ket{\Psi(t)}=\cos\alpha\ket{\Uparrow} \\
+i \sin\alpha\sum_\eta \sum_{k<p\atop k,p\in B_\eta}e^{-i \varepsilon(k,p)t}c^\ast _{-1,0}(k,p)\ket{k,p;\eta}
\end{multline}
where  $B_\eta=\{(k,p)|e^{i L k}=\eta S(p,k)\wedge e^{i L p}=\eta S(k,p)\}$ are the sets of momenta satisfying the Bethe equations, and $\varepsilon(k,p)=\varepsilon(k)+\varepsilon(p)$ is the two-particle excitation energy over the ground state. 
Since the dynamics are confined in a $(L^2-L+1)$-dimensional space, we can work out the reduced density matrix by explicitly taking the partial trace over the rest of the system. 
We will detail the calculation in a separate investigation, but we point out that the result can be put in the form
\begin{equation}
\rho_A=\sum_{s=\pm}A_s\ket{\varphi_s}\bra{\varphi_s}+C_s\ket{s_A}\bra{s_A}+\hat P_{s}\, .
\end{equation}
Here $\ket{\varphi_\pm}$  and $\ket{\pm_A}$ are pure states in $A$ and $\ket{\pm_A}$ have all the spins aligned upwards and downwards, respectively.  
In the limit of large time and subsystem length, they capture the contribution from zero and two particles inside the subsystem. In the same limit, the remaining terms $\hat{P}_\pm$  capture the remaining contribution in which a single particle is inside the subsystem.
The operators  $\hat{P}_\pm$  asymptotically  commute with the rest of the operators. The only significant overlap among the remaining vectors is between $\ket{\varphi_+}$ and $\ket{+_A}$, which equals $\cos\alpha/\sqrt{A_+}$. Furthermore, the rank of $\hat{P}_\pm$, which turns out to be $2$ in the non-interacting case, remains finite as long as the number of nonzero Fourier coefficients of $Z(p,q)$ is finite. Even in cases where there are infinitely many nonzero coefficients, a truncation to a finite number is justified when $Z(p,q)$ is a smooth function over $[-\pi, \pi] \times [-\pi, \pi]$\footnote{This smoothness condition is expected to suffice, given that bound-state excitations in the two-particle sector---associated with complex momenta---are not semilocal with respect to order parameters (they correspond to propagating nearby domain walls); consequently, the dominant contribution to the quantum Fisher information is expected to come from excitations with real momenta.}.

\begin{figure}[t]
\includegraphics[width=0.48\textwidth]{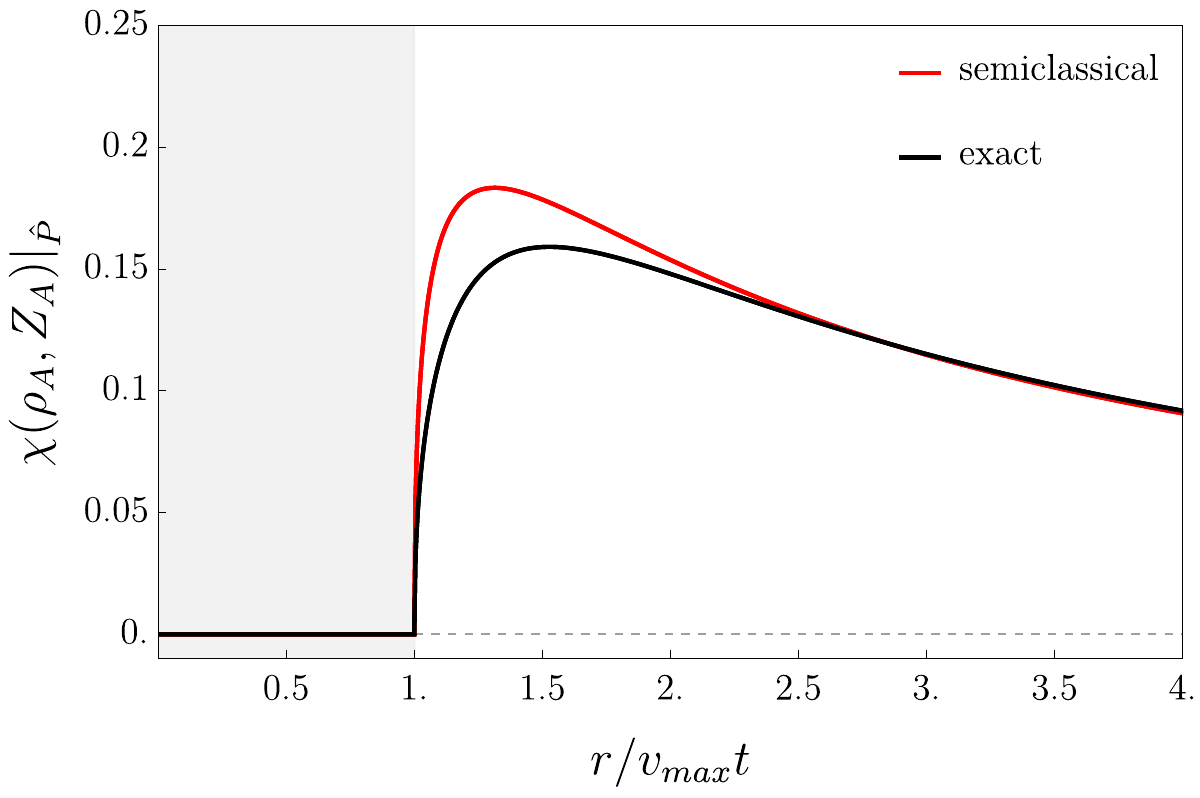}
\caption{Contribution of the one-particle sector to the normalized QFI, $\chi(\rho_A,Z_A)|_{\hat P}$, of a spin block $A = \llbracket -r, r\rrbracket$ under the protocol described in section \ref{ss:beyond} in the limit $r, t\to \infty$ with $r/t$ fixed. The red line corresponds to the semiclassical approximation and the black line to the exact asymptotic expression~\eqref{eq:QFI_beyondsc}. The shaded region corresponds to times at which the subsystem is asymptotically pure.}
\label{f:QFI_beyondsc}
\end{figure}

Already in the noninteracting case, in which $S(p,q)=-1$, this simplified model highlights some limitations of the semiclassical approximation that we presented. First, Section~\ref{s:scQFI} overlooked the mixing of the sectors with zero and two particles (angle $\alpha$), which should be expected after a spin flip also in the TFIM with $h\neq 0$. Second, it  overlooked some interference terms coming from the exchange of  particles. To explain this latter effect, we exhibit the asymptotics of $\hat P_+$ in the thermodynamic limit and for large time and subsystem length when  $S(p,q)=-1$. We can treat this term of $\rho_A$---corresponding to  the processes $(c_1)$ and $(c_2)$ in Fig.~\ref{f:local_perturbation}---independently from the others when computing the normalized QFI of $\rho_A$ because it asymptotically commutes with them and because the matrix elements of the order parameter $Z_A\equiv \sum_{\ell\in A}\sigma_\ell^z$  between the eigenstates of $\hat P_+$ and the other eigenvectors of the reduced density matrix approach zero. We carried out the calculation and obtained
\begin{multline}\label{eq:Pplus}
\hat P_+\sim\mathcal L_0(t) (\ket{\tilde\Psi_0(t)}\bra{\tilde \Psi_0(t)}+\ket{\tilde\Psi_1(t)}\bra{\tilde \Psi_1(t)})\\
-\mathcal L_1(t) \ket{\tilde\Psi_0(t)}\bra{\tilde \Psi_1(t)}-\mathcal L_{-1}(t)\ket{\tilde\Psi_1(t)}\bra{\tilde\Psi_0(t)}
\end{multline}
where $\mathcal L_n(t)=\int_{-\pi}^\pi\frac{d k}{2\pi}e^{i n k}\theta(l-v(k)t)$, with $v(k)=\varepsilon'(k)$, and $\ket{\tilde\Psi_j(t)}$ are non-normalized states in $A\equiv \llbracket l,r\rrbracket$ with overlaps
\begin{multline}
\braket{\tilde\Psi_n(t)|\tilde\Psi_m(t)}\sim \mathcal A_{m-n}(t)= \\
\int_{-\pi}^{\pi}\frac{dk}{2\pi}\theta(r-v(k)t)
\theta(v(k)t-l)e^{i(m-n)k} \, .
\end{multline}
The  matrix elements of $Z_A$ and $Z_A^2$ have instead the following asymptotic behavior
\begin{multline}
\frac{1}{t}\braket{\tilde \Psi_n(t)|Z_A|\tilde\Psi_m(t)}\sim \mathcal O_{m-n}(t)= \\
\int_{-\pi}^{\pi}\frac{dk}{2\pi}\theta(r-v(k)t)
\theta(v(k)t-l)(\tfrac{r+l}{t}-2 v(k))e^{i(m-n)k}
\end{multline}
\begin{multline}
\frac{1}{t^2}\braket{\tilde\Psi_n(t)|Z_A^2|\tilde\Psi_m(t)}\sim \mathcal O_{m-n}^{(2)}(t)= \\
\int_{-\pi}^{\pi}\frac{dk}{2\pi}\theta(r-v(k)t)
\theta(v(k)t-l)(\tfrac{r+l}{t}-2v(k))^2e^{i(m-n)k}\, .
\end{multline}
The states $\ket{\tilde\Psi_0(t)}$ and $\ket{\tilde\Psi_1(t)}$ describe the two events depicted in Fig.~\ref{f:local_perturbation} in green and red; the index $n$ in $\ket{\tilde\Psi_n(t)}$ can be interpreted as (minus) the initial position of the excitation. 
The semiclassical approximation corresponds to ignoring the overlap between $\ket{\tilde\Psi_0(t)}$ and $\ket{\tilde\Psi_1(t)}$, the off-diagonal terms in \eqref{eq:Pplus}, as well as the off-diagonal matrix elements of  the order parameter. It is evident that this approximation is not a priori justified after a spin flip, but, for more general localized perturbations, the underlying error would approach zero as the inverse of the distance between the excitations created by the perturbation. 

We now compare the semiclassical approximation with the exact calculation of the $\hat P_+$ contribution to the normalized QFI. 
To that aim, it is convenient to exploit the low rank of $\hat P_+$ and represent the operators by $2$-by-$2$ matrices. Choosing $\bra{\tilde\Psi_0}=\mathcal A_0^{1/2}\begin{pmatrix}1
&0
\end{pmatrix}$ and $\bra{\tilde\Psi_1}=\mathcal A_0^{-1/2}\begin{pmatrix}
A_{-1}&\sqrt{\mathcal A_0^2-|\mathcal A_1|^2}
\end{pmatrix}$ we find
\begin{multline} 
\hat P_+\sim \mathbb P=\\
\begin{bmatrix}
\mathcal L_0\frac{\mathcal A_0^2+|\mathcal A_1|^2}{\mathcal A_0}-2\mathrm{Re}[\mathcal A_1\mathcal L_{-1}]&\frac{\sqrt{\mathcal A_0^2-|\mathcal A_1|^2}(\mathcal A_1\mathcal L_{0}-\mathcal A_0\mathcal L_{1})}{\mathcal A_0}\\
\frac{\sqrt{\mathcal A_0^2-|\mathcal A_1|^2}(\mathcal A_{-1}\mathcal L_{0}-\mathcal A_0\mathcal L_{-1})}{A_0}&\mathcal L_0\frac{\mathcal A_0^2-|\mathcal A_1|^2}{\mathcal A_0}
\end{bmatrix}\\
\xrightarrow{s.c.}\mathcal L_0\mathcal A_0 \mathbb I
\end{multline}
\begin{multline}
\tfrac{1}{t}Z_A\sim \mathbb O=\\
\begin{bmatrix}
\frac{\mathcal O_{0}}{\mathcal A_0}&\frac{\mathcal A_0\mathcal O_{1}-\mathcal A_{1}\mathcal O_0}{\mathcal A_0\sqrt{\mathcal A_0^2-|\mathcal A_1|^2}}\\
\frac{\mathcal A_0\mathcal O_{-1}-\mathcal A_{-1}\mathcal O_0}{\mathcal A_0\sqrt{\mathcal A_0^2-|\mathcal A_1|^2}}&\frac{(\mathcal A_0^2+|\mathcal A_1|^2)\mathcal O_0-2\mathcal A_0\mathrm{Re}[\mathcal A_1\mathcal O_{-1}]}{\mathcal A_0(\mathcal A_0^2-|\mathcal A_1|^2)}
\end{bmatrix}\\
\xrightarrow{s.c.}\tfrac{\mathcal O_0}{\mathcal A_0}\mathbb I
\end{multline}
\begin{multline}
\tfrac{1}{t^2}Z_A^2\sim\mathbb O^{(2)}=\\
\begin{bmatrix}
\frac{\mathcal O^{(2)}_{0}}{\mathcal A_0}&\frac{\mathcal A_0\mathcal O^{(2)}_{1}-\mathcal A_{1}\mathcal O^{(2)}_0}{\mathcal A_0\sqrt{\mathcal A_0^2-|\mathcal A_1|^2}}\\
\frac{\mathcal A_0\mathcal O^{(2)}_{-1}-\mathcal A_{-1}\mathcal O_0^{(2)}}{\mathcal A_0\sqrt{\mathcal A_0^2-|\mathcal A_1|^2}}&\frac{(\mathcal A_0^2+|\mathcal A_1|^2)\mathcal O_0^{(2)}-2\mathcal A_0\mathrm{Re}[\mathcal A_1\mathcal O_{-1}^{(2)}]}{\mathcal A_0(\mathcal A_0^2-|\mathcal A_1|^2)}
\end{bmatrix}\\
\xrightarrow{s.c.}\tfrac{\mathcal O^{(2)}_0}{\mathcal A_0}\mathbb I
\end{multline}
where we also exhibited the result of the semiclassical ($s.c.$) approximation.
Specializing  Eq.~\eqref{eq:chialt} to $2$-by-$2$
 matrices gives
\begin{multline}\label{eq:QFI_beyondsc}
\chi(\rho_A,Z_A)|_{\hat P^+}=\tfrac{t^2}{|A|^2}\Bigl(\mathrm{tr}[\mathbb P \mathbb O^{(2)}]-\tfrac{\mathrm{tr}[\mathbb P \mathbb O]^2}{\mathrm{tr}[\mathbb P]}\\
+\tfrac{(\mathrm{tr}[\mathbb P^2]-\mathrm{tr}[\mathbb P]^2)(\mathrm{tr}[(\mathbb P-\frac{1}{2}\mathrm{tr}[\mathbb P])\mathbb O]^2+\frac{1}{2}\mathrm{tr}[[\mathbb P ,\mathbb O][ \mathbb O,\mathbb P]])}{\mathrm{tr}[\mathbb P](\mathrm{tr}[\mathbb P^2]-\frac{1}{2}\mathrm{tr}[\mathbb P]^2)}\Bigr)\, .
\end{multline}
In the semiclassical approximation only the first line is nonzero and equals  $\frac{2t^2}{|A|^2}\mathcal L_0 (\mathcal O_0^{(2)}-\frac{\mathcal O_0^2}{\mathcal A_0} )$. 
The contribution from the other one-particle term, $\hat{P}_-$, can be obtained by reflecting the system, as it accounts for events where a particle lies to the right of the subsystem. Specifically, that contribution is given by Eq.~\eqref{eq:QFI_beyondsc}, with the substitutions $l \rightarrow -r$, $r \rightarrow -l$, and $v(k) \rightarrow -v(-k)$.
Figure \ref{f:QFI_beyondsc} shows a comparison between exact asymptotics and semiclassical approximation of the one-particle contribution  to the normalized QFI in a noninteracting model with the  dispersion relation of the TFIM with $h=\frac{1}{2}$.



\section{Global quench}\label{s:global}%

A quantum quench is considered global when the initial state has a substantial overlap with an exponentially large number of excited states. Typically, the initial state is an equilibrium state of a different Hamiltonian, and the quench corresponds to a sudden change in a Hamiltonian parameter, such as a magnetic field~\cite{Essler2016Quench}.  
Semiclassical theories have proven highly effective in describing correlations and entanglement entropy dynamics following global quenches---see., e.g., Refs~\cite{Evangelisti2013Semi-classical,Kormos2016Quantum,Hybrid2017Moca,Kormos2018Semiclassical,Alba2019Entanglement,Bertini2019Transport,Horvath2022Inhomogeneous}. In integrable systems, these theories rely on a quasiparticle picture, where particles move independently with an effective (mean-field) velocity determined by the local properties of the state. A key aspect of this framework is the structure of correlations between quasiparticles in the initial state, which are often assumed to be correlated in pairs.
One of the first studies to highlight this emergent pair structure was Ref.~\cite{Calabrese_2005}, which examined global quenches in $(1+1)$-dimensional conformal field theories and spin chains with a free fermion representation. Subsequent works confirmed the pair structure in 1D free scalar theories~\cite{Cotler_2016}, interacting integrable spin chains~\cite{Alba_2017, Piroli_2019}, and even non-integrable models~\cite{Kormos_2016}. Additionally, the same structure has been identified in higher dimensions~\cite{Casini_2016}.

Given its broad applicability, we will assume a pair structure in our analysis. However, exceptions do exist---see, for example, Ref.~\cite{Bertini_2018, Bastianello2020Entanglement}---and, in some cases, can still be captured within a semiclassical framework.
In reflection symmetric models the pairs consist of quasiparticles with opposite momenta.
The usual starting point is to assume that the global quench generates independent pairs everywhere with a density $\varrho(k)$,  which in translationally invariant systems is independent of the position. As discussed earlier, for a post-quench Hamiltonian with the ground state in a ferromagnetic phase in which a $\mathbb Z_2$ symmetry is broken, a semiclassical quasiparticle crossing the position of an order parameter has the effect of reversing its sign. The semiclassical predictions for one- and two-point functions of order parameters are then readily obtained
\begin{align}
\braket{O_\ell(t)}\sim &\braket{+|O_\bullet|+}e^{-2\mathcal N[C_\ell(t)]}\\
    \braket{O_\ell(t)O_n(t)} \sim& \braket{+|O_\bullet|+}^2 e^{-2\mathcal{N}[C_\ell(t)\sqcup C_n(t)]},
\end{align}
where $C_j(t)$ is the phase-space region associated with the light-cone of $j$, i.e. $(x_0, k)\in C_j(t) \Leftrightarrow \mathbb{I}_{C_j(t)}(x_0, k)=1 \Leftrightarrow x_0\in(j-|v_k|t, j+|v_k|t)$, and 
$C_i(t)\sqcup C_j(t)$ denotes the disjoint union of the region;
$\mathcal{N}[R]$ is the average number of excitations inside region $R$, i.e.
\begin{multline}
    \mathcal{N}[C_j(t)] = \int_{-\infty}^{+\infty}dx_0\int_{0}^{\pi}dk\, \varrho(k) \mathbb{I}_{L_C(j, t)}(x_0, k)=\\ 2\int_0^\pi dk\, \rho(k)|v_k| t,
\end{multline}
\begin{multline}
    \mathcal{N}[C_i(t)\sqcup C_j(t)] = \\
    \mathcal{N}[C_i(t)] + \mathcal{N}[C_j(t)] - 2 \mathcal{N}[C_i(t)\cap C_j(t)]=\\
    2\int_0^\pi dk\, \varrho(k)\max(2|v_k|t, |i-j|).
\end{multline}
Such a quasiparticle picture is also very effective to predict the behavior of the entanglement entropy of a subsystem. The pair structure implies indeed that the entropy is proportional to the number of pairs in which one quasiparticle is inside the subsystem while the other is outside, referred to in the following as ``isolated''. 

From this perspective, the quantum Fisher information gives  complementary information.  Specifically, within the semiclassical approximation, the eigenstates of the reduced density matrix consist of the coherent superposition of all the pairs in the subsystem for given configuration of isolated quasiparticles within it. Thus, we can regard the eigenstates of the reduced density matrix $\rho_A$ describing the ensemble of neighbouring spins $A=\llbracket l,r\rrbracket$ as being characterized by the same  content of pairs but different content of isolated quasiparticles
\begin{equation}
\rho_A\sim \sum_{b=\pm 1}\sum_{x_l, \cdots, x_r=0}^1 p_A({\underline x_b})\ket{\Psi_A,\underline x_b}\bra{\Psi_A,\underline x_b}\, .
\end{equation}
Here the coordinate $x_j\in \{0,1\}$ of vector $\underline x$ is $1$ if and only if there is an isolated quasiparticle at position $j$; $p_A(\underline x_b)$ is the probability of  the configuration of isolated quasiparticles associated with $\underline x$, given that the number of particles on the right hand side of the subsystem has parity $b$; $\ket{\Psi_A,\underline x_b}$ denotes the state corresponding to that configuration of isolated quasiparticles and all possible pair configurations inside $A$. The order parameter does not have nonzero matrix elements between different $\ket{\Psi_A,\underline x_b}$, hence we can use
equality~\eqref{eq:FeqI} and follow the same procedure described in Section~\ref{s:scQFI}. Specifically, the normalized QFI reads 
\begin{multline}
\chi(\rho_A, O)\sim\tfrac{\mathrm{tr}[\rho_A O^2]}{\parallel O\parallel^2}-\tfrac{1}{\parallel O\parallel^2} \sum_{i, j\in A}\sum_{x_l, \cdots, x_r=0}^1 p_A({\underline x})\Bigl[\\
\braket{\Psi_A,\underline x_+|O_i|\Psi_A,\underline x\_+}\braket{\Psi_A,\underline x_+|O_j|\Psi_A,\underline x_+}\Bigr]\, ,
\end{multline}
where $p_A({\underline x})=p_A(\underline x_+)+p_A(\underline x_-)$ and we used that $\ket{\Psi_A,\underline x_\pm}$ are mapped to one another by the spin flip transformation, under which the variance in invariant.  Importantly, since the order parameter simply changes sign when is crossed by a quasiparticle, contributions from the pairs and the isolated quasiparticles factorize in the one and two-point functions
\begin{multline}
    \braket{\Psi_A,\underline x_+|O_i|\Psi_A,\underline x_+} =\\
    \braket{\Psi_A,\underline 0_+|O_i|\Psi_A,\underline 0_+}\mathcal{P}_i(\underline x),
\end{multline}
and
\begin{multline}
    \braket{\Psi_A,\underline x_+|O_i O_j|\Psi_A,\underline x_+} =\\
    \braket{\Psi_A,\underline 0_+|O_i O_j|\Psi_A,\underline 0_+}\mathcal{P}_i(\underline x)\mathcal{P}_j(\underline x),
\end{multline}
where $\underline 0$ stands for the vector of zeroes and we defined
\begin{equation}\label{eq:Pix}
    \mathcal{P}_i(\underline x) = \prod_{j=l}^r(-1)^{x_j\theta(j-i)}
\end{equation}
(the ambiguity in the definition of $\theta(0)$ cannot be lifted within the semiclassical theory, and our choice $\theta(0)=1$ is purely conventional).  
 Since $\ket{\Psi_A,\underline 0_+}$  consists only of pairs, we have ($l$ and $r$ are at the edges of the subsystem)
\begin{equation}\label{eq:boundarypaironepoint}
\braket{\Psi_A,\underline 0_+|O_{l, r}|\Psi_A,\underline 0_+} =\braket{+|O_\bullet|+}
\end{equation}
\begin{multline}\label{eq:boundarypairtwopoint}
    \braket{\Psi_A,\underline 0_+|O_{l, r}O_i|\Psi_A,\underline 0_+} =\\\braket{+|O_\bullet|+}
    \braket{\Psi_A,\underline 0_+|O_i|\Psi_A,\underline 0_+}\, ,
\end{multline}
and hence
\begin{multline}\label{eq:sum_onepoints_quench}
    \sum_{x_l, \cdots, x_r=0}^1 p_A({\underline x})\Bigl[\\
\braket{\Psi_A,\underline x_+|O_i|\Psi_A,\underline x_+}\braket{\Psi_A,\underline x_+|O_j|\Psi_A,\underline x_+}\Bigr]=\\
\tfrac{\braket{\Psi_A,\underline 0_+|O_i O_r|\Psi_A,\underline 0_+}\braket{\Psi_A,\underline 0_+|O_l O_j|\Psi_A,\underline 0_+}}{\braket{+|O_\bullet|+}^2}\\
\times \underbrace{\sum_{x_l, \cdots, x_r=0}^1 p_A(\underline x)\mathcal{P}_i(\underline x)\mathcal{P}_j(\underline x)}_{\equiv \langle \mathcal{P}_i\mathcal{P}_j \rangle}.
\end{multline}
It is convenient to compare this expression with the two-point function, which reads
\begin{equation}\label{eq:2pointquench}
    \mathrm{tr}[\rho_A O_i O_j] = \braket{\Psi_A,\underline 0_+|O_i O_j|\Psi_A,\underline 0_+}\langle \mathcal{P}_i\mathcal{P}_j \rangle.
\end{equation}
This allows us to express \eqref{eq:sum_onepoints_quench} as 
\begin{equation}\label{eq:sum_onepoints_quench1}
=\tfrac{\mathrm{tr}[\rho_A O_i O_r]\mathrm{tr}[\rho_A O_l O_j]}{\braket{+|O_\bullet|+}^2}\tfrac{\langle \mathcal{P}_i\mathcal{P}_j \rangle}{\langle \mathcal{P}_l\mathcal{P}_j \rangle \langle \mathcal{P}_i\mathcal{P}_r \rangle}.
\end{equation}
The pair factorization of the initial state implies that isolated quasiparticles are independent
\begin{equation}
    p_A(\underline x) = \prod_{j=l}^r p_j^{x_j}(1-p_j)^{1-x_j}\, ,
\end{equation}
where $p_n$ is the probability to have an isolated quasiparticle at site $n$. This allows us to simplify $\langle \mathcal{P}_i\mathcal{P}_j \rangle$
\begin{multline}
    \langle \mathcal{P}_i\mathcal{P}_j \rangle = \sum_{x_l, \cdots, x_r=0}^1 \prod_{n= l}^r p_n^{x_n}(1-p_n)^{1-x_n}\\(-1)^{\theta(n-i)x_n}(-1)^{\theta(n-j)x_n} \\
    \underset{i\le j}{=}\prod_{n=l}^r(1-2p_nx_n\theta(n-i)\theta(j-1-n))\\
    =\prod_{n=i}^{j-1}(1-2p_nx_n),
\end{multline}
where we used 
\begin{equation}
(-1)^{\theta(n-i)x_n+\theta(n-j)x_n} \underset{i\le j}{=} (-1)^{[\theta(n-i)\theta(j-1-n)]x_n } \, . 
\end{equation}
This factorization implies that, for any $i\leq \ell\leq j \in A$, $\langle \mathcal{P}_i\mathcal{P}_j \rangle=\langle \mathcal{P}_i\mathcal{P}_\ell \rangle\langle \mathcal{P}_\ell\mathcal{P}_j\rangle$. Thus we can simplify the last factor in \eqref{eq:sum_onepoints_quench1} and obtain 
\begin{multline}
\chi(\rho_A, O)\sim\\
\tfrac{\mathrm{tr}[\rho_A O^2]}{\parallel O\parallel^2}-\tfrac{1}{\parallel O\parallel^2} \sum_{i, j\in A}\tfrac{\langle O_l O_{\max(i, j)}\rangle \langle O_{\min(i, j)} O_r\rangle}{\braket{+|O_\bullet|+}^2 \langle \mathcal{P}_l\mathcal{P}_r\rangle }. 
\end{multline}
Finally, the denominator in the sum can be expressed in terms of the two-point function using equations \eqref{eq:boundarypaironepoint}, \eqref{eq:boundarypairtwopoint}, and \eqref{eq:2pointquench}, which gives
\begin{multline}\label{eq:qfi_pred_quench_final}
\chi(\rho_A, O)\sim\\
\tfrac{\mathrm{tr}[\rho_A O^2]}{\parallel O\parallel^2}-\tfrac{1}{\parallel O\parallel^2} \sum_{i, j\in A}\tfrac{\langle O_l O_{\max(i, j)}\rangle \langle O_{\min(i, j)} O_r\rangle}{\langle O_l O_r\rangle }\, .
\end{multline}
Eq.~\eqref{eq:predictionquench} is an a posteriori refinement of \eqref{eq:qfi_pred_quench_final} coming from the numerical analysis (the shift $r\to r+1$ and $l\to l-1$ is an $\mathcal{O}(1/|A|)$ correction beyond the semiclassical theory).

\begin{figure}[t]
\includegraphics[width=0.48\textwidth]{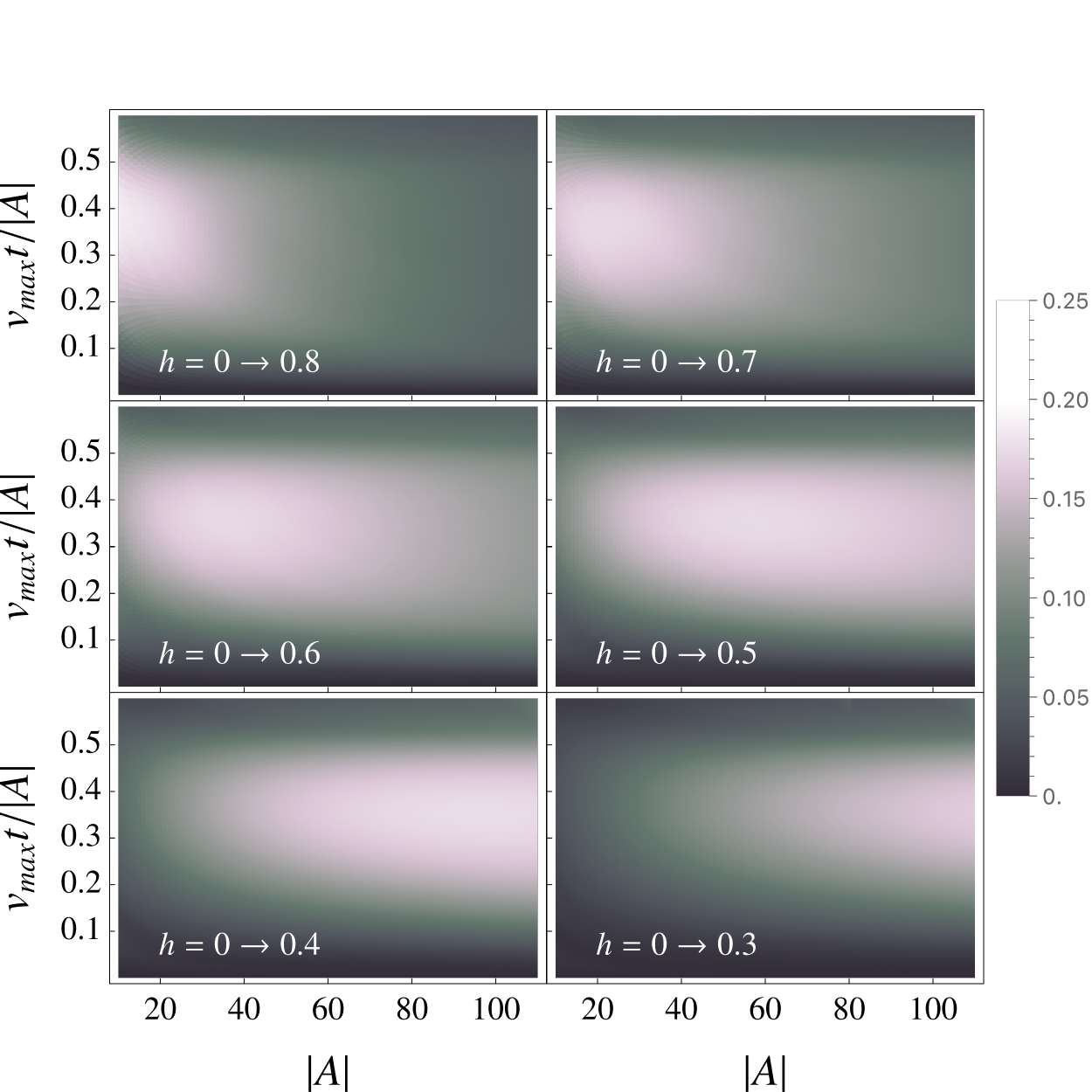}
\caption{The prediction for $\chi(\rho_A,X)$~\eqref{eq:chi} of a spin block $A$  with respect to the order parameter $X_A=\sum_{\ell\in A}\sigma_\ell^x$ as a function of $|A|$ and the rescaled time $\bar v_{M}t/|A|$ after six global quenches within the transverse-field Ising model. Subsystems develop multipartite entanglement more or less within the same interval of rescaled times; the smaller the quench, the larger the typical length of the subsystems in macroscopic quantum states.}
\label{f:severalquenches}
\end{figure}

After a quench in the transverse-field Ising model in which the magnetic field is changed from $h_0$ to $h$, we can use the asymptotic formula derived in Ref.~\cite{Calabrese2012Quantum1} to predict the asymptotic behavior of the quantum Fisher information in the limit of large distance and time. We find
\begin{multline}\label{eq:quenchasmyp}
\frac{1}{4}F_Q(\rho_A,\sum_{\ell\in A}\sigma_\ell^x)\sim 2\mathcal C_{FF}(\zeta)t^2\int_0^{\zeta}d u(\zeta -u) e^{-t \nu_x(u)}\\
- 4t^2\int_{\frac{\zeta}{2}}^{\zeta}du\int_{\zeta-u}^{u}d ve^{-t(\nu_x(u)+\nu_x(v)-\nu_x(\zeta))}
\end{multline}
where $\zeta=\frac{|A|}{t}$, 
\begin{equation}
\nu_x(\zeta)=-\int_0^\pi\frac{\mathrm d k}{\pi}\log\cos\Delta_k\min(2 v_k,\zeta)
\end{equation}
and $\Delta$ is the difference between the Bogoliubov angles before and after the quench, which in the TFIM reads
\begin{equation}
\cos\Delta_k=\frac{h_0 h-(h_0+h)\cos k+1}{\sqrt{1+h_0^2-2 h_0 \cos k}\sqrt{1+h^2-2 h \cos k}}
\end{equation}
If one were to uncritically adopt the assumptions of Ref.~\cite{Calabrese2012Quantum1}, the function  $\mathcal C_{FF}(\zeta)$ would be independent of $\zeta$ and given by
\begin{equation}
\mathcal C_{FF}(\zeta)\rightarrow \mathcal C_{FF}=\frac{1-h_0 h +\sqrt{(1-h_0^2)(1-h^2)}}{2\sqrt{1-h_0 h}(1-h_0^2)^{\frac{1}{4}}}
\end{equation}
However, based on the observations in Ref.\cite{Granet2020Finite},  we infer that this    can only be an excellent approximation, so we expect small corrections to \eqref{eq:quenchasmyp} after the replacement $\mathcal C_{FF}(\zeta)\rightarrow \mathcal C_{FF}$.

Figure~\ref{f:severalquenches} shows the typical behavior of the semiclassical prediction \eqref{eq:qfi_pred_quench_final} after a quench in the TFIM. Remarkably, almost irrespectively of the quench parameters, there exists a space-time frame where the normalized QFI stabilizes around $\sim 0.2$. However, for large quenches, $\chi$ remains significantly nonzero only in small subsystems and over short time scales. Extensive multipartite entanglement is witnessed by $\chi$ only in the case of small quenches.   

We conclude by noting that our semiclassical analysis does not rely on translational invariance. Therefore, we expect Eq.~\eqref{eq:qfi_pred_quench_final} to remain valid whenever the pair structure is effectively realized locally, as in the systems considered in Refs~\cite{Bertini2018Entanglement, Alba2019Entanglement}, where the initial state is inhomogeneous. 

\section{Discussion}\label{s:conclusion}

\begin{figure}[t]
\includegraphics[width=0.48\textwidth]{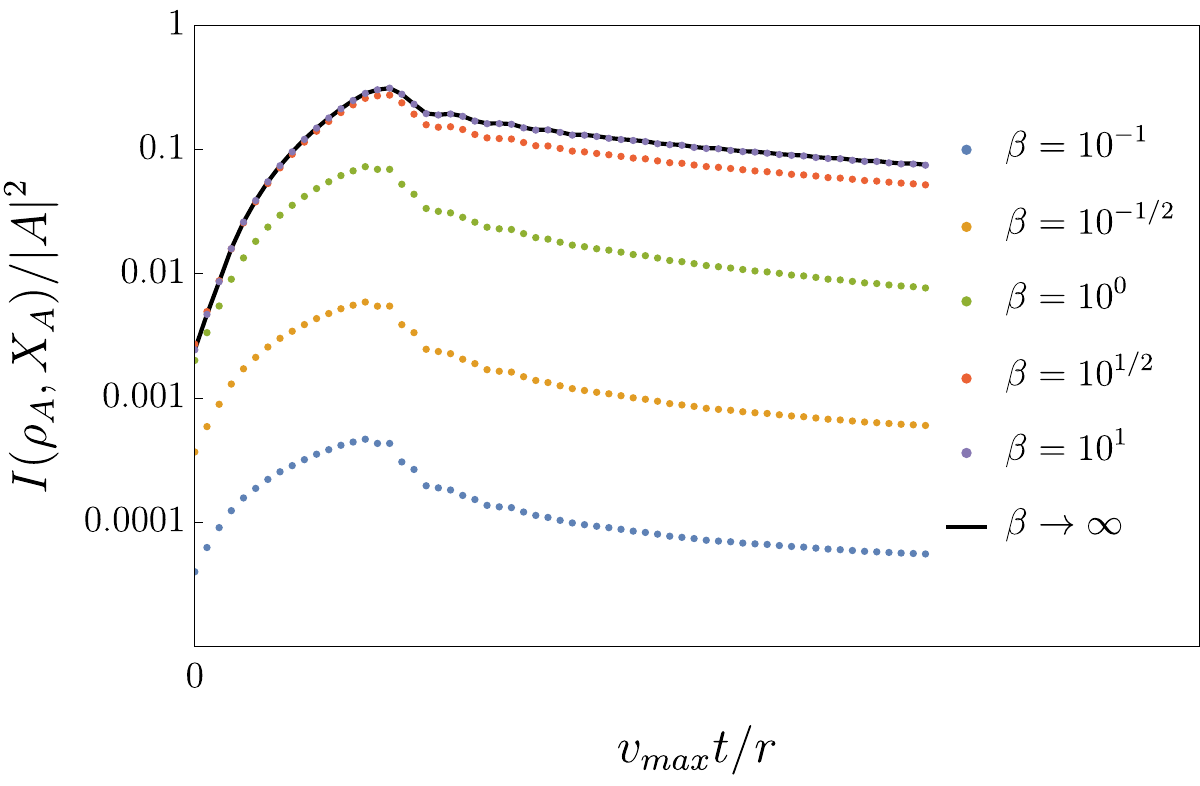}
\caption{$I(\rho_A,X_A)/|A|^2$~\eqref{eq:chi} of a spin block $A=\llbracket -15, 15\rrbracket$  with respect to the order parameter $X_A=\sum_{\ell\in A}\sigma_\ell^x$ as a function of the rescaled time $\bar v_{M}t/|A|$ after a single spin flip over the thermal state $\rho_\beta$ at different inverse temperatures $\beta$. Subsystems develop multipartite entanglement even at non-zero temperature.
}
\label{f:SF_thermal}
\end{figure}

We have explored the behavior of quantum Fisher information in subsystems of quantum spin chains out of equilibrium, aiming to identify nonequilibrium protocols and time scales in which subsystems fall into macroscopic quantum states. Our findings reveal a surprisingly rich variety of situations in which large subsystems exhibit multipartite entanglement in the thermodynamic limit. On the one hand, we generalized the results of Ref.~\cite{Pappalardi2017Multipartite}  to subsystems in the thermodynamic limit, showing that a small quench within a ferromagnetic phase can be used to generate multipartite entanglement in subsystems as large as the inverse of the density of excitations. However, such effect is temporary, indeed at late times the quantum Fisher information of extensive observables returns to be proportional to the subsystem's length. On the other hand, we showed that localized perturbations in a ferromagnetic phase can significantly enhance subsystem quantum Fisher information, which witnesses macroscopic entanglement in arbitrarily large subsystems within a time window proportional to the subsystem's length. Remarkably, by periodically applying localized perturbations, we can sustain this enhancement indefinitely in subsystems whose length is proportional to the driving period.

We developed semiclassical predictions for the quantum Fisher information and validated them against exact numerical results in noninteracting spin chains. Specifically, we computed both the Wigner-Yanase-Dyson skew information and the quantum Fisher information using free-fermion techniques and an arguably novel identity---Eq. \eqref{eq:QFIidentity}---that directly relates the two. Additionally, we analyzed the limitations of the semiclassical approximation and proposed a framework to extend the study of the normalized quantum Fisher information to systems with interactions.

Our findings offer a promising outlook from both the perspective of quantum resource theory and the characterization of quantum many-body systems. On one hand, we have identified novel pathways for engineering macroscopic quantum states out of equilibrium. On the other, we have demonstrated that nonequilibrium quantum systems can exhibit nontrivial quantum properties on macroscopic scales.

Additionally, we take this opportunity to highlight an important aspect not explicitly emphasized in the main text. While spontaneous symmetry breaking is essential for a nonzero order parameter and the restoration of clustering properties in the initial state, the quantum Fisher information remains largely unaffected by symmetry breaking. Notably, our results hold even at low but nonzero temperatures, as evidenced by Fig.~\ref{f:SF_thermal}.

Beyond these conceptual insights, our work also yields several remarkable formulas. For quantum quenches within a ferromagnetic phase, we have established a curious relationship between the subsystem quantum Fisher information with respect to an order parameter and the corresponding correlation functions—Eq.~\eqref{eq:predictionquench}. While its precise range of validity remains uncertain, both sides of the equation can be computed numerically, even in interacting systems, using tensor network algorithms for time evolution. The quantum Fisher information can be evaluated via the method proposed in Ref.~\cite{PhysRevLett.127.260501} or potentially through a novel approach based on identity \eqref{eq:QFIidentity}, while the correlation functions can be obtained using standard techniques.
A similar observation applies to the analogous formula for the kicking protocol—Eq.~\eqref{eq:qfi_pred_DW}—which one could try to express in terms of correlation functions by inverting Eq.~\eqref{eq:correlations}. This opens the door to further investigations in interacting systems, particularly integrable ones, where the assumptions underlying our derivations are most likely to hold. 

Another intriguing direction of research is to consider analogous protocols of nonequilibrium time evolution starting  from scars of quantum many-body systems rather than from symmetry-breaking ground states. One could indeed argue that separable scars and SSB ground states have a lot in common, and it was already observed, on the one hand, that quantum many-body scars can exhibit extensive multipartite entanglement~\cite{Desaules2022Extensive}, and, on the other hands, that localized perturbations on separable scars produce macroscopic effects~\cite{Bocini2023Growing}. 

One crucial aspect that warrants thorough investigation is the role of dimensionality in the emergence of macroscopic quantum states. In that context, it would be valuable to revisit the studies of Refs.\cite{Balducci_2022, Balducci_2023} on the 2D quantum Ising model, as well as to explore the higher-dimensional generalization of quantum jamming effects examined by Ref.\cite{Fagotti2024Quantum} in quantum spin chains with kinetic constraints.

\begin{acknowledgments}
We thank Sara Murciano, Luca Capizzi, Giuseppe Del Vecchio Del Vecchio, Vanja Mari\'c, and Leonardo Mazza for discussions. 

This work was supported by the European Research Council under the Starting Grant No. 805252 LoCoMacro.
\end{acknowledgments}

\appendix

\bibliography{references.bib}

\begin{thebibliography}{93}%
\makeatletter
\providecommand \@ifxundefined [1]{%
 \@ifx{#1\undefined}
}%
\providecommand \@ifnum [1]{%
 \ifnum #1\expandafter \@firstoftwo
 \else \expandafter \@secondoftwo
 \fi
}%
\providecommand \@ifx [1]{%
 \ifx #1\expandafter \@firstoftwo
 \else \expandafter \@secondoftwo
 \fi
}%
\providecommand \natexlab [1]{#1}%
\providecommand \enquote  [1]{``#1''}%
\providecommand \bibnamefont  [1]{#1}%
\providecommand \bibfnamefont [1]{#1}%
\providecommand \citenamefont [1]{#1}%
\providecommand \href@noop [0]{\@secondoftwo}%
\providecommand \href [0]{\begingroup \@sanitize@url \@href}%
\providecommand \@href[1]{\@@startlink{#1}\@@href}%
\providecommand \@@href[1]{\endgroup#1\@@endlink}%
\providecommand \@sanitize@url [0]{\catcode `\\12\catcode `\$12\catcode `\&12\catcode `\#12\catcode `\^12\catcode `\_12\catcode `\%12\relax}%
\providecommand \@@startlink[1]{}%
\providecommand \@@endlink[0]{}%
\providecommand \url  [0]{\begingroup\@sanitize@url \@url }%
\providecommand \@url [1]{\endgroup\@href {#1}{\urlprefix }}%
\providecommand \urlprefix  [0]{URL }%
\providecommand \Eprint [0]{\href }%
\providecommand \doibase [0]{https://doi.org/}%
\providecommand \selectlanguage [0]{\@gobble}%
\providecommand \bibinfo  [0]{\@secondoftwo}%
\providecommand \bibfield  [0]{\@secondoftwo}%
\providecommand \translation [1]{[#1]}%
\providecommand \BibitemOpen [0]{}%
\providecommand \bibitemStop [0]{}%
\providecommand \bibitemNoStop [0]{.\EOS\space}%
\providecommand \EOS [0]{\spacefactor3000\relax}%
\providecommand \BibitemShut  [1]{\csname bibitem#1\endcsname}%
\let\auto@bib@innerbib\@empty
\bibitem [{\citenamefont {Bennett}\ \emph {et~al.}(1996)\citenamefont {Bennett}, \citenamefont {Bernstein}, \citenamefont {Popescu},\ and\ \citenamefont {Schumacher}}]{Bennett1996Concentrating}%
  \BibitemOpen
  \bibfield  {author} {\bibinfo {author} {\bibfnamefont {C.~H.}\ \bibnamefont {Bennett}}, \bibinfo {author} {\bibfnamefont {H.~J.}\ \bibnamefont {Bernstein}}, \bibinfo {author} {\bibfnamefont {S.}~\bibnamefont {Popescu}},\ and\ \bibinfo {author} {\bibfnamefont {B.}~\bibnamefont {Schumacher}},\ }\bibfield  {title} {\bibinfo {title} {{Concentrating partial entanglement by local operations}},\ }\href {https://doi.org/10.1103/PhysRevA.53.2046} {\bibfield  {journal} {\bibinfo  {journal} {Phys. Rev. A}\ }\textbf {\bibinfo {volume} {53}},\ \bibinfo {pages} {2046} (\bibinfo {year} {1996})}\BibitemShut {NoStop}%
\bibitem [{\citenamefont {Eisert}\ and\ \citenamefont {Plenio}(1999)}]{Eisert1999A}%
  \BibitemOpen
  \bibfield  {author} {\bibinfo {author} {\bibfnamefont {J.}~\bibnamefont {Eisert}}\ and\ \bibinfo {author} {\bibfnamefont {M.~B.}\ \bibnamefont {Plenio}},\ }\bibfield  {title} {\bibinfo {title} {{A comparison of entanglement measures}},\ }\href {https://doi.org/10.1080/09500349908231260} {\bibfield  {journal} {\bibinfo  {journal} {Journal of Modern Optics}\ }\textbf {\bibinfo {volume} {46}},\ \bibinfo {pages} {145} (\bibinfo {year} {1999})}\BibitemShut {NoStop}%
\bibitem [{\citenamefont {Plenio}\ and\ \citenamefont {Virmani}(2014)}]{Plenio2014An}%
  \BibitemOpen
  \bibfield  {author} {\bibinfo {author} {\bibfnamefont {M.~B.}\ \bibnamefont {Plenio}}\ and\ \bibinfo {author} {\bibfnamefont {S.~S.}\ \bibnamefont {Virmani}},\ }\bibinfo {title} {An introduction to entanglement theory},\ in\ \href {https://doi.org/10.1007/978-3-319-04063-9_8} {\emph {\bibinfo {booktitle} {Quantum Information and Coherence}}},\ \bibinfo {editor} {edited by\ \bibinfo {editor} {\bibfnamefont {E.}~\bibnamefont {Andersson}}\ and\ \bibinfo {editor} {\bibfnamefont {P.}~\bibnamefont {{\"O}hberg}}}\ (\bibinfo  {publisher} {Springer International Publishing},\ \bibinfo {address} {Cham},\ \bibinfo {year} {2014})\ pp.\ \bibinfo {pages} {173--209}\BibitemShut {NoStop}%
\bibitem [{\citenamefont {Latorre}\ \emph {et~al.}(2004)\citenamefont {Latorre}, \citenamefont {Rico},\ and\ \citenamefont {Vidal}}]{Latorre2004Ground}%
  \BibitemOpen
  \bibfield  {author} {\bibinfo {author} {\bibfnamefont {J.~I.}\ \bibnamefont {Latorre}}, \bibinfo {author} {\bibfnamefont {E.}~\bibnamefont {Rico}},\ and\ \bibinfo {author} {\bibfnamefont {G.}~\bibnamefont {Vidal}},\ }\bibfield  {title} {\bibinfo {title} {{Ground state entanglement in quantum spin chains}},\ }\href {https://doi.org/10.26421/QIC4.1-4} {\bibfield  {journal} {\bibinfo  {journal} {Quantum Information and Computation}\ }\textbf {\bibinfo {volume} {4}},\ \bibinfo {pages} {048} (\bibinfo {year} {2004})}\BibitemShut {NoStop}%
\bibitem [{\citenamefont {Kitaev}\ and\ \citenamefont {Preskill}(2006)}]{Kitaev2006Topological}%
  \BibitemOpen
  \bibfield  {author} {\bibinfo {author} {\bibfnamefont {A.}~\bibnamefont {Kitaev}}\ and\ \bibinfo {author} {\bibfnamefont {J.}~\bibnamefont {Preskill}},\ }\bibfield  {title} {\bibinfo {title} {{Topological Entanglement Entropy}},\ }\href {https://doi.org/10.1103/PhysRevLett.96.110404} {\bibfield  {journal} {\bibinfo  {journal} {Phys. Rev. Lett.}\ }\textbf {\bibinfo {volume} {96}},\ \bibinfo {pages} {110404} (\bibinfo {year} {2006})}\BibitemShut {NoStop}%
\bibitem [{\citenamefont {Hastings}(2007)}]{Hastings2007An}%
  \BibitemOpen
  \bibfield  {author} {\bibinfo {author} {\bibfnamefont {M.~B.}\ \bibnamefont {Hastings}},\ }\bibfield  {title} {\bibinfo {title} {{An area law for one-dimensional quantum systems}},\ }\href {https://doi.org/10.1088/1742-5468/2007/08/P08024} {\bibfield  {journal} {\bibinfo  {journal} {Journal of Statistical Mechanics: Theory and Experiment}\ }\textbf {\bibinfo {volume} {2007}},\ \bibinfo {pages} {P08024} (\bibinfo {year} {2007})}\BibitemShut {NoStop}%
\bibitem [{\citenamefont {Amico}\ \emph {et~al.}(2008)\citenamefont {Amico}, \citenamefont {Fazio}, \citenamefont {Osterloh},\ and\ \citenamefont {Vedral}}]{Amico2008Entanglement}%
  \BibitemOpen
  \bibfield  {author} {\bibinfo {author} {\bibfnamefont {L.}~\bibnamefont {Amico}}, \bibinfo {author} {\bibfnamefont {R.}~\bibnamefont {Fazio}}, \bibinfo {author} {\bibfnamefont {A.}~\bibnamefont {Osterloh}},\ and\ \bibinfo {author} {\bibfnamefont {V.}~\bibnamefont {Vedral}},\ }\bibfield  {title} {\bibinfo {title} {{Entanglement in many-body systems}},\ }\href {https://doi.org/10.1103/RevModPhys.80.517} {\bibfield  {journal} {\bibinfo  {journal} {Rev. Mod. Phys.}\ }\textbf {\bibinfo {volume} {80}},\ \bibinfo {pages} {517} (\bibinfo {year} {2008})}\BibitemShut {NoStop}%
\bibitem [{\citenamefont {Caraglio}\ and\ \citenamefont {Gliozzi}(2008)}]{Caraglio2008Entanglement}%
  \BibitemOpen
  \bibfield  {author} {\bibinfo {author} {\bibfnamefont {M.}~\bibnamefont {Caraglio}}\ and\ \bibinfo {author} {\bibfnamefont {F.}~\bibnamefont {Gliozzi}},\ }\bibfield  {title} {\bibinfo {title} {{Entanglement entropy and twist fields}},\ }\href {https://doi.org/10.1088/1126-6708/2008/11/076} {\bibfield  {journal} {\bibinfo  {journal} {Journal of High Energy Physics}\ }\textbf {\bibinfo {volume} {2008}},\ \bibinfo {pages} {076} (\bibinfo {year} {2008})}\BibitemShut {NoStop}%
\bibitem [{\citenamefont {Cardy}\ \emph {et~al.}(2008)\citenamefont {Cardy}, \citenamefont {Castro-Alvaredo},\ and\ \citenamefont {Doyon}}]{Cardy2008Form}%
  \BibitemOpen
  \bibfield  {author} {\bibinfo {author} {\bibfnamefont {J.~L.}\ \bibnamefont {Cardy}}, \bibinfo {author} {\bibfnamefont {O.~A.}\ \bibnamefont {Castro-Alvaredo}},\ and\ \bibinfo {author} {\bibfnamefont {B.}~\bibnamefont {Doyon}},\ }\bibfield  {title} {\bibinfo {title} {{Form Factors of Branch-Point Twist Fields in Quantum Integrable Models and Entanglement Entropy}},\ }\href {https://doi.org/10.1007/s10955-007-9422-x} {\bibfield  {journal} {\bibinfo  {journal} {Journal of Statistical Physics}\ }\textbf {\bibinfo {volume} {130}},\ \bibinfo {pages} {129} (\bibinfo {year} {2008})}\BibitemShut {NoStop}%
\bibitem [{\citenamefont {Calabrese}\ and\ \citenamefont {Cardy}(2009)}]{Calabrese2009Entanglement}%
  \BibitemOpen
  \bibfield  {author} {\bibinfo {author} {\bibfnamefont {P.}~\bibnamefont {Calabrese}}\ and\ \bibinfo {author} {\bibfnamefont {J.}~\bibnamefont {Cardy}},\ }\bibfield  {title} {\bibinfo {title} {{Entanglement entropy and conformal field theory}},\ }\href {https://doi.org/10.1088/1751-8113/42/50/504005} {\bibfield  {journal} {\bibinfo  {journal} {Journal of Physics A: Mathematical and Theoretical}\ }\textbf {\bibinfo {volume} {42}},\ \bibinfo {pages} {504005} (\bibinfo {year} {2009})}\BibitemShut {NoStop}%
\bibitem [{\citenamefont {Casini}\ and\ \citenamefont {Huerta}(2009)}]{Casini2009Entanglement}%
  \BibitemOpen
  \bibfield  {author} {\bibinfo {author} {\bibfnamefont {H.}~\bibnamefont {Casini}}\ and\ \bibinfo {author} {\bibfnamefont {M.}~\bibnamefont {Huerta}},\ }\bibfield  {title} {\bibinfo {title} {{Entanglement entropy in free quantum field theory}},\ }\href {https://doi.org/10.1088/1751-8113/42/50/504007} {\bibfield  {journal} {\bibinfo  {journal} {Journal of Physics A: Mathematical and Theoretical}\ }\textbf {\bibinfo {volume} {42}},\ \bibinfo {pages} {504007} (\bibinfo {year} {2009})}\BibitemShut {NoStop}%
\bibitem [{\citenamefont {Ercolessi}\ \emph {et~al.}(2011)\citenamefont {Ercolessi}, \citenamefont {Evangelisti}, \citenamefont {Franchini},\ and\ \citenamefont {Ravanini}}]{Ercolessi2011Essential}%
  \BibitemOpen
  \bibfield  {author} {\bibinfo {author} {\bibfnamefont {E.}~\bibnamefont {Ercolessi}}, \bibinfo {author} {\bibfnamefont {S.}~\bibnamefont {Evangelisti}}, \bibinfo {author} {\bibfnamefont {F.}~\bibnamefont {Franchini}},\ and\ \bibinfo {author} {\bibfnamefont {F.}~\bibnamefont {Ravanini}},\ }\bibfield  {title} {\bibinfo {title} {{Essential singularity in the Renyi entanglement entropy of the one-dimensional $\mathit{XYZ}$ spin-$\frac{1}{2}$ chain}},\ }\href {https://doi.org/10.1103/PhysRevB.83.012402} {\bibfield  {journal} {\bibinfo  {journal} {Phys. Rev. B}\ }\textbf {\bibinfo {volume} {83}},\ \bibinfo {pages} {012402} (\bibinfo {year} {2011})}\BibitemShut {NoStop}%
\bibitem [{\citenamefont {Calabrese}\ \emph {et~al.}(2012{\natexlab{a}})\citenamefont {Calabrese}, \citenamefont {Cardy},\ and\ \citenamefont {Tonni}}]{Calabrese2012Entanglement}%
  \BibitemOpen
  \bibfield  {author} {\bibinfo {author} {\bibfnamefont {P.}~\bibnamefont {Calabrese}}, \bibinfo {author} {\bibfnamefont {J.}~\bibnamefont {Cardy}},\ and\ \bibinfo {author} {\bibfnamefont {E.}~\bibnamefont {Tonni}},\ }\bibfield  {title} {\bibinfo {title} {{Entanglement Negativity in Quantum Field Theory}},\ }\href {https://doi.org/10.1103/PhysRevLett.109.130502} {\bibfield  {journal} {\bibinfo  {journal} {Phys. Rev. Lett.}\ }\textbf {\bibinfo {volume} {109}},\ \bibinfo {pages} {130502} (\bibinfo {year} {2012}{\natexlab{a}})}\BibitemShut {NoStop}%
\bibitem [{\citenamefont {Eisler}\ and\ \citenamefont {Zimborás}(2015)}]{Eisler2015On}%
  \BibitemOpen
  \bibfield  {author} {\bibinfo {author} {\bibfnamefont {V.}~\bibnamefont {Eisler}}\ and\ \bibinfo {author} {\bibfnamefont {Z.}~\bibnamefont {Zimborás}},\ }\bibfield  {title} {\bibinfo {title} {{On the partial transpose of fermionic Gaussian states}},\ }\href {https://doi.org/10.1088/1367-2630/17/5/053048} {\bibfield  {journal} {\bibinfo  {journal} {New Journal of Physics}\ }\textbf {\bibinfo {volume} {17}},\ \bibinfo {pages} {053048} (\bibinfo {year} {2015})}\BibitemShut {NoStop}%
\bibitem [{\citenamefont {Alba}\ and\ \citenamefont {Calabrese}(2017{\natexlab{a}})}]{Alba2017Entanglement}%
  \BibitemOpen
  \bibfield  {author} {\bibinfo {author} {\bibfnamefont {V.}~\bibnamefont {Alba}}\ and\ \bibinfo {author} {\bibfnamefont {P.}~\bibnamefont {Calabrese}},\ }\bibfield  {title} {\bibinfo {title} {{Entanglement and thermodynamics after a quantum quench in integrable systems}},\ }\href {https://doi.org/10.1073/pnas.1703516114} {\bibfield  {journal} {\bibinfo  {journal} {Proceedings of the National Academy of Sciences}\ }\textbf {\bibinfo {volume} {114}},\ \bibinfo {pages} {7947} (\bibinfo {year} {2017}{\natexlab{a}})},\ \Eprint {https://arxiv.org/abs/https://www.pnas.org/doi/pdf/10.1073/pnas.1703516114} {https://www.pnas.org/doi/pdf/10.1073/pnas.1703516114} \BibitemShut {NoStop}%
\bibitem [{\citenamefont {Alba}\ \emph {et~al.}(2019)\citenamefont {Alba}, \citenamefont {Bertini},\ and\ \citenamefont {Fagotti}}]{Alba2019Entanglement}%
  \BibitemOpen
  \bibfield  {author} {\bibinfo {author} {\bibfnamefont {V.}~\bibnamefont {Alba}}, \bibinfo {author} {\bibfnamefont {B.}~\bibnamefont {Bertini}},\ and\ \bibinfo {author} {\bibfnamefont {M.}~\bibnamefont {Fagotti}},\ }\bibfield  {title} {\bibinfo {title} {{Entanglement evolution and generalised hydrodynamics: interacting integrable systems}},\ }\href {https://doi.org/10.21468/SciPostPhys.7.1.005} {\bibfield  {journal} {\bibinfo  {journal} {SciPost Phys.}\ }\textbf {\bibinfo {volume} {7}},\ \bibinfo {pages} {005} (\bibinfo {year} {2019})}\BibitemShut {NoStop}%
\bibitem [{\citenamefont {Bertini}\ \emph {et~al.}(2022)\citenamefont {Bertini}, \citenamefont {Klobas}, \citenamefont {Alba}, \citenamefont {Lagnese},\ and\ \citenamefont {Calabrese}}]{Bertini2022Growth}%
  \BibitemOpen
  \bibfield  {author} {\bibinfo {author} {\bibfnamefont {B.}~\bibnamefont {Bertini}}, \bibinfo {author} {\bibfnamefont {K.}~\bibnamefont {Klobas}}, \bibinfo {author} {\bibfnamefont {V.}~\bibnamefont {Alba}}, \bibinfo {author} {\bibfnamefont {G.}~\bibnamefont {Lagnese}},\ and\ \bibinfo {author} {\bibfnamefont {P.}~\bibnamefont {Calabrese}},\ }\bibfield  {title} {\bibinfo {title} {{Growth of R\'enyi Entropies in Interacting Integrable Models and the Breakdown of the Quasiparticle Picture}},\ }\href {https://doi.org/10.1103/PhysRevX.12.031016} {\bibfield  {journal} {\bibinfo  {journal} {Phys. Rev. X}\ }\textbf {\bibinfo {volume} {12}},\ \bibinfo {pages} {031016} (\bibinfo {year} {2022})}\BibitemShut {NoStop}%
\bibitem [{\citenamefont {Dalmonte}\ \emph {et~al.}(2022)\citenamefont {Dalmonte}, \citenamefont {Eisler}, \citenamefont {Falconi},\ and\ \citenamefont {Vermersch}}]{Dalmonte2022Entanglement}%
  \BibitemOpen
  \bibfield  {author} {\bibinfo {author} {\bibfnamefont {M.}~\bibnamefont {Dalmonte}}, \bibinfo {author} {\bibfnamefont {V.}~\bibnamefont {Eisler}}, \bibinfo {author} {\bibfnamefont {M.}~\bibnamefont {Falconi}},\ and\ \bibinfo {author} {\bibfnamefont {B.}~\bibnamefont {Vermersch}},\ }\bibfield  {title} {\bibinfo {title} {{Entanglement Hamiltonians: From Field Theory to Lattice Models and Experiments}},\ }\href {https://doi.org/https://doi.org/10.1002/andp.202200064} {\bibfield  {journal} {\bibinfo  {journal} {Annalen der Physik}\ }\textbf {\bibinfo {volume} {534}},\ \bibinfo {pages} {2200064} (\bibinfo {year} {2022})},\ \Eprint {https://arxiv.org/abs/https://onlinelibrary.wiley.com/doi/pdf/10.1002/andp.202200064} {https://onlinelibrary.wiley.com/doi/pdf/10.1002/andp.202200064} \BibitemShut {NoStop}%
\bibitem [{\citenamefont {Mari{\'{c}}}\ \emph {et~al.}(2024)\citenamefont {Mari{\'{c}}}, \citenamefont {Bocini},\ and\ \citenamefont {Fagotti}}]{Maric2024Entanglement}%
  \BibitemOpen
  \bibfield  {author} {\bibinfo {author} {\bibfnamefont {V.}~\bibnamefont {Mari{\'{c}}}}, \bibinfo {author} {\bibfnamefont {S.}~\bibnamefont {Bocini}},\ and\ \bibinfo {author} {\bibfnamefont {M.}~\bibnamefont {Fagotti}},\ }\bibfield  {title} {\bibinfo {title} {{Entanglement entropy of two disjoint intervals and spin structures in interacting chains in and out of equilibrium}},\ }\href {https://doi.org/10.1007/JHEP03(2024)044} {\bibfield  {journal} {\bibinfo  {journal} {Journal of High Energy Physics}\ }\textbf {\bibinfo {volume} {2024}},\ \bibinfo {pages} {44} (\bibinfo {year} {2024})}\BibitemShut {NoStop}%
\bibitem [{\citenamefont {Chitambar}\ and\ \citenamefont {Gour}(2019)}]{Chitambar2019Quantum}%
  \BibitemOpen
  \bibfield  {author} {\bibinfo {author} {\bibfnamefont {E.}~\bibnamefont {Chitambar}}\ and\ \bibinfo {author} {\bibfnamefont {G.}~\bibnamefont {Gour}},\ }\bibfield  {title} {\bibinfo {title} {{Quantum resource theories}},\ }\href {https://doi.org/10.1103/RevModPhys.91.025001} {\bibfield  {journal} {\bibinfo  {journal} {Rev. Mod. Phys.}\ }\textbf {\bibinfo {volume} {91}},\ \bibinfo {pages} {025001} (\bibinfo {year} {2019})}\BibitemShut {NoStop}%
\bibitem [{\citenamefont {Tan}\ \emph {et~al.}(2021)\citenamefont {Tan}, \citenamefont {Narasimhachar},\ and\ \citenamefont {Regula}}]{Tan2021Fisher}%
  \BibitemOpen
  \bibfield  {author} {\bibinfo {author} {\bibfnamefont {K.~C.}\ \bibnamefont {Tan}}, \bibinfo {author} {\bibfnamefont {V.}~\bibnamefont {Narasimhachar}},\ and\ \bibinfo {author} {\bibfnamefont {B.}~\bibnamefont {Regula}},\ }\bibfield  {title} {\bibinfo {title} {Fisher information universally identifies quantum resources},\ }\href {https://doi.org/10.1103/PhysRevLett.127.200402} {\bibfield  {journal} {\bibinfo  {journal} {Phys. Rev. Lett.}\ }\textbf {\bibinfo {volume} {127}},\ \bibinfo {pages} {200402} (\bibinfo {year} {2021})}\BibitemShut {NoStop}%
\bibitem [{\citenamefont {Braunstein}\ and\ \citenamefont {Caves}(1994)}]{PhysRevLett.72.3439}%
  \BibitemOpen
  \bibfield  {author} {\bibinfo {author} {\bibfnamefont {S.~L.}\ \bibnamefont {Braunstein}}\ and\ \bibinfo {author} {\bibfnamefont {C.~M.}\ \bibnamefont {Caves}},\ }\bibfield  {title} {\bibinfo {title} {Statistical distance and the geometry of quantum states},\ }\href {https://doi.org/10.1103/PhysRevLett.72.3439} {\bibfield  {journal} {\bibinfo  {journal} {Phys. Rev. Lett.}\ }\textbf {\bibinfo {volume} {72}},\ \bibinfo {pages} {3439} (\bibinfo {year} {1994})}\BibitemShut {NoStop}%
\bibitem [{\citenamefont {Helstrom}(1969)}]{cramer_rao}%
  \BibitemOpen
  \bibfield  {author} {\bibinfo {author} {\bibfnamefont {C.}~\bibnamefont {Helstrom}},\ }\bibfield  {title} {\bibinfo {title} {Quantum detection and estimation theory},\ }\href {https://doi.org/https://doi.org/10.1007/BF01007479} {\bibfield  {journal} {\bibinfo  {journal} {J Stat Phys}\ }\textbf {\bibinfo {volume} {122}},\ \bibinfo {pages} {231252} (\bibinfo {year} {1969})}\BibitemShut {NoStop}%
\bibitem [{\citenamefont {Pezz\'e}\ and\ \citenamefont {Smerzi}(2009)}]{Pezze2009Entanglement}%
  \BibitemOpen
  \bibfield  {author} {\bibinfo {author} {\bibfnamefont {L.}~\bibnamefont {Pezz\'e}}\ and\ \bibinfo {author} {\bibfnamefont {A.}~\bibnamefont {Smerzi}},\ }\bibfield  {title} {\bibinfo {title} {{Entanglement, Nonlinear Dynamics, and the Heisenberg Limit}},\ }\href {https://doi.org/10.1103/PhysRevLett.102.100401} {\bibfield  {journal} {\bibinfo  {journal} {Phys. Rev. Lett.}\ }\textbf {\bibinfo {volume} {102}},\ \bibinfo {pages} {100401} (\bibinfo {year} {2009})}\BibitemShut {NoStop}%
\bibitem [{\citenamefont {Hyllus}\ \emph {et~al.}(2012)\citenamefont {Hyllus}, \citenamefont {Laskowski}, \citenamefont {Krischek}, \citenamefont {Schwemmer}, \citenamefont {Wieczorek}, \citenamefont {Weinfurter}, \citenamefont {Pezz\'e},\ and\ \citenamefont {Smerzi}}]{Hyllus2012Fisher}%
  \BibitemOpen
  \bibfield  {author} {\bibinfo {author} {\bibfnamefont {P.}~\bibnamefont {Hyllus}}, \bibinfo {author} {\bibfnamefont {W.}~\bibnamefont {Laskowski}}, \bibinfo {author} {\bibfnamefont {R.}~\bibnamefont {Krischek}}, \bibinfo {author} {\bibfnamefont {C.}~\bibnamefont {Schwemmer}}, \bibinfo {author} {\bibfnamefont {W.}~\bibnamefont {Wieczorek}}, \bibinfo {author} {\bibfnamefont {H.}~\bibnamefont {Weinfurter}}, \bibinfo {author} {\bibfnamefont {L.}~\bibnamefont {Pezz\'e}},\ and\ \bibinfo {author} {\bibfnamefont {A.}~\bibnamefont {Smerzi}},\ }\bibfield  {title} {\bibinfo {title} {{Fisher information and multiparticle entanglement}},\ }\href {https://doi.org/10.1103/PhysRevA.85.022321} {\bibfield  {journal} {\bibinfo  {journal} {Phys. Rev. A}\ }\textbf {\bibinfo {volume} {85}},\ \bibinfo {pages} {022321} (\bibinfo {year} {2012})}\BibitemShut {NoStop}%
\bibitem [{\citenamefont {T\'oth}(2012)}]{Toth2012Multipartite}%
  \BibitemOpen
  \bibfield  {author} {\bibinfo {author} {\bibfnamefont {G.}~\bibnamefont {T\'oth}},\ }\bibfield  {title} {\bibinfo {title} {{Multipartite entanglement and high-precision metrology}},\ }\href {https://doi.org/10.1103/PhysRevA.85.022322} {\bibfield  {journal} {\bibinfo  {journal} {Phys. Rev. A}\ }\textbf {\bibinfo {volume} {85}},\ \bibinfo {pages} {022322} (\bibinfo {year} {2012})}\BibitemShut {NoStop}%
\bibitem [{\citenamefont {Fr\"owis}\ and\ \citenamefont {D\"ur}(2012)}]{Froewis2012Measures}%
  \BibitemOpen
  \bibfield  {author} {\bibinfo {author} {\bibfnamefont {F.}~\bibnamefont {Fr\"owis}}\ and\ \bibinfo {author} {\bibfnamefont {W.}~\bibnamefont {D\"ur}},\ }\bibfield  {title} {\bibinfo {title} {{Measures of macroscopicity for quantum spin systems}},\ }\href {https://doi.org/10.1088/1367-2630/14/9/093039} {\bibfield  {journal} {\bibinfo  {journal} {New Journal of Physics}\ }\textbf {\bibinfo {volume} {14}},\ \bibinfo {pages} {093039} (\bibinfo {year} {2012})}\BibitemShut {NoStop}%
\bibitem [{\citenamefont {Wigner}\ and\ \citenamefont {Yanase}(1963)}]{Wigner1963INFORMATION}%
  \BibitemOpen
  \bibfield  {author} {\bibinfo {author} {\bibfnamefont {E.~P.}\ \bibnamefont {Wigner}}\ and\ \bibinfo {author} {\bibfnamefont {M.~M.}\ \bibnamefont {Yanase}},\ }\bibfield  {title} {\bibinfo {title} {{INFORMATION CONTENTS OF DISTRIBUTIONS}},\ }\href {https://doi.org/10.1073/pnas.49.6.910} {\bibfield  {journal} {\bibinfo  {journal} {Proceedings of the National Academy of Sciences}\ }\textbf {\bibinfo {volume} {49}},\ \bibinfo {pages} {910} (\bibinfo {year} {1963})},\ \Eprint {https://arxiv.org/abs/https://www.pnas.org/doi/pdf/10.1073/pnas.49.6.910} {https://www.pnas.org/doi/pdf/10.1073/pnas.49.6.910} \BibitemShut {NoStop}%
\bibitem [{\citenamefont {Rath}\ \emph {et~al.}(2021)\citenamefont {Rath}, \citenamefont {Branciard}, \citenamefont {Minguzzi},\ and\ \citenamefont {Vermersch}}]{PhysRevLett.127.260501}%
  \BibitemOpen
  \bibfield  {author} {\bibinfo {author} {\bibfnamefont {A.}~\bibnamefont {Rath}}, \bibinfo {author} {\bibfnamefont {C.}~\bibnamefont {Branciard}}, \bibinfo {author} {\bibfnamefont {A.}~\bibnamefont {Minguzzi}},\ and\ \bibinfo {author} {\bibfnamefont {B.}~\bibnamefont {Vermersch}},\ }\bibfield  {title} {\bibinfo {title} {Quantum fisher information from randomized measurements},\ }\href {https://doi.org/10.1103/PhysRevLett.127.260501} {\bibfield  {journal} {\bibinfo  {journal} {Phys. Rev. Lett.}\ }\textbf {\bibinfo {volume} {127}},\ \bibinfo {pages} {260501} (\bibinfo {year} {2021})}\BibitemShut {NoStop}%
\bibitem [{\citenamefont {T\'oth}\ and\ \citenamefont {Petz}(2013)}]{Toth2013Extremal}%
  \BibitemOpen
  \bibfield  {author} {\bibinfo {author} {\bibfnamefont {G.}~\bibnamefont {T\'oth}}\ and\ \bibinfo {author} {\bibfnamefont {D.}~\bibnamefont {Petz}},\ }\bibfield  {title} {\bibinfo {title} {{Extremal properties of the variance and the quantum Fisher information}},\ }\href {https://doi.org/10.1103/PhysRevA.87.032324} {\bibfield  {journal} {\bibinfo  {journal} {Phys. Rev. A}\ }\textbf {\bibinfo {volume} {87}},\ \bibinfo {pages} {032324} (\bibinfo {year} {2013})}\BibitemShut {NoStop}%
\bibitem [{\citenamefont {Yu}(2013)}]{yu2013quantum}%
  \BibitemOpen
  \bibfield  {author} {\bibinfo {author} {\bibfnamefont {S.}~\bibnamefont {Yu}},\ }\bibfield  {title} {\bibinfo {title} {{Quantum Fisher Information as the Convex Roof of Variance}},\ }\Eprint {https://arxiv.org/abs/1302.5311} {arXiv:1302.5311}  (\bibinfo {year} {2013})\BibitemShut {NoStop}%
\bibitem [{\citenamefont {Fr\'erot}\ and\ \citenamefont {Roscilde}(2016)}]{Frerot2016Quantum}%
  \BibitemOpen
  \bibfield  {author} {\bibinfo {author} {\bibfnamefont {I.}~\bibnamefont {Fr\'erot}}\ and\ \bibinfo {author} {\bibfnamefont {T.}~\bibnamefont {Roscilde}},\ }\bibfield  {title} {\bibinfo {title} {{Quantum variance: A measure of quantum coherence and quantum correlations for many-body systems}},\ }\href {https://doi.org/10.1103/PhysRevB.94.075121} {\bibfield  {journal} {\bibinfo  {journal} {Phys. Rev. B}\ }\textbf {\bibinfo {volume} {94}},\ \bibinfo {pages} {075121} (\bibinfo {year} {2016})}\BibitemShut {NoStop}%
\bibitem [{\citenamefont {Leggett}(1980)}]{Leggett1980Macroscopic}%
  \BibitemOpen
  \bibfield  {author} {\bibinfo {author} {\bibfnamefont {A.~J.}\ \bibnamefont {Leggett}},\ }\bibfield  {title} {\bibinfo {title} {{Macroscopic Quantum Systems and the Quantum Theory of Measurement}},\ }\href {https://doi.org/10.1143/PTP.69.80} {\bibfield  {journal} {\bibinfo  {journal} {Progress of Theoretical Physics Supplement}\ }\textbf {\bibinfo {volume} {69}},\ \bibinfo {pages} {80} (\bibinfo {year} {1980})}\BibitemShut {NoStop}%
\bibitem [{\citenamefont {Zauner}\ \emph {et~al.}(2015)\citenamefont {Zauner}, \citenamefont {Ganahl}, \citenamefont {Evertz},\ and\ \citenamefont {Nishino}}]{Zauner2015Time}%
  \BibitemOpen
  \bibfield  {author} {\bibinfo {author} {\bibfnamefont {V.}~\bibnamefont {Zauner}}, \bibinfo {author} {\bibfnamefont {M.}~\bibnamefont {Ganahl}}, \bibinfo {author} {\bibfnamefont {H.~G.}\ \bibnamefont {Evertz}},\ and\ \bibinfo {author} {\bibfnamefont {T.}~\bibnamefont {Nishino}},\ }\bibfield  {title} {\bibinfo {title} {{Time evolution within a comoving window: scaling of signal fronts and magnetization plateaus after a local quench in quantum spin chains}},\ }\href {https://doi.org/10.1088/0953-8984/27/42/425602} {\bibfield  {journal} {\bibinfo  {journal} {Journal of Physics: Condensed Matter}\ }\textbf {\bibinfo {volume} {27}},\ \bibinfo {pages} {425602} (\bibinfo {year} {2015})}\BibitemShut {NoStop}%
\bibitem [{\citenamefont {Eisler}\ and\ \citenamefont {Maislinger}(2020)}]{Eisler2020Front}%
  \BibitemOpen
  \bibfield  {author} {\bibinfo {author} {\bibfnamefont {V.}~\bibnamefont {Eisler}}\ and\ \bibinfo {author} {\bibfnamefont {F.}~\bibnamefont {Maislinger}},\ }\bibfield  {title} {\bibinfo {title} {{Front dynamics in the XY chain after local excitations}},\ }\href {https://doi.org/10.21468/SciPostPhys.8.3.037} {\bibfield  {journal} {\bibinfo  {journal} {SciPost Phys.}\ }\textbf {\bibinfo {volume} {8}},\ \bibinfo {pages} {037} (\bibinfo {year} {2020})}\BibitemShut {NoStop}%
\bibitem [{\citenamefont {Delfino}\ and\ \citenamefont {Sorba}(2022)}]{Delfino2022Space}%
  \BibitemOpen
  \bibfield  {author} {\bibinfo {author} {\bibfnamefont {G.}~\bibnamefont {Delfino}}\ and\ \bibinfo {author} {\bibfnamefont {M.}~\bibnamefont {Sorba}},\ }\bibfield  {title} {\bibinfo {title} {{Space of initial conditions and universality in nonequilibrium quantum dynamics}},\ }\href {https://doi.org/https://doi.org/10.1016/j.nuclphysb.2022.115910} {\bibfield  {journal} {\bibinfo  {journal} {Nuclear Physics B}\ }\textbf {\bibinfo {volume} {983}},\ \bibinfo {pages} {115910} (\bibinfo {year} {2022})}\BibitemShut {NoStop}%
\bibitem [{\citenamefont {Haake}\ \emph {et~al.}(2018)\citenamefont {Haake}, \citenamefont {Gnutzmann},\ and\ \citenamefont {Ku\'s}}]{Haake2018Quantum}%
  \BibitemOpen
  \bibfield  {author} {\bibinfo {author} {\bibfnamefont {F.}~\bibnamefont {Haake}}, \bibinfo {author} {\bibfnamefont {S.}~\bibnamefont {Gnutzmann}},\ and\ \bibinfo {author} {\bibfnamefont {M.}~\bibnamefont {Ku\'s}},\ }\href {https://doi.org/https://doi.org/10.1007/978-3-319-97580-1} {\emph {\bibinfo {title} {{Quantum Signatures of Chaos}}}}\ (\bibinfo  {publisher} {Springer Cham},\ \bibinfo {year} {2018})\BibitemShut {NoStop}%
\bibitem [{\citenamefont {Prosen}(1998)}]{Prosen1998Time}%
  \BibitemOpen
  \bibfield  {author} {\bibinfo {author} {\bibfnamefont {T.}~\bibnamefont {Prosen}},\ }\bibfield  {title} {\bibinfo {title} {{Time Evolution of a Quantum Many-Body System: Transition from Integrability to Ergodicity in the Thermodynamic Limit}},\ }\href {https://doi.org/10.1103/PhysRevLett.80.1808} {\bibfield  {journal} {\bibinfo  {journal} {Phys. Rev. Lett.}\ }\textbf {\bibinfo {volume} {80}},\ \bibinfo {pages} {1808} (\bibinfo {year} {1998})}\BibitemShut {NoStop}%
\bibitem [{\citenamefont {Prosen}(2002)}]{Prosen2002General}%
  \BibitemOpen
  \bibfield  {author} {\bibinfo {author} {\bibfnamefont {T.}~\bibnamefont {Prosen}},\ }\bibfield  {title} {\bibinfo {title} {{General relation between quantum ergodicity and fidelity of quantum dynamics}},\ }\href {https://doi.org/10.1103/PhysRevE.65.036208} {\bibfield  {journal} {\bibinfo  {journal} {Phys. Rev. E}\ }\textbf {\bibinfo {volume} {65}},\ \bibinfo {pages} {036208} (\bibinfo {year} {2002})}\BibitemShut {NoStop}%
\bibitem [{\citenamefont {Pezz\`e}\ \emph {et~al.}(2017)\citenamefont {Pezz\`e}, \citenamefont {Gabbrielli}, \citenamefont {Lepori},\ and\ \citenamefont {Smerzi}}]{Pezze2017Multipartite}%
  \BibitemOpen
  \bibfield  {author} {\bibinfo {author} {\bibfnamefont {L.}~\bibnamefont {Pezz\`e}}, \bibinfo {author} {\bibfnamefont {M.}~\bibnamefont {Gabbrielli}}, \bibinfo {author} {\bibfnamefont {L.}~\bibnamefont {Lepori}},\ and\ \bibinfo {author} {\bibfnamefont {A.}~\bibnamefont {Smerzi}},\ }\bibfield  {title} {\bibinfo {title} {{Multipartite Entanglement in Topological Quantum Phases}},\ }\href {https://doi.org/10.1103/PhysRevLett.119.250401} {\bibfield  {journal} {\bibinfo  {journal} {Phys. Rev. Lett.}\ }\textbf {\bibinfo {volume} {119}},\ \bibinfo {pages} {250401} (\bibinfo {year} {2017})}\BibitemShut {NoStop}%
\bibitem [{\citenamefont {Dell'Anna}\ \emph {et~al.}(2023)\citenamefont {Dell'Anna}, \citenamefont {Pradhan}, \citenamefont {Boschi},\ and\ \citenamefont {Ercolessi}}]{DellAnna2023Quantum}%
  \BibitemOpen
  \bibfield  {author} {\bibinfo {author} {\bibfnamefont {F.}~\bibnamefont {Dell'Anna}}, \bibinfo {author} {\bibfnamefont {S.}~\bibnamefont {Pradhan}}, \bibinfo {author} {\bibfnamefont {C.~D.~E.}\ \bibnamefont {Boschi}},\ and\ \bibinfo {author} {\bibfnamefont {E.}~\bibnamefont {Ercolessi}},\ }\bibfield  {title} {\bibinfo {title} {{Quantum Fisher information and multipartite entanglement in spin-1 chains}},\ }\href {https://doi.org/10.1103/PhysRevB.108.144414} {\bibfield  {journal} {\bibinfo  {journal} {Phys. Rev. B}\ }\textbf {\bibinfo {volume} {108}},\ \bibinfo {pages} {144414} (\bibinfo {year} {2023})}\BibitemShut {NoStop}%
\bibitem [{\citenamefont {Qu}\ \emph {et~al.}(2025)\citenamefont {Qu}, \citenamefont {Xu}, \citenamefont {Guo},\ and\ \citenamefont {Sun}}]{Qu2025Quantum}%
  \BibitemOpen
  \bibfield  {author} {\bibinfo {author} {\bibfnamefont {S.}~\bibnamefont {Qu}}, \bibinfo {author} {\bibfnamefont {F.-Q.}\ \bibnamefont {Xu}}, \bibinfo {author} {\bibfnamefont {B.}~\bibnamefont {Guo}},\ and\ \bibinfo {author} {\bibfnamefont {Z.-Y.}\ \bibnamefont {Sun}},\ }\bibfield  {title} {\bibinfo {title} {{Quantum Fisher information in one-dimensional translation-invariant quantum systems: Large-N limit analysis}},\ }\href {https://doi.org/https://doi.org/10.1016/j.physleta.2024.130103} {\bibfield  {journal} {\bibinfo  {journal} {Physics Letters A}\ }\textbf {\bibinfo {volume} {529}},\ \bibinfo {pages} {130103} (\bibinfo {year} {2025})}\BibitemShut {NoStop}%
\bibitem [{\citenamefont {Lieb}(1973{\natexlab{a}})}]{Lieb1973Convex}%
  \BibitemOpen
  \bibfield  {author} {\bibinfo {author} {\bibfnamefont {E.~H.}\ \bibnamefont {Lieb}},\ }\bibfield  {title} {\bibinfo {title} {{Convex trace functions and the Wigner-Yanase-Dyson conjecture}},\ }\href {https://doi.org/https://doi.org/10.1016/0001-8708(73)90011-X} {\bibfield  {journal} {\bibinfo  {journal} {Advances in Mathematics}\ }\textbf {\bibinfo {volume} {11}},\ \bibinfo {pages} {267} (\bibinfo {year} {1973}{\natexlab{a}})}\BibitemShut {NoStop}%
\bibitem [{\citenamefont {Luo}(2004)}]{luoshulongwigvsqfi}%
  \BibitemOpen
  \bibfield  {author} {\bibinfo {author} {\bibfnamefont {S.}~\bibnamefont {Luo}},\ }\bibfield  {title} {\bibinfo {title} {Wigner-yanase skew information vs. quantum fisher information},\ }\href {https://doi.org/10.2307/1194711} {\bibfield  {journal} {\bibinfo  {journal} {Proceedings of the American Mathematical Society}\ }\textbf {\bibinfo {volume} {132}},\ \bibinfo {pages} {885} (\bibinfo {year} {2004})}\BibitemShut {NoStop}%
\bibitem [{\citenamefont {Fisher}(1925)}]{Fisher_1925}%
  \BibitemOpen
  \bibfield  {author} {\bibinfo {author} {\bibfnamefont {R.~A.}\ \bibnamefont {Fisher}},\ }\bibfield  {title} {\bibinfo {title} {Theory of statistical estimation},\ }\href {https://doi.org/10.1017/S0305004100009580} {\bibfield  {journal} {\bibinfo  {journal} {Mathematical Proceedings of the Cambridge Philosophical Society}\ }\textbf {\bibinfo {volume} {22}},\ \bibinfo {pages} {700–725} (\bibinfo {year} {1925})}\BibitemShut {NoStop}%
\bibitem [{\citenamefont {Girolami}(2014)}]{Girolami_2014}%
  \BibitemOpen
  \bibfield  {author} {\bibinfo {author} {\bibfnamefont {D.}~\bibnamefont {Girolami}},\ }\bibfield  {title} {\bibinfo {title} {Observable measure of quantum coherence in finite dimensional systems},\ }\bibfield  {journal} {\bibinfo  {journal} {Physical Review Letters}\ }\textbf {\bibinfo {volume} {113}},\ \href {https://doi.org/10.1103/physrevlett.113.170401} {10.1103/physrevlett.113.170401} (\bibinfo {year} {2014})\BibitemShut {NoStop}%
\bibitem [{\citenamefont {Lieb}(1973{\natexlab{b}})}]{LIEB1973267}%
  \BibitemOpen
  \bibfield  {author} {\bibinfo {author} {\bibfnamefont {E.~H.}\ \bibnamefont {Lieb}},\ }\bibfield  {title} {\bibinfo {title} {Convex trace functions and the wigner-yanase-dyson conjecture},\ }\href {https://doi.org/https://doi.org/10.1016/0001-8708(73)90011-X} {\bibfield  {journal} {\bibinfo  {journal} {Advances in Mathematics}\ }\textbf {\bibinfo {volume} {11}},\ \bibinfo {pages} {267} (\bibinfo {year} {1973}{\natexlab{b}})}\BibitemShut {NoStop}%
\bibitem [{\citenamefont {Lambert}\ and\ \citenamefont {S\o{}rensen}(2019)}]{Lambert2019Estimates}%
  \BibitemOpen
  \bibfield  {author} {\bibinfo {author} {\bibfnamefont {J.}~\bibnamefont {Lambert}}\ and\ \bibinfo {author} {\bibfnamefont {E.~S.}\ \bibnamefont {S\o{}rensen}},\ }\bibfield  {title} {\bibinfo {title} {Estimates of the quantum fisher information in the $s=1$ antiferromagnetic heisenberg spin chain with uniaxial anisotropy},\ }\href {https://doi.org/10.1103/PhysRevB.99.045117} {\bibfield  {journal} {\bibinfo  {journal} {Phys. Rev. B}\ }\textbf {\bibinfo {volume} {99}},\ \bibinfo {pages} {045117} (\bibinfo {year} {2019})}\BibitemShut {NoStop}%
\bibitem [{\citenamefont {Liu}\ \emph {et~al.}(2013)\citenamefont {Liu}, \citenamefont {Ma},\ and\ \citenamefont {Wang}}]{Liu2013Quantum}%
  \BibitemOpen
  \bibfield  {author} {\bibinfo {author} {\bibfnamefont {W.-F.}\ \bibnamefont {Liu}}, \bibinfo {author} {\bibfnamefont {J.}~\bibnamefont {Ma}},\ and\ \bibinfo {author} {\bibfnamefont {X.}~\bibnamefont {Wang}},\ }\bibfield  {title} {\bibinfo {title} {Quantum fisher information and spin squeezing in the ground state of the xy model},\ }\href {https://doi.org/10.1088/1751-8113/46/4/045302} {\bibfield  {journal} {\bibinfo  {journal} {Journal of Physics A: Mathematical and Theoretical}\ }\textbf {\bibinfo {volume} {46}},\ \bibinfo {pages} {045302} (\bibinfo {year} {2013})}\BibitemShut {NoStop}%
\bibitem [{\citenamefont {Hauke}\ \emph {et~al.}(2016)\citenamefont {Hauke}, \citenamefont {Heyl}, \citenamefont {Tagliacozzo},\ and\ \citenamefont {Zoller}}]{Hauke2016Measuring}%
  \BibitemOpen
  \bibfield  {author} {\bibinfo {author} {\bibfnamefont {P.}~\bibnamefont {Hauke}}, \bibinfo {author} {\bibfnamefont {M.}~\bibnamefont {Heyl}}, \bibinfo {author} {\bibfnamefont {L.}~\bibnamefont {Tagliacozzo}},\ and\ \bibinfo {author} {\bibfnamefont {P.}~\bibnamefont {Zoller}},\ }\bibfield  {title} {\bibinfo {title} {Measuring multipartite entanglement through dynamic susceptibilities},\ }\href {https://doi.org/10.1038/nphys3700} {\bibfield  {journal} {\bibinfo  {journal} {Nature Physics}\ }\textbf {\bibinfo {volume} {12}},\ \bibinfo {pages} {778–782} (\bibinfo {year} {2016})}\BibitemShut {NoStop}%
\bibitem [{\citenamefont {Vidal}\ \emph {et~al.}(2021)\citenamefont {Vidal}, \citenamefont {Bera}, \citenamefont {Riera}, \citenamefont {Lewenstein},\ and\ \citenamefont {Bera}}]{Vidal2021Quantum}%
  \BibitemOpen
  \bibfield  {author} {\bibinfo {author} {\bibfnamefont {N.~T.}\ \bibnamefont {Vidal}}, \bibinfo {author} {\bibfnamefont {M.~L.}\ \bibnamefont {Bera}}, \bibinfo {author} {\bibfnamefont {A.}~\bibnamefont {Riera}}, \bibinfo {author} {\bibfnamefont {M.}~\bibnamefont {Lewenstein}},\ and\ \bibinfo {author} {\bibfnamefont {M.~N.}\ \bibnamefont {Bera}},\ }\bibfield  {title} {\bibinfo {title} {{Quantum operations in an information theory for fermions}},\ }\href {https://doi.org/10.1103/PhysRevA.104.032411} {\bibfield  {journal} {\bibinfo  {journal} {Phys. Rev. A}\ }\textbf {\bibinfo {volume} {104}},\ \bibinfo {pages} {032411} (\bibinfo {year} {2021})}\BibitemShut {NoStop}%
\bibitem [{\citenamefont {Fagotti}\ \emph {et~al.}(2024)\citenamefont {Fagotti}, \citenamefont {Mari\ifmmode~\acute{c}\else \'{c}\fi{}},\ and\ \citenamefont {Zadnik}}]{Fagotti2024Nonequilibrium}%
  \BibitemOpen
  \bibfield  {author} {\bibinfo {author} {\bibfnamefont {M.}~\bibnamefont {Fagotti}}, \bibinfo {author} {\bibfnamefont {V.}~\bibnamefont {Mari\ifmmode~\acute{c}\else \'{c}\fi{}}},\ and\ \bibinfo {author} {\bibfnamefont {L.}~\bibnamefont {Zadnik}},\ }\bibfield  {title} {\bibinfo {title} {{Nonequilibrium symmetry-protected topological order: Emergence of semilocal Gibbs ensembles}},\ }\href {https://doi.org/10.1103/PhysRevB.109.115117} {\bibfield  {journal} {\bibinfo  {journal} {Phys. Rev. B}\ }\textbf {\bibinfo {volume} {109}},\ \bibinfo {pages} {115117} (\bibinfo {year} {2024})}\BibitemShut {NoStop}%
\bibitem [{\citenamefont {Zeng}\ \emph {et~al.}(2019)\citenamefont {Zeng}, \citenamefont {Chen}, \citenamefont {Zhou},\ and\ \citenamefont {Wen}}]{Zeng2019book}%
  \BibitemOpen
  \bibfield  {author} {\bibinfo {author} {\bibfnamefont {B.}~\bibnamefont {Zeng}}, \bibinfo {author} {\bibfnamefont {X.}~\bibnamefont {Chen}}, \bibinfo {author} {\bibfnamefont {D.-L.}\ \bibnamefont {Zhou}},\ and\ \bibinfo {author} {\bibfnamefont {X.-G.}\ \bibnamefont {Wen}},\ }\href {https://doi.org/https://doi.org/10.1007/978-1-4939-9084-9} {\emph {\bibinfo {title} {Quantum Information Meets Quantum Matter}}}\ (\bibinfo  {publisher} {Springer, New York},\ \bibinfo {year} {2019})\BibitemShut {NoStop}%
\bibitem [{\citenamefont {Fr\"owis}\ \emph {et~al.}(2018)\citenamefont {Fr\"owis}, \citenamefont {Sekatski}, \citenamefont {D\"ur}, \citenamefont {Gisin},\ and\ \citenamefont {Sangouard}}]{Frowis2018Macroscopic}%
  \BibitemOpen
  \bibfield  {author} {\bibinfo {author} {\bibfnamefont {F.}~\bibnamefont {Fr\"owis}}, \bibinfo {author} {\bibfnamefont {P.}~\bibnamefont {Sekatski}}, \bibinfo {author} {\bibfnamefont {W.}~\bibnamefont {D\"ur}}, \bibinfo {author} {\bibfnamefont {N.}~\bibnamefont {Gisin}},\ and\ \bibinfo {author} {\bibfnamefont {N.}~\bibnamefont {Sangouard}},\ }\bibfield  {title} {\bibinfo {title} {{Macroscopic quantum states: Measures, fragility, and implementations}},\ }\href {https://doi.org/10.1103/RevModPhys.90.025004} {\bibfield  {journal} {\bibinfo  {journal} {Rev. Mod. Phys.}\ }\textbf {\bibinfo {volume} {90}},\ \bibinfo {pages} {025004} (\bibinfo {year} {2018})}\BibitemShut {NoStop}%
\bibitem [{\citenamefont {Chu}\ \emph {et~al.}(2023)\citenamefont {Chu}, \citenamefont {Li},\ and\ \citenamefont {Cai}}]{Chu2023Strong}%
  \BibitemOpen
  \bibfield  {author} {\bibinfo {author} {\bibfnamefont {Y.}~\bibnamefont {Chu}}, \bibinfo {author} {\bibfnamefont {X.}~\bibnamefont {Li}},\ and\ \bibinfo {author} {\bibfnamefont {J.}~\bibnamefont {Cai}},\ }\bibfield  {title} {\bibinfo {title} {{Strong Quantum Metrological Limit from Many-Body Physics}},\ }\href {https://doi.org/10.1103/PhysRevLett.130.170801} {\bibfield  {journal} {\bibinfo  {journal} {Phys. Rev. Lett.}\ }\textbf {\bibinfo {volume} {130}},\ \bibinfo {pages} {170801} (\bibinfo {year} {2023})}\BibitemShut {NoStop}%
\bibitem [{\citenamefont {Alba}\ and\ \citenamefont {Fagotti}(2017)}]{Alba2017Prethermalization}%
  \BibitemOpen
  \bibfield  {author} {\bibinfo {author} {\bibfnamefont {V.}~\bibnamefont {Alba}}\ and\ \bibinfo {author} {\bibfnamefont {M.}~\bibnamefont {Fagotti}},\ }\bibfield  {title} {\bibinfo {title} {{Prethermalization at Low Temperature: The Scent of Long-Range Order}},\ }\href {https://doi.org/10.1103/PhysRevLett.119.010601} {\bibfield  {journal} {\bibinfo  {journal} {Phys. Rev. Lett.}\ }\textbf {\bibinfo {volume} {119}},\ \bibinfo {pages} {010601} (\bibinfo {year} {2017})}\BibitemShut {NoStop}%
\bibitem [{\citenamefont {Pappalardi}\ \emph {et~al.}(2017)\citenamefont {Pappalardi}, \citenamefont {Russomanno}, \citenamefont {Silva},\ and\ \citenamefont {Fazio}}]{Pappalardi2017Multipartite}%
  \BibitemOpen
  \bibfield  {author} {\bibinfo {author} {\bibfnamefont {S.}~\bibnamefont {Pappalardi}}, \bibinfo {author} {\bibfnamefont {A.}~\bibnamefont {Russomanno}}, \bibinfo {author} {\bibfnamefont {A.}~\bibnamefont {Silva}},\ and\ \bibinfo {author} {\bibfnamefont {R.}~\bibnamefont {Fazio}},\ }\bibfield  {title} {\bibinfo {title} {{Multipartite entanglement after a quantum quench}},\ }\href {https://doi.org/10.1088/1742-5468/aa6809} {\bibfield  {journal} {\bibinfo  {journal} {Journal of Statistical Mechanics: Theory and Experiment}\ }\textbf {\bibinfo {volume} {2017}},\ \bibinfo {pages} {053104} (\bibinfo {year} {2017})}\BibitemShut {NoStop}%
\bibitem [{\citenamefont {Collura}(2019)}]{Collura2019Relaxation}%
  \BibitemOpen
  \bibfield  {author} {\bibinfo {author} {\bibfnamefont {M.}~\bibnamefont {Collura}},\ }\bibfield  {title} {\bibinfo {title} {{Relaxation of the order-parameter statistics in the Ising quantum chain}},\ }\href {https://doi.org/10.21468/SciPostPhys.7.6.072} {\bibfield  {journal} {\bibinfo  {journal} {SciPost Phys.}\ }\textbf {\bibinfo {volume} {7}},\ \bibinfo {pages} {072} (\bibinfo {year} {2019})}\BibitemShut {NoStop}%
\bibitem [{\citenamefont {Fagotti}(2024)}]{Fagotti2024Quantum}%
  \BibitemOpen
  \bibfield  {author} {\bibinfo {author} {\bibfnamefont {M.}~\bibnamefont {Fagotti}},\ }\bibfield  {title} {\bibinfo {title} {{Quantum Jamming Brings Quantum Mechanics to Macroscopic Scales}},\ }\href {https://doi.org/10.1103/PhysRevX.14.021015} {\bibfield  {journal} {\bibinfo  {journal} {Phys. Rev. X}\ }\textbf {\bibinfo {volume} {14}},\ \bibinfo {pages} {021015} (\bibinfo {year} {2024})}\BibitemShut {NoStop}%
\bibitem [{\citenamefont {Bocini}\ and\ \citenamefont {Fagotti}(2024{\natexlab{a}})}]{Bocini2023Growing}%
  \BibitemOpen
  \bibfield  {author} {\bibinfo {author} {\bibfnamefont {S.}~\bibnamefont {Bocini}}\ and\ \bibinfo {author} {\bibfnamefont {M.}~\bibnamefont {Fagotti}},\ }\bibfield  {title} {\bibinfo {title} {{Growing Schr\"odinger's cat states by local unitary time evolution of product states}},\ }\href {https://doi.org/10.1103/PhysRevResearch.6.033108} {\bibfield  {journal} {\bibinfo  {journal} {Phys. Rev. Res.}\ }\textbf {\bibinfo {volume} {6}},\ \bibinfo {pages} {033108} (\bibinfo {year} {2024}{\natexlab{a}})}\BibitemShut {NoStop}%
\bibitem [{\citenamefont {Fagotti}(2022)}]{Fagotti2022Global}%
  \BibitemOpen
  \bibfield  {author} {\bibinfo {author} {\bibfnamefont {M.}~\bibnamefont {Fagotti}},\ }\bibfield  {title} {\bibinfo {title} {{Global Quenches after Localized Perturbations}},\ }\href {https://doi.org/10.1103/PhysRevLett.128.110602} {\bibfield  {journal} {\bibinfo  {journal} {Phys. Rev. Lett.}\ }\textbf {\bibinfo {volume} {128}},\ \bibinfo {pages} {110602} (\bibinfo {year} {2022})}\BibitemShut {NoStop}%
\bibitem [{\citenamefont {Mari\'c}\ \emph {et~al.}(2024{\natexlab{a}})\citenamefont {Mari\'c}, \citenamefont {Ferro},\ and\ \citenamefont {Fagotti}}]{Maric2024Macroscopic}%
  \BibitemOpen
  \bibfield  {author} {\bibinfo {author} {\bibfnamefont {V.}~\bibnamefont {Mari\'c}}, \bibinfo {author} {\bibfnamefont {F.}~\bibnamefont {Ferro}},\ and\ \bibinfo {author} {\bibfnamefont {M.}~\bibnamefont {Fagotti}},\ }\bibfield  {title} {\bibinfo {title} {{Macroscopic Quantum States and Universal Correlations in a Disorder-Order Interface Propagating over a 1D Ground State}},\ }\Eprint {https://arxiv.org/abs/2410.10645} {arXiv:2410.10645}  (\bibinfo {year} {2024}{\natexlab{a}})\BibitemShut {NoStop}%
\bibitem [{\citenamefont {Mari\'c}\ \emph {et~al.}(2024{\natexlab{b}})\citenamefont {Mari\'c}, \citenamefont {Ferro},\ and\ \citenamefont {Fagotti}}]{Maric2024Disorder}%
  \BibitemOpen
  \bibfield  {author} {\bibinfo {author} {\bibfnamefont {V.}~\bibnamefont {Mari\'c}}, \bibinfo {author} {\bibfnamefont {F.}~\bibnamefont {Ferro}},\ and\ \bibinfo {author} {\bibfnamefont {M.}~\bibnamefont {Fagotti}},\ }\bibfield  {title} {\bibinfo {title} {{Disorder-Order Interface Propagating over the Ferromagnetic Ground State in the Transverse Field Ising Chain}},\ }\Eprint {https://arxiv.org/abs/2411.04089} {arXiv:2411.04089}  (\bibinfo {year} {2024}{\natexlab{b}})\BibitemShut {NoStop}%
\bibitem [{\citenamefont {Capizzi}\ and\ \citenamefont {Mazzoni}(2024)}]{capizzi2024entanglementcontentkinkexcitations}%
  \BibitemOpen
  \bibfield  {author} {\bibinfo {author} {\bibfnamefont {L.}~\bibnamefont {Capizzi}}\ and\ \bibinfo {author} {\bibfnamefont {M.}~\bibnamefont {Mazzoni}},\ }\bibfield  {title} {\bibinfo {title} {Entanglement content of kink excitations},\ }\Eprint {https://arxiv.org/abs/2409.03048} {arXiv:2409.03048 [cond-mat.str-el]}  (\bibinfo {year} {2024})\BibitemShut {NoStop}%
\bibitem [{\citenamefont {Lieb}\ and\ \citenamefont {Robinson}(1972)}]{Lieb1972The}%
  \BibitemOpen
  \bibfield  {author} {\bibinfo {author} {\bibfnamefont {E.~H.}\ \bibnamefont {Lieb}}\ and\ \bibinfo {author} {\bibfnamefont {D.~W.}\ \bibnamefont {Robinson}},\ }\bibfield  {title} {\bibinfo {title} {{The finite group velocity of quantum spin systems}},\ }\href {https://doi.org/10.1007/BF01645779} {\bibfield  {journal} {\bibinfo  {journal} {Commun. Math. Phys.}\ }\textbf {\bibinfo {volume} {28}},\ \bibinfo {pages} {251} (\bibinfo {year} {1972})}\BibitemShut {NoStop}%
\bibitem [{\citenamefont {Alba}\ and\ \citenamefont {Calabrese}(2017{\natexlab{b}})}]{Alba_2017}%
  \BibitemOpen
  \bibfield  {author} {\bibinfo {author} {\bibfnamefont {V.}~\bibnamefont {Alba}}\ and\ \bibinfo {author} {\bibfnamefont {P.}~\bibnamefont {Calabrese}},\ }\bibfield  {title} {\bibinfo {title} {Entanglement and thermodynamics after a quantum quench in integrable systems},\ }\href {https://doi.org/10.1073/pnas.1703516114} {\bibfield  {journal} {\bibinfo  {journal} {Proceedings of the National Academy of Sciences}\ }\textbf {\bibinfo {volume} {114}},\ \bibinfo {pages} {7947–7951} (\bibinfo {year} {2017}{\natexlab{b}})}\BibitemShut {NoStop}%
\bibitem [{\citenamefont {Piroli}\ \emph {et~al.}(2019)\citenamefont {Piroli}, \citenamefont {Vernier}, \citenamefont {Calabrese},\ and\ \citenamefont {Pozsgay}}]{Piroli_2019}%
  \BibitemOpen
  \bibfield  {author} {\bibinfo {author} {\bibfnamefont {L.}~\bibnamefont {Piroli}}, \bibinfo {author} {\bibfnamefont {E.}~\bibnamefont {Vernier}}, \bibinfo {author} {\bibfnamefont {P.}~\bibnamefont {Calabrese}},\ and\ \bibinfo {author} {\bibfnamefont {B.}~\bibnamefont {Pozsgay}},\ }\bibfield  {title} {\bibinfo {title} {Integrable quenches in nested spin chains i: the exact steady states},\ }\href {https://doi.org/10.1088/1742-5468/ab1c51} {\bibfield  {journal} {\bibinfo  {journal} {Journal of Statistical Mechanics: Theory and Experiment}\ }\textbf {\bibinfo {volume} {2019}},\ \bibinfo {pages} {063103} (\bibinfo {year} {2019})}\BibitemShut {NoStop}%
\bibitem [{\citenamefont {Calabrese}\ \emph {et~al.}(2012{\natexlab{b}})\citenamefont {Calabrese}, \citenamefont {Essler},\ and\ \citenamefont {Fagotti}}]{Calabrese2012Quantum1}%
  \BibitemOpen
  \bibfield  {author} {\bibinfo {author} {\bibfnamefont {P.}~\bibnamefont {Calabrese}}, \bibinfo {author} {\bibfnamefont {F.~H.~L.}\ \bibnamefont {Essler}},\ and\ \bibinfo {author} {\bibfnamefont {M.}~\bibnamefont {Fagotti}},\ }\bibfield  {title} {\bibinfo {title} {{Quantum quench in the transverse field Ising chain: I. Time evolution of order parameter correlators}},\ }\href {https://doi.org/10.1088/1742-5468/2012/07/P07016} {\bibfield  {journal} {\bibinfo  {journal} {Journal of Statistical Mechanics: Theory and Experiment}\ }\textbf {\bibinfo {volume} {2012}},\ \bibinfo {pages} {P07016} (\bibinfo {year} {2012}{\natexlab{b}})}\BibitemShut {NoStop}%
\bibitem [{\citenamefont {Granet}\ \emph {et~al.}(2020)\citenamefont {Granet}, \citenamefont {Fagotti},\ and\ \citenamefont {Essler}}]{Granet2020Finite}%
  \BibitemOpen
  \bibfield  {author} {\bibinfo {author} {\bibfnamefont {E.}~\bibnamefont {Granet}}, \bibinfo {author} {\bibfnamefont {M.}~\bibnamefont {Fagotti}},\ and\ \bibinfo {author} {\bibfnamefont {F.~H.~L.}\ \bibnamefont {Essler}},\ }\bibfield  {title} {\bibinfo {title} {{Finite temperature and quench dynamics in the Transverse Field Ising Model from form factor expansions}},\ }\href {https://doi.org/10.21468/SciPostPhys.9.3.033} {\bibfield  {journal} {\bibinfo  {journal} {SciPost Phys.}\ }\textbf {\bibinfo {volume} {9}},\ \bibinfo {pages} {033} (\bibinfo {year} {2020})}\BibitemShut {NoStop}%
\bibitem [{\citenamefont {Fagotti}\ and\ \citenamefont {Calabrese}(2010)}]{Fagotti2010disjoint}%
  \BibitemOpen
  \bibfield  {author} {\bibinfo {author} {\bibfnamefont {M.}~\bibnamefont {Fagotti}}\ and\ \bibinfo {author} {\bibfnamefont {P.}~\bibnamefont {Calabrese}},\ }\bibfield  {title} {\bibinfo {title} {{Entanglement entropy of two disjoint blocks in XY chains}},\ }\href {https://doi.org/10.1088/1742-5468/2010/04/p04016} {\bibfield  {journal} {\bibinfo  {journal} {Journal of Statistical Mechanics: Theory and Experiment}\ }\textbf {\bibinfo {volume} {2010}},\ \bibinfo {pages} {P04016} (\bibinfo {year} {2010})}\BibitemShut {NoStop}%
\bibitem [{\citenamefont {Sachdev}\ and\ \citenamefont {Young}(1997)}]{PhysRevLett.78.2220}%
  \BibitemOpen
  \bibfield  {author} {\bibinfo {author} {\bibfnamefont {S.}~\bibnamefont {Sachdev}}\ and\ \bibinfo {author} {\bibfnamefont {A.~P.}\ \bibnamefont {Young}},\ }\bibfield  {title} {\bibinfo {title} {{Low Temperature Relaxational Dynamics of the Ising Chain in a Transverse Field}},\ }\href {https://doi.org/10.1103/PhysRevLett.78.2220} {\bibfield  {journal} {\bibinfo  {journal} {Phys. Rev. Lett.}\ }\textbf {\bibinfo {volume} {78}},\ \bibinfo {pages} {2220} (\bibinfo {year} {1997})}\BibitemShut {NoStop}%
\bibitem [{\citenamefont {Bocini}\ and\ \citenamefont {Fagotti}(2024{\natexlab{b}})}]{Bocini2024No}%
  \BibitemOpen
  \bibfield  {author} {\bibinfo {author} {\bibfnamefont {S.}~\bibnamefont {Bocini}}\ and\ \bibinfo {author} {\bibfnamefont {M.}~\bibnamefont {Fagotti}},\ }\bibfield  {title} {\bibinfo {title} {{No eigenstate of the critical transverse-field Ising chain satisfies the area law}},\ }\href {https://doi.org/10.1103/PhysRevB.109.L201116} {\bibfield  {journal} {\bibinfo  {journal} {Phys. Rev. B}\ }\textbf {\bibinfo {volume} {109}},\ \bibinfo {pages} {L201116} (\bibinfo {year} {2024}{\natexlab{b}})}\BibitemShut {NoStop}%
\bibitem [{\citenamefont {Rieger}\ and\ \citenamefont {Iglói}(2011)}]{Rieger_2011}%
  \BibitemOpen
  \bibfield  {author} {\bibinfo {author} {\bibfnamefont {H.}~\bibnamefont {Rieger}}\ and\ \bibinfo {author} {\bibfnamefont {F.}~\bibnamefont {Iglói}},\ }\bibfield  {title} {\bibinfo {title} {Semiclassical theory for quantum quenches in finite transverse ising chains},\ }\bibfield  {journal} {\bibinfo  {journal} {Physical Review B}\ }\textbf {\bibinfo {volume} {84}},\ \href {https://doi.org/10.1103/physrevb.84.165117} {10.1103/physrevb.84.165117} (\bibinfo {year} {2011})\BibitemShut {NoStop}%
\bibitem [{\citenamefont {Eisler}\ \emph {et~al.}(2016)\citenamefont {Eisler}, \citenamefont {Maislinger},\ and\ \citenamefont {Evertz}}]{Eisler2016Universal}%
  \BibitemOpen
  \bibfield  {author} {\bibinfo {author} {\bibfnamefont {V.}~\bibnamefont {Eisler}}, \bibinfo {author} {\bibfnamefont {F.}~\bibnamefont {Maislinger}},\ and\ \bibinfo {author} {\bibfnamefont {H.~G.}\ \bibnamefont {Evertz}},\ }\bibfield  {title} {\bibinfo {title} {Universal front propagation in the quantum ising chain with domain-wall initial states},\ }\bibfield  {journal} {\bibinfo  {journal} {SciPost Physics}\ }\textbf {\bibinfo {volume} {1}},\ \href {https://doi.org/10.21468/scipostphys.1.2.014} {10.21468/scipostphys.1.2.014} (\bibinfo {year} {2016})\BibitemShut {NoStop}%
\bibitem [{\citenamefont {Bethe}(1931)}]{Bethe1931}%
  \BibitemOpen
  \bibfield  {author} {\bibinfo {author} {\bibfnamefont {H.}~\bibnamefont {Bethe}},\ }\bibfield  {title} {\bibinfo {title} {{Zur Theorie der Metalle}},\ }\href {https://doi.org/10.1007/BF01341708} {\bibfield  {journal} {\bibinfo  {journal} {Zeitschrift f{\"u}r Physik}\ }\textbf {\bibinfo {volume} {71}},\ \bibinfo {pages} {205} (\bibinfo {year} {1931})}\BibitemShut {NoStop}%
\bibitem [{\citenamefont {Essler}\ \emph {et~al.}(2005)\citenamefont {Essler}, \citenamefont {Frahm}, \citenamefont {Göhmann}, \citenamefont {Klümper},\ and\ \citenamefont {Korepin}}]{Essler2005book}%
  \BibitemOpen
  \bibfield  {author} {\bibinfo {author} {\bibfnamefont {F.~H.~L.}\ \bibnamefont {Essler}}, \bibinfo {author} {\bibfnamefont {H.}~\bibnamefont {Frahm}}, \bibinfo {author} {\bibfnamefont {F.}~\bibnamefont {Göhmann}}, \bibinfo {author} {\bibfnamefont {A.}~\bibnamefont {Klümper}},\ and\ \bibinfo {author} {\bibfnamefont {V.~E.}\ \bibnamefont {Korepin}},\ }\href@noop {} {\emph {\bibinfo {title} {The One-Dimensional Hubbard Model}}}\ (\bibinfo  {publisher} {Cambridge University Press},\ \bibinfo {year} {2005})\BibitemShut {NoStop}%
\bibitem [{\citenamefont {Essler}\ and\ \citenamefont {Fagotti}(2016)}]{Essler2016Quench}%
  \BibitemOpen
  \bibfield  {author} {\bibinfo {author} {\bibfnamefont {F.~H.~L.}\ \bibnamefont {Essler}}\ and\ \bibinfo {author} {\bibfnamefont {M.}~\bibnamefont {Fagotti}},\ }\bibfield  {title} {\bibinfo {title} {{Quench dynamics and relaxation in isolated integrable quantum spin chains}},\ }\href {https://doi.org/10.1088/1742-5468/2016/06/064002} {\bibfield  {journal} {\bibinfo  {journal} {Journal of Statistical Mechanics: Theory and Experiment}\ }\textbf {\bibinfo {volume} {2016}},\ \bibinfo {pages} {064002} (\bibinfo {year} {2016})}\BibitemShut {NoStop}%
\bibitem [{\citenamefont {Evangelisti}(2013)}]{Evangelisti2013Semi-classical}%
  \BibitemOpen
  \bibfield  {author} {\bibinfo {author} {\bibfnamefont {S.}~\bibnamefont {Evangelisti}},\ }\bibfield  {title} {\bibinfo {title} {{Semi-classical theory for quantum quenches in the O(3) non-linear sigma model}},\ }\href {https://doi.org/10.1088/1742-5468/2013/04/P04003} {\bibfield  {journal} {\bibinfo  {journal} {Journal of Statistical Mechanics: Theory and Experiment}\ }\textbf {\bibinfo {volume} {2013}},\ \bibinfo {pages} {P04003} (\bibinfo {year} {2013})}\BibitemShut {NoStop}%
\bibitem [{\citenamefont {Kormos}\ and\ \citenamefont {Zar\'and}(2016)}]{Kormos2016Quantum}%
  \BibitemOpen
  \bibfield  {author} {\bibinfo {author} {\bibfnamefont {M.}~\bibnamefont {Kormos}}\ and\ \bibinfo {author} {\bibfnamefont {G.}~\bibnamefont {Zar\'and}},\ }\bibfield  {title} {\bibinfo {title} {{Quantum quenches in the sine-Gordon model: A semiclassical approach}},\ }\href {https://doi.org/10.1103/PhysRevE.93.062101} {\bibfield  {journal} {\bibinfo  {journal} {Phys. Rev. E}\ }\textbf {\bibinfo {volume} {93}},\ \bibinfo {pages} {062101} (\bibinfo {year} {2016})}\BibitemShut {NoStop}%
\bibitem [{\citenamefont {Moca}\ \emph {et~al.}(2017)\citenamefont {Moca}, \citenamefont {Kormos},\ and\ \citenamefont {Zar\'and}}]{Hybrid2017Moca}%
  \BibitemOpen
  \bibfield  {author} {\bibinfo {author} {\bibfnamefont {C.~P.}\ \bibnamefont {Moca}}, \bibinfo {author} {\bibfnamefont {M.}~\bibnamefont {Kormos}},\ and\ \bibinfo {author} {\bibfnamefont {G.}~\bibnamefont {Zar\'and}},\ }\bibfield  {title} {\bibinfo {title} {{Hybrid Semiclassical Theory of Quantum Quenches in One-Dimensional Systems}},\ }\href {https://doi.org/10.1103/PhysRevLett.119.100603} {\bibfield  {journal} {\bibinfo  {journal} {Phys. Rev. Lett.}\ }\textbf {\bibinfo {volume} {119}},\ \bibinfo {pages} {100603} (\bibinfo {year} {2017})}\BibitemShut {NoStop}%
\bibitem [{\citenamefont {Kormos}\ \emph {et~al.}(2018)\citenamefont {Kormos}, \citenamefont {Moca},\ and\ \citenamefont {Zar\'and}}]{Kormos2018Semiclassical}%
  \BibitemOpen
  \bibfield  {author} {\bibinfo {author} {\bibfnamefont {M.}~\bibnamefont {Kormos}}, \bibinfo {author} {\bibfnamefont {C.~P.}\ \bibnamefont {Moca}},\ and\ \bibinfo {author} {\bibfnamefont {G.}~\bibnamefont {Zar\'and}},\ }\bibfield  {title} {\bibinfo {title} {{Semiclassical theory of front propagation and front equilibration following an inhomogeneous quantum quench}},\ }\href {https://doi.org/10.1103/PhysRevE.98.032105} {\bibfield  {journal} {\bibinfo  {journal} {Phys. Rev. E}\ }\textbf {\bibinfo {volume} {98}},\ \bibinfo {pages} {032105} (\bibinfo {year} {2018})}\BibitemShut {NoStop}%
\bibitem [{\citenamefont {Bertini}\ \emph {et~al.}(2019)\citenamefont {Bertini}, \citenamefont {Piroli},\ and\ \citenamefont {Kormos}}]{Bertini2019Transport}%
  \BibitemOpen
  \bibfield  {author} {\bibinfo {author} {\bibfnamefont {B.}~\bibnamefont {Bertini}}, \bibinfo {author} {\bibfnamefont {L.}~\bibnamefont {Piroli}},\ and\ \bibinfo {author} {\bibfnamefont {M.}~\bibnamefont {Kormos}},\ }\bibfield  {title} {\bibinfo {title} {{Transport in the sine-Gordon field theory: From generalized hydrodynamics to semiclassics}},\ }\href {https://doi.org/10.1103/PhysRevB.100.035108} {\bibfield  {journal} {\bibinfo  {journal} {Phys. Rev. B}\ }\textbf {\bibinfo {volume} {100}},\ \bibinfo {pages} {035108} (\bibinfo {year} {2019})}\BibitemShut {NoStop}%
\bibitem [{\citenamefont {Horváth}\ \emph {et~al.}(2022)\citenamefont {Horváth}, \citenamefont {Sotiriadis}, \citenamefont {Kormos},\ and\ \citenamefont {Takács}}]{Horvath2022Inhomogeneous}%
  \BibitemOpen
  \bibfield  {author} {\bibinfo {author} {\bibfnamefont {D.~X.}\ \bibnamefont {Horváth}}, \bibinfo {author} {\bibfnamefont {S.}~\bibnamefont {Sotiriadis}}, \bibinfo {author} {\bibfnamefont {M.}~\bibnamefont {Kormos}},\ and\ \bibinfo {author} {\bibfnamefont {G.}~\bibnamefont {Takács}},\ }\bibfield  {title} {\bibinfo {title} {{Inhomogeneous quantum quenches in the sine-Gordon theory}},\ }\href {https://doi.org/10.21468/SciPostPhys.12.5.144} {\bibfield  {journal} {\bibinfo  {journal} {SciPost Phys.}\ }\textbf {\bibinfo {volume} {12}},\ \bibinfo {pages} {144} (\bibinfo {year} {2022})}\BibitemShut {NoStop}%
\bibitem [{\citenamefont {Calabrese}\ and\ \citenamefont {Cardy}(2005)}]{Calabrese_2005}%
  \BibitemOpen
  \bibfield  {author} {\bibinfo {author} {\bibfnamefont {P.}~\bibnamefont {Calabrese}}\ and\ \bibinfo {author} {\bibfnamefont {J.}~\bibnamefont {Cardy}},\ }\bibfield  {title} {\bibinfo {title} {Evolution of entanglement entropy in one-dimensional systems},\ }\href {https://doi.org/10.1088/1742-5468/2005/04/p04010} {\bibfield  {journal} {\bibinfo  {journal} {Journal of Statistical Mechanics: Theory and Experiment}\ }\textbf {\bibinfo {volume} {2005}},\ \bibinfo {pages} {P04010} (\bibinfo {year} {2005})}\BibitemShut {NoStop}%
\bibitem [{\citenamefont {Cotler}\ \emph {et~al.}(2016)\citenamefont {Cotler}, \citenamefont {Hertzberg}, \citenamefont {Mezei},\ and\ \citenamefont {Mueller}}]{Cotler_2016}%
  \BibitemOpen
  \bibfield  {author} {\bibinfo {author} {\bibfnamefont {J.~S.}\ \bibnamefont {Cotler}}, \bibinfo {author} {\bibfnamefont {M.~P.}\ \bibnamefont {Hertzberg}}, \bibinfo {author} {\bibfnamefont {M.}~\bibnamefont {Mezei}},\ and\ \bibinfo {author} {\bibfnamefont {M.~T.}\ \bibnamefont {Mueller}},\ }\bibfield  {title} {\bibinfo {title} {Entanglement growth after a global quench in free scalar field theory},\ }\bibfield  {journal} {\bibinfo  {journal} {Journal of High Energy Physics}\ }\textbf {\bibinfo {volume} {2016}},\ \href {https://doi.org/10.1007/jhep11(2016)166} {10.1007/jhep11(2016)166} (\bibinfo {year} {2016})\BibitemShut {NoStop}%
\bibitem [{\citenamefont {Kormos}\ \emph {et~al.}(2016)\citenamefont {Kormos}, \citenamefont {Collura}, \citenamefont {Takács},\ and\ \citenamefont {Calabrese}}]{Kormos_2016}%
  \BibitemOpen
  \bibfield  {author} {\bibinfo {author} {\bibfnamefont {M.}~\bibnamefont {Kormos}}, \bibinfo {author} {\bibfnamefont {M.}~\bibnamefont {Collura}}, \bibinfo {author} {\bibfnamefont {G.}~\bibnamefont {Takács}},\ and\ \bibinfo {author} {\bibfnamefont {P.}~\bibnamefont {Calabrese}},\ }\bibfield  {title} {\bibinfo {title} {Real-time confinement following a quantum quench to a non-integrable model},\ }\href {https://doi.org/10.1038/nphys3934} {\bibfield  {journal} {\bibinfo  {journal} {Nature Physics}\ }\textbf {\bibinfo {volume} {13}},\ \bibinfo {pages} {246–249} (\bibinfo {year} {2016})}\BibitemShut {NoStop}%
\bibitem [{\citenamefont {Casini}\ \emph {et~al.}(2016)\citenamefont {Casini}, \citenamefont {Liu},\ and\ \citenamefont {Mezei}}]{Casini_2016}%
  \BibitemOpen
  \bibfield  {author} {\bibinfo {author} {\bibfnamefont {H.}~\bibnamefont {Casini}}, \bibinfo {author} {\bibfnamefont {H.}~\bibnamefont {Liu}},\ and\ \bibinfo {author} {\bibfnamefont {M.}~\bibnamefont {Mezei}},\ }\bibfield  {title} {\bibinfo {title} {Spread of entanglement and causality},\ }\bibfield  {journal} {\bibinfo  {journal} {Journal of High Energy Physics}\ }\textbf {\bibinfo {volume} {2016}},\ \href {https://doi.org/10.1007/jhep07(2016)077} {10.1007/jhep07(2016)077} (\bibinfo {year} {2016})\BibitemShut {NoStop}%
\bibitem [{\citenamefont {Bertini}\ \emph {et~al.}(2018{\natexlab{a}})\citenamefont {Bertini}, \citenamefont {Tartaglia},\ and\ \citenamefont {Calabrese}}]{Bertini_2018}%
  \BibitemOpen
  \bibfield  {author} {\bibinfo {author} {\bibfnamefont {B.}~\bibnamefont {Bertini}}, \bibinfo {author} {\bibfnamefont {E.}~\bibnamefont {Tartaglia}},\ and\ \bibinfo {author} {\bibfnamefont {P.}~\bibnamefont {Calabrese}},\ }\bibfield  {title} {\bibinfo {title} {Entanglement and diagonal entropies after a quench with no pair structure},\ }\href {https://doi.org/10.1088/1742-5468/aac73f} {\bibfield  {journal} {\bibinfo  {journal} {Journal of Statistical Mechanics: Theory and Experiment}\ }\textbf {\bibinfo {volume} {2018}},\ \bibinfo {pages} {063104} (\bibinfo {year} {2018}{\natexlab{a}})}\BibitemShut {NoStop}%
\bibitem [{\citenamefont {Bastianello}\ and\ \citenamefont {Collura}(2020)}]{Bastianello2020Entanglement}%
  \BibitemOpen
  \bibfield  {author} {\bibinfo {author} {\bibfnamefont {A.}~\bibnamefont {Bastianello}}\ and\ \bibinfo {author} {\bibfnamefont {M.}~\bibnamefont {Collura}},\ }\bibfield  {title} {\bibinfo {title} {Entanglement spreading and quasiparticle picture beyond the pair structure},\ }\bibfield  {journal} {\bibinfo  {journal} {SciPost Physics}\ }\textbf {\bibinfo {volume} {8}},\ \href {https://doi.org/10.21468/scipostphys.8.3.045} {10.21468/scipostphys.8.3.045} (\bibinfo {year} {2020})\BibitemShut {NoStop}%
\bibitem [{\citenamefont {Bertini}\ \emph {et~al.}(2018{\natexlab{b}})\citenamefont {Bertini}, \citenamefont {Fagotti}, \citenamefont {Piroli},\ and\ \citenamefont {Calabrese}}]{Bertini2018Entanglement}%
  \BibitemOpen
  \bibfield  {author} {\bibinfo {author} {\bibfnamefont {B.}~\bibnamefont {Bertini}}, \bibinfo {author} {\bibfnamefont {M.}~\bibnamefont {Fagotti}}, \bibinfo {author} {\bibfnamefont {L.}~\bibnamefont {Piroli}},\ and\ \bibinfo {author} {\bibfnamefont {P.}~\bibnamefont {Calabrese}},\ }\bibfield  {title} {\bibinfo {title} {{Entanglement evolution and generalised hydrodynamics: noninteracting systems}},\ }\href {https://doi.org/10.1088/1751-8121/aad82e} {\bibfield  {journal} {\bibinfo  {journal} {Journal of Physics A: Mathematical and Theoretical}\ }\textbf {\bibinfo {volume} {51}},\ \bibinfo {pages} {39LT01} (\bibinfo {year} {2018}{\natexlab{b}})}\BibitemShut {NoStop}%
\bibitem [{\citenamefont {Desaules}\ \emph {et~al.}(2022)\citenamefont {Desaules}, \citenamefont {Pietracaprina}, \citenamefont {Papi\ifmmode~\acute{c}\else \'{c}\fi{}}, \citenamefont {Goold},\ and\ \citenamefont {Pappalardi}}]{Desaules2022Extensive}%
  \BibitemOpen
  \bibfield  {author} {\bibinfo {author} {\bibfnamefont {J.-Y.}\ \bibnamefont {Desaules}}, \bibinfo {author} {\bibfnamefont {F.}~\bibnamefont {Pietracaprina}}, \bibinfo {author} {\bibfnamefont {Z.}~\bibnamefont {Papi\ifmmode~\acute{c}\else \'{c}\fi{}}}, \bibinfo {author} {\bibfnamefont {J.}~\bibnamefont {Goold}},\ and\ \bibinfo {author} {\bibfnamefont {S.}~\bibnamefont {Pappalardi}},\ }\bibfield  {title} {\bibinfo {title} {{Extensive Multipartite Entanglement from su(2) Quantum Many-Body Scars}},\ }\href {https://doi.org/10.1103/PhysRevLett.129.020601} {\bibfield  {journal} {\bibinfo  {journal} {Phys. Rev. Lett.}\ }\textbf {\bibinfo {volume} {129}},\ \bibinfo {pages} {020601} (\bibinfo {year} {2022})}\BibitemShut {NoStop}%
\bibitem [{\citenamefont {Balducci}\ \emph {et~al.}(2022)\citenamefont {Balducci}, \citenamefont {Gambassi}, \citenamefont {Lerose}, \citenamefont {Scardicchio},\ and\ \citenamefont {Vanoni}}]{Balducci_2022}%
  \BibitemOpen
  \bibfield  {author} {\bibinfo {author} {\bibfnamefont {F.}~\bibnamefont {Balducci}}, \bibinfo {author} {\bibfnamefont {A.}~\bibnamefont {Gambassi}}, \bibinfo {author} {\bibfnamefont {A.}~\bibnamefont {Lerose}}, \bibinfo {author} {\bibfnamefont {A.}~\bibnamefont {Scardicchio}},\ and\ \bibinfo {author} {\bibfnamefont {C.}~\bibnamefont {Vanoni}},\ }\bibfield  {title} {\bibinfo {title} {Localization and melting of interfaces in the two-dimensional quantum ising model},\ }\bibfield  {journal} {\bibinfo  {journal} {Physical Review Letters}\ }\textbf {\bibinfo {volume} {129}},\ \href {https://doi.org/10.1103/physrevlett.129.120601} {10.1103/physrevlett.129.120601} (\bibinfo {year} {2022})\BibitemShut {NoStop}%
\bibitem [{\citenamefont {Balducci}\ \emph {et~al.}(2023)\citenamefont {Balducci}, \citenamefont {Gambassi}, \citenamefont {Lerose}, \citenamefont {Scardicchio},\ and\ \citenamefont {Vanoni}}]{Balducci_2023}%
  \BibitemOpen
  \bibfield  {author} {\bibinfo {author} {\bibfnamefont {F.}~\bibnamefont {Balducci}}, \bibinfo {author} {\bibfnamefont {A.}~\bibnamefont {Gambassi}}, \bibinfo {author} {\bibfnamefont {A.}~\bibnamefont {Lerose}}, \bibinfo {author} {\bibfnamefont {A.}~\bibnamefont {Scardicchio}},\ and\ \bibinfo {author} {\bibfnamefont {C.}~\bibnamefont {Vanoni}},\ }\bibfield  {title} {\bibinfo {title} {Interface dynamics in the two-dimensional quantum ising model},\ }\bibfield  {journal} {\bibinfo  {journal} {Physical Review B}\ }\textbf {\bibinfo {volume} {107}},\ \href {https://doi.org/10.1103/physrevb.107.024306} {10.1103/physrevb.107.024306} (\bibinfo {year} {2023})\BibitemShut {NoStop}%
\end{thebibliography}%
\end{document}